\documentclass[aps,preprint,floats,epsf,epsfig,nofootinbib,letter]{revtex4}
\usepackage{epsfig}
\usepackage{subfigure}

\begin{document}
\def\be{\begin{eqnarray}}
\def\en{\end{eqnarray}}
\def\non{\nonumber}
\def\ov{\overline}
\def\la{\langle}
\def\ra{\rangle}
\def\B{{\cal B}}
\def\3bar{{\bf \bar 3}}
\def\6bar{{\bf \bar 6}}
\def\10bar{{\bf \ov{10}}}
\def\pr{{\sl Phys. Rev.}~}
\def\prl{{\sl Phys. Rev. Lett.}~}
\def\pl{{\sl Phys. Lett.}~}
\def\np{{\sl Nucl. Phys.}~}
\def\zp{{\sl Z. Phys.}~}
\def\lsim{ {\ \lower-1.2pt\vbox{\hbox{\rlap{$<$}\lower5pt\vbox{\hbox{$\sim$}
}}}\ } }
\def\gsim{ {\ \lower-1.2pt\vbox{\hbox{\rlap{$>$}\lower5pt\vbox{\hbox{$\sim$}
}}}\ } }

\font\el=cmbx10 scaled \magstep2{\obeylines\hfill Dec. 2018} 

\vskip 1.5 cm

\centerline{\large\bf Color-allowed Bottom Baryon to Charmed Baryon}
\centerline{\large\bf non-leptonic Decays}

\vskip 1.2 cm

\bigskip
\bigskip
\centerline{\bf Chun-Khiang Chua}
\medskip
\medskip
\centerline{Department of Physics and Center for High Energy Physics}
\centerline{Chung Yuan Christian University}
\centerline{Chung-Li, Taiwan 320, Republic of China}

\bigskip
\bigskip
\centerline{\bf Abstract}
We study color allowed $\Lambda_b\to \Lambda^{(*,**)}_c M^-$, $\Xi_b\to\Xi_c^{(**)} M^-$ and $\Omega_b\to\Omega^{(*)}_c M^-$ decays with $M=\pi, K,\rho, K^*, D, D_s, D^*, D^*_s, a_1$, $\Lambda^{(*,**)}_c=\Lambda_c, \Lambda_c(2595), \Lambda_c(2765), \Lambda_c(2940)$, $\Xi_c^{(**)}=\Xi_c, \Xi_c(2790)$ and $\Omega^{(*)}_c=\Omega_c, \Omega_c(3090)$, in this work. There are three types of transitions, namely ${\cal B}_b({\bf \bar 3_f},1/2^+)$ to ${\cal B}_c({\bf \bar 3_f},1/2^+)$, ${\cal B}_b({\bf 6_f},1/2^+)$ to ${\cal B}_c({\bf 6_f},1/2^+)$ and ${\cal B}_b({\bf \bar 3_f},1/2^+)$ to ${\cal B}_c({\bf \bar 3_f},1/2^-)$ transitions. The bottom baryon to charmed baryon form factors are calculated using the light-front quark model. Decay rates and up-down asymmetries are predicted using na\"{i}ve factorization and can be checked experimentally. We find that in ${\cal B}_b\to {\cal B}_cP$ decays, rates in ${\cal B}_b({\bf 6_f},1/2^+)\to {\cal B}_c({\bf 6_f},1/2^+)$ [type~(ii)] transition are smaller than those in ${\cal B}_b({\bf \bar 3_f},1/2^+)\to {\cal B}_c({\bf \bar 3_f},1/2^+)$ [type (i)] transition, but similar to those in ${\cal B}_b({\bf \bar 3_f},1/2^+)\to {\cal B}_c({\bf \bar 3_f},1/2^-)$ [type (iii)] transition, while in $\B_b\to\B_c V, \B_c A$ decays, rates in ${\cal B}_b({\bf 6_f},1/2^+)\to {\cal B}_c({\bf 6_f},1/2^+)$ [type~(ii)] transition are much smaller than those in ${\cal B}_b({\bf \bar 3_f},1/2^+)\to {\cal B}_c({\bf \bar 3_f},1/2^+)$ [type (i)] transition and are also smaller than those in ${\cal B}_b({\bf \bar 3_f},1/2^+)\to {\cal B}_c({\bf \bar 3_f},1/2^-)$ [type (iii)] transition. For the up-down asymmetries, the signs are mostly negative, except for those in the $\B_b({\bf 6_f},1/2^+)\to\B_c({\bf 6_f},1/2^+)$ [type (ii)] transition. Most of these asymmetries are large in sizes. The study on these decay modes may shed light on the quantum numbers of some of the charmed baryons as the decays depend on the bottom baryon to charmed baryon form factors, which are sensitive to the configurations of the final state charmed baryons.

\bigskip
\small

\pagebreak


\section{Introduction}

As noted in the review of Particle Data Group (PDG) (see the review by C.G. Wohl in~\cite{PDG}),
there are 24 singly charmed baryons and nine singly bottom baryons.~\footnote{In the review by Wohl, particles in the same isospin-multiplet, such as $\Sigma_c^{++,+,0}$, are not counted separately.}
Among them $\Lambda_c(2864)$ and five $\Omega_c$ states, namely 
$\Omega_c(3000)^0$, 
$\Omega_c(3050)^0$, 
$\Omega_c(3066)^0$, 
$\Omega_c(3090)^0$ and
$\Omega_c(3119)^0$, 
are newly discovered by LHCb in year 2017 \cite{Aaij:2017vbw,Aaij:2017nav}.
The quantum numbers of 9 out of the 24 charmed baryons are unspecified. 
These include the above five $\Omega_c$ states, $\Sigma_c(2800)^{++,+,0}$, $\Xi_c(3055)^{+,0}$, $\Xi_c(3080)^{+,0}$ and $\Xi_c(2970)^{+,0}$ baryons.
Note that, in addition to the above states, some other states, including $\Lambda_c(2765)^+$ (or $\Sigma_c(2765)$), 
$\Xi_c(2930)^0$ and $\Xi_c(3123)^+$, 
are not included in the short review and their quantum numbers remain unspecified as well.
Furthermore PDG stated that $\frac{3}{2}^-$  is the favored quantum number of $\Lambda_c(2940)^+$, but it is not certain~\cite{PDG}, while the authors of ref.~\cite{Cheng:2017ove} argued that it should be a $\frac{1}{2}^-$ state.
It is not surprising that there are various suggestions on the quantum numbers of the newly discovered $\Omega_c$ states, see for example~\cite{Agaev:2017jyt,Chen:2017sci,Karliner:2017kfm,Wang:2017hej,Padmanath:2017lng,Wang:2017zjw,Chen:2017gnu,Cheng:2017ove}. 
It is, therefore, of great important to identify the
quantum numbers of these states and understand their
properties. 

Among low lying singly bottom baryons, only $\Lambda_b$, $\Xi_b$ and $\Omega_b$ decay weakly~\cite{PDG}.
Several color allowed $\Lambda_b\to\Lambda_c P$ decay rates with $P=\pi, K, D, D_s$ were reported by 
LHCb in year 2014~\cite{Aaij:2014jyk,Aaij:2014lpa,Aaij:2014pha}.
We expect more to come in the near future.
It will be interesting and timely to study weak decays of singly bottom baryons to final states involving singly charmed baryons. 
In general, baryon decays are complicate processes. 
Nevertheless, when the transition only involve the heavy quarks, namely $b\to c$ transition, 
while the light quarks are spectating, the decay processes are easier.  
Accordingly we will study color allowed $\Lambda^0_b\to \Lambda^{(*,**)}_c M$, $\Xi_b\to\Xi_c^{(**)} M$ and $\Omega_b\to\Omega^{(*)}_c M$ decays with $M=\pi, K,\rho, K^*,D, D_s, D^*, D^*_s, a_1$,
$\Lambda^{(*,**)}_c=\Lambda_c, \Lambda_c(2595), \Lambda_c(2765), \Lambda_c(2940)$,
$\Xi_c^{(*,**)}=\Xi_c, \Xi_c(2790)$ and
$\Omega^{(*)}_c=\Omega_c, \Omega_c(3090)$. 
In this work, we follow ref.~\cite{Cheng:2017ove} to take 
$\Lambda_c(2765)$, $\Lambda_c(2940)$
and $\Omega_c(3090)$
as
a radial excited $s$-wave $1/2^+$ state, a radial excited $p$-wave $1/2^-$ state and a radial excite $s$-wave $1/2^+$ state, respectively.
There are other quantum number assignments.
For example, as noted in the previous paragraph, PDG and LHCb prefer $3/2^-$ for quantum number of $\Lambda_c(2940)$~\cite{PDG, Aaij:2017vbw} and several authors consider $\Omega_c(3090)$ as a candidate of a $p$-wave state, usually with a spin higher than $1/2$~\cite{Chen:2017sci,Karliner:2017kfm,Wang:2017hej,Padmanath:2017lng,Wang:2017zjw,Chen:2017gnu}. 
It should be noted that some authors also consider $\Omega_c(3090)$ as a $1/2^+$ state~\cite{Agaev:2017jyt,Karliner:2017kfm}.
The study on these $\B_b\to\B_c M$ decays may shed light on the quantum numbers of $\Lambda_c(2765)$, $\Lambda_c(2940)$
and $\Omega_c(3090)$,
as the decays depend on the bottom baryon to charmed baryon form factors, which are sensitive to the configurations of the final state charmed baryons.
We will use the light-front quark model to calculate the form factors.
The formalism is similar to the one in ref.~\cite{Cheng:2004cc}, which was used to study a different problem.
For some other studies on some of the above modes or on some related form factors in various approaches, one is referred to
\cite{
Mannel:1992ti, 
Cheng97a,
Ivanov:1997ra,
Ivanov:1997hi,
Fayyazuddin:1998ap,
Mohanta:1998iu,
Giri:1997te,
Shih:1999yh,
Albertus:2004wj, 
Ke:2007tg, 
Ke:2012wa, 
Detmold:2015aaa, 
Gutsche:2018utw,
Zhao:2018zcb, 
Zhu:2018jet}.

We begin with a brief review of the spectroscopy of charmed and bottom baryon states and discuss their possible spin-parity quantum numbers and inner structure in Sec. 2. 
In Sec. 3 we work out the formulas for form factors in the light-front quark model. 
We present our numerical results for form factors, decay rates and up-down asymmetries in Sec. 4.
Sec. 5 comes to our conclusions. 
Appendix~A is prepared to give some details of the derivations of the vertex functions, 
while some discussions on the technical issue of obtaining form factors are collected in Appendix~B.

\section{Spectroscopy of singly charmed and bottom baryons}

In this section we briefly review the spectroscopy of singly charmed and bottom baryons.
Our discussion follows closely to those in \cite{Cheng:2006dk,Cheng:2015naa}.
The singly charmed or bottom baryon is composed of a charmed quark or a bottom quark and
two light quarks. 
We will discuss the allowed quantum numbers for the light quark system before the brief review.

\subsection{Allowed quantum numbers for the light quark system}

From Fermi statistics the wave function of the light quarks needs to be antisymmetry under permutation.
As the charm or bottom quark is a color triplet ${\bf 3_c}$, the diquark system, consists of the two light quarks, can only be an anti-color triplet 
${\bf \bar 3_c}$ state, which is anti-symmetric (denoted as $({\bf \bar 3_c})_A$) under permutation of the two light quarks.
The remaining part of the diquark wave function consists of 
\be
\bigg(\psi({\rm space})\times \psi({\rm flavor}) \times \psi({\rm spin})\bigg)_S,
\en
must be symmetry under permutation.

The spin of the light quarks can be in a symmetric triplet state $({\bf 3_{sp}})_S$ ($S_l=1$) or an anti-symmetric singlet state
$({\bf 1_{sp}})_A$ ($S_l=0$). Under permutation, the spin wave function picks up an phase factor
\be
\psi({\rm spin})\to (-)^{S_l+1} \psi({\rm spin}).
\en
Given that each light quark is a triplet of the flavor SU(3)
and ${\bf 3_f}\times{\bf 3_f}=({\bf \bar 3_f})_A+({\bf 6_f})_S$, 
there are two different SU(3) multiplets of charmed or bottom baryons: a symmetric sextet $({\bf 6_f})_S$ and an
antisymmetric antitriplet $({\bf \bar 3_f})_A$. 
The iso-singlet $\Lambda_Q$ and iso-doublet 
$\Xi_Q$ form a $({\bf \bar 3_f})_A$ representation,
while the $\Omega_Q$, iso-doublet $\Xi'_Q$, 
and iso-triplet $\Sigma_Q$ form a  $({\bf 6}_f)_S$ representation.~\footnote{
We have followed the Particle
Data Group's convention \cite{PDG} to use a prime to distinguish the
iso-doublet 
in the ${\bf 6_f}$ from the one in the ${\bf \bar 3_f}$.}
Under permutation, the flavor wave function picks up an phase factor
\be
\psi({\rm flavor})\to (-)^{N_f}\psi({\rm flavor}),
\en
with $N_f=3, 6$ for ${\bf \bar 3_f}$, ${\bf 6_f}$, respectively.

In the quark model, the orbital angular momentum of the light
diquark can be decomposed into ${\bf L}_\ell={\bf L}_k+{\bf
L}_K$, where ${\bf L}_k$ is the orbital angular momentum
between the two light quarks and ${\bf L}_K$ the orbital
angular momentum between the diquark (the light quark pair) and the heavy quark. 
Roughly speaking, we have
\be
\psi({\rm space})\sim R_n(|\vec K|)\times Y_{L_k m_k}(\vec k) Y_{L_K m_K}(\vec K), 
\label{eq: psi space}
\en
where $R_n$ is the radial wave function, $Y_{lm}$ is the spherical harmonics, 
$\vec k$ is basically the relative momentum of the two light quarks
and
$\vec K$ is the relative momentum of the heavy quark and the diquark system.~\footnote{
Explicitly, we have $k^\mu\equiv (p_1-p_2)^\mu- (p_1+p_2)^\mu [(p_1+p_2)\cdot (p_1-p_2)/(p_1+p_2)^2]$
and $K^\mu\equiv (p_1+p_2-p_3)^\mu-P^\mu [P\cdot (p_1+p_2-p_3)/P^2]$ 
with $p_1, p_2$ and $p_3$ the momenta of the light quarks and the heavy quark, respectively, and $P\equiv p_1+p_2+p_3$.
Note the above constructions in $K$ and $k$ are to make sure that 
in the rest frame of the light-quark system and in the whole baryon system, 
we have $k=(0,\vec k)$ and $K=(0,\vec K)$, respectively.}
In the above equation, we do not show explicitly the Clebsch-Gordan coefficient and the Wigner rotation (see later discussion), as the rest frame of the whole system and the rest frame of the diquark system are not identical. 
Nevertheless the above wave function can still be used as an book keeping devise for working out the allowed quantum numbers. 

The angular momentum of the diquark system, without taking into account the orbital momentum of the $Q-[qq]$ system, is
\be
\vec S_{[qq]}=\vec L_k+\vec S_{qq},
\en
with $S_{[qq]}$ given by $|L_k-S_{qq}|,\dots, L_k+S_{qq}$. 
Note that $\vec S_{qq}$ is the spin of the light quark pair without taking into account the orbital momentum between them.
The combination of $\vec S_{[qq]}$ is better when viewing the diquark as a sub-system, i.e. one may have scalar diquark, axial-vector diquark and so on.
The angular momentum of the diquark system, with the orbital momentum of the $Q-[qq]$ system, is
\be
\vec J_l=\vec S_{[qq]}+\vec L_K,
\en
with $J_l$ given by $|S_{[qq]}-L_k|,\dots, S_{[qq]}+L_k$.
Consequently, the total angular momentum is
\be
\vec J_{Qqq}=\vec S_Q+\vec L_k+\vec S_{qq}+\vec L_K
=\vec S_Q+\vec S_{[qq]}+\vec L_K
=\vec S_Q+\vec J_l.
\en

\begin{table}[t]
\caption{Allowed quantum number of the diquark system satisfying Fermi statistics. 
$\vec S_{[qq]}=\vec L_k+\vec S_{qq}$ is the angular momentum of the diquark system, without taking into account the orbital momentum of the $Q-[qq]$ system ($\vec L_K$).
The parity eigenvalues of the diquark $[qq]$ system and the $Qqq$ system are given by $(-)^{L_k}$ and $(-)^{L_k+L_K}$, respectively.}
 \label{tab:quantum numbers}
 \begin{center}
\begin{tabular}{| c c c  c c c c c|}
\hline
$\psi({\rm color})$
    & $L_K$
    & $L_k$
    & $\psi({\rm flavor})$
    & $\psi({\rm spin})$
    & $S_{[qq]}$ 
    & parity($[qq]$)
    & parity($Qqq$)
    \\
    \hline
$({\bf \bar 3_c})_A$    
    &even    
    & even
    & $({\bf \bar 3_f})_A$
    &  $({\bf 1_{sp}})_A$
    & $L_k$
    & $+$
    & $+$
    \\    
$({\bf \bar 3_c})_A$    
    &even
    & even
    & $({\bf 6_f})_S$
    & $({\bf 3_{sp}})_S$
    & $|L_k-1|,\cdots, L_k+1$
    & $+$
    & $+$
    \\
$({\bf \bar 3_c})_A$    
    &odd
    & odd
    & $({\bf \bar 3_f})_A$
    & $({\bf 3_{sp}})_S$
    & $L_k-1,L_k, L_k+1$
    & $-$
    & $+$
    \\    
$({\bf \bar 3_c})_A$    
    &odd
    & odd
    & $({\bf 6_f})_S$
    & $({\bf 1_{sp}})_A$
    & $L_k$
    & $-$
    & $+$
    \\ 
\hline           
$({\bf \bar 3_c})_A$    
    &even
    & odd
    & $({\bf \bar 3_f})_A$
    & $({\bf 3_{sp}})_S$
    & $L_k-1, L_k, L_k+1$
    & $-$
    & $-$
    \\    
$({\bf \bar 3_c})_A$    
    &even
    & odd
    & $({\bf 6_f})_S$
    & $({\bf 1_{sp}})_A$
    & $L_k$
    & $-$
    & $-$
    \\
$({\bf \bar 3_c})_A$    
    &odd   
    & even
    & $({\bf \bar 3_f})_A$
    & $({\bf 1_{sp}})_A$
    & $L_k$
    & $+$
    & $-$
    \\    
$({\bf \bar 3_c})_A$    
    &odd
    & even
    & $({\bf 6_f})_S$
    & $({\bf 3_{sp}})_S$
    & $|L_k-1|,\cdots, L_k+1$
    & $+$
    & $-$
    \\
    \hline
\end{tabular}
\end{center}
\end{table}

Under permutation of the light quark momenta, $p_1\leftrightarrow p_2$, we have $\vec k\to -\vec k$ and $\vec K\to\vec K$, 
while under party we have $\vec k\to -\vec k$ and $\vec K\to -\vec K$.
Consequently, using the well known symmetry property of $Y_{lm}$, under the permutation, 
the space part wave function, see Eq.~(\ref{eq: psi space}), transforms as
\be
\psi({\rm space})\to (-)^{L_k} \psi({\rm space}),
\en
while under parity, it transforms as
\be
\psi({\rm space})\to (-)^{L_k+L_K}\psi({\rm space}).
\en
The parity eigenvalues of the $[qq]$ diquark and the whole $Qqq$ systems are given by $(-)^{L_k}$ and $(-)^{L_k+L_K}$, respectively.

Putting all of these together, under permutation of the light quarks, we have
\be
&&\psi({\rm color})\times\psi({\rm space})\times \psi({\rm flavor}) \times \psi({\rm spin})
\non\\
&&\qquad\qquad\to -(-)^{L_k}(-)^{N_f}(-)^{S_l+1}
\psi({\rm color})\times\psi({\rm space})\times \psi({\rm flavor})\times \psi({\rm spin}).
\en
Fermi statistics requires the wave function to be antisymmetric giving the following constraint:
\be
(-)^{L_k+N_f+S_l}=-1.
\en
The quantum numbers of all possible allowed configurations of the diquark system satisfying the Fermi statistic 
are shown in Table~\ref{tab:quantum numbers}. 
The corresponding parity eigenvalues of the diquark and the heavy baryons are also shown.

\begin{table}[t!]
\caption{Mass spectra and widths (in units of MeV unless specified) of
charmed baryons. 
Experimental values and $J^P$ are taken from the Particle
Data Group \cite{PDG}. 
The quantum number of $\Lambda_c(2940)$ can be different from the one shown in the table, 
see text for more details.
}
\label{tab:spectrumC}
\begin{center}
\scriptsize{
\begin{tabular}{|c|cc cc c c c c|c|} \hline \hline
State 
   & $J^P$
   & $n$
   & $(L_K,L_k)$ 
   & $S_{[qq]}^P$
   & $J_\ell^{P_\ell}$ 
   & Mass 
   & Width 
   & Decay modes
   \\   
\hline
$\Lambda_c^+$ 
   & ${1\over 2}^+$ 
   & 1
   & (0,0) 
   & $0^+$
   & $0^+$ 
   & $2286.46\pm0.14$ 
   & 
   & weak  
   \\
 \hline
 $\Lambda_c(2595)^+$ 
   & ${1\over 2}^-$ 
   & 1
   & (1,0) 
   & $0^+$
   & $1^-$ 
   & $2592.25\pm0.28$ 
   & $2.6\pm0.6$ 
   & $\Lambda_c\pi\pi,\Sigma_c\pi$ 
   \\
 \hline
 $\Lambda_c(2625)^+$ 
   & ${3\over 2}^-$ 
   & 1
   & (1,0) 
   & $0^+$
   & $1^-$ 
   & $2628.11\pm0.19$ 
   &$<0.97$ 
   & $\Lambda_c\pi\pi,\Sigma_c\pi$ 
   \\
 \hline
 $\Lambda_c(2765)^+$ 
   & $?^?$ 
   & ?
   & ? 
   & ? 
   & $?$ 
   & $2766.6\pm2.4$ 
   & $50$ 
   & $\Sigma_c\pi,\Lambda_c\pi\pi$ 
   \\
 \hline
  $\Lambda_c(2860)^+$ 
   & $\frac{3}{2}^+$ 
   & 1
   & (2,0)
   & $0^+$
   & $2^+$ 
   & $2856.1^{+2.3}_{-6.0}$ 
   & $68^{+12}_{-22}$ 
   & $\Sigma^{(*)}_c\pi,D^0p, D^+ n$ 
   \\
 \hline
 $\Lambda_c(2880)^+$ 
   & ${5\over 2}^+$ 
   & 1
   & (2,0)
   & $0^+$
   & $2^+$ 
   & $2881.63\pm0.24$ 
   & $5.6^{+0.8}_{-0.6}$
   & $\Sigma_c^{(*)}\pi,\Lambda_c\pi\pi,D^0p$ 
   \\
 \hline
 $\Lambda_c(2940)^+$ 
   & $\frac{3}{2}^-$
   & 2
   & (1,0) 
   & $0^+$
   & $1^-$ 
   & $2939.6^{+1.3}_{-1.5}$ 
   & $20^{+6}_{-5}$ 
   & $\Sigma_c^{(*)}\pi,\Lambda_c\pi\pi,D^0p$ 
   \\ 
 \hline
 $\Sigma_c(2455)^{++}$ 
   & ${1\over 2}^+$ 
   & 1
   & $(0,0)$ 
   & $1^+$
   & $1^+$ 
   & $2453.97\pm0.14$ 
   & $1.89^{+0.09}_{-0.18}$ 
   & $\Lambda_c\pi$ 
   \\
 \hline
 $\Sigma_c(2455)^{+}$ 
   & ${1\over 2}^+$ 
   & 1
   & $(0,0)$ 
   & $1^+$
   & $1^+$     
   & $2452.9\pm0.4$ 
   & $<4.6$ 
   & $\Lambda_c\pi$
   \\
 \hline
 $\Sigma_c(2455)^{0}$ 
   & ${1\over 2}^+$ 
   & 1
   & $(0,0)$ 
   & $1^+$
   & $1^+$ 
   & $2453.75\pm0.14$
   & $1.83^{+0.11}_{-0.19}$ 
   & $\Lambda_c\pi$ 
   \\
 \hline
 $\Sigma_c(2520)^{++}$ 
   & ${3\over 2}^+$ 
   & 1
   & $(0,0)$ 
   & $1^+$
   & $1^+$ 
   & $2518.41^{+0.21}_{-0.19}$
   & $14.78^{+0.30}_{-0.40}$ 
   & $\Lambda_c\pi$
   \\
 \hline
 $\Sigma_c(2520)^{+}$ 
   & ${3\over 2}^+$ 
   & 1
   & $(0,0)$ 
   & $1^+$
   & $1^+$     
   & $2517.5\pm2.3$
   & $<17$ 
   & $\Lambda_c\pi$ 
   \\
 \hline
 $\Sigma_c(2520)^{0}$ 
   & ${3\over 2}^+$ 
   & 1
   & $(0,0)$ 
   & $1^+$
   & $1^+$ 
   & $2518.48\pm0.20$
   & $15.3^{+0.4}_{-0.5}$ 
   & $\Lambda_c\pi$ 
   \\
 \hline
 $\Sigma_c(2800)^{++}$ 
   & $?^?$ 
   & ?
   & ? 
   & ?
   & ? 
   & $2801^{+4}_{-6}$ 
   & $75^{+22}_{-17}$ 
   & $\Lambda_c\pi,\Sigma_c^{(*)}\pi,\Lambda_c\pi\pi$ 
   \\
 \hline
 $\Sigma_c(2800)^{+}$ 
   & $?^?$ 
   & ?
   & ? 
   & ? 
   & ?
   & $2792^{+14}_{-~5}$ 
   & $62^{+60}_{-40}$ 
   & $\Lambda_c\pi,\Sigma_c^{(*)}\pi,\Lambda_c\pi\pi$ 
   \\
 \hline
 $\Sigma_c(2800)^{0}$ 
   & $?^?$ 
   & ?
   & ? 
   & ? 
   & ?
   & $2806^{+5}_{-7}$ 
   & $72^{+22}_{-15}$ 
   & $\Lambda_c\pi,\Sigma_c^{(*)}\pi,\Lambda_c\pi\pi$
   \\
 \hline
 $\Xi_c^+$ 
   & ${1\over 2}^+$ 
   & 1
   & (0,0) 
   & $0^+$
   & $0^+$ 
   & $2467.87\pm 0.30$ 
   & 
   & weak 
   \\ 
 \hline
 $\Xi_c^0$ 
   & ${1\over 2}^+$ 
   & 1
   & (0,0) 
   & $0^+$
   & $0^+$  
   & $2470.87^{+0.28}_{-0.31}$ 
   & 
   & weak 
   \\ 
\hline
 $\Xi'^+_c$ 
   & ${1\over 2}^+$ 
   & 1
   & (0,0) 
   & $1^+$
   & $1^+$ 
   & $2577.4\pm1.2$ 
   & 
   & $\Xi_c\gamma$ 
   \\ 
\hline
 $\Xi'^0_c$ 
   & ${1\over 2}^+$ 
   & 1
   & (0,0) 
   & $1^+$
   & $1^+$  
   & $2578.8\pm0.5$ 
   & 
   & $\Xi_c\gamma$ 
   \\ 
\hline
 $\Xi_c(2645)^+$ 
   & ${3\over 2}^+$ 
   & 1
   & (0,0) 
   & $1^+$
   & $1^+$  
   & $2645.53\pm 0.31$ 
   & $2.14\pm0.19$ 
   & $\Xi_c\pi$ 
   \\
\hline
 $\Xi_c(2645)^0$ 
   & ${3\over 2}^+$ 
   & 1
   & (0,0) 
   & $1^+$
   & $1^+$  
   & $2646.32\pm0.31$ 
   & $2.35\pm0.22$ 
   & $\Xi_c\pi$ 
   \\
 \hline
 $\Xi_c(2790)^+$ 
   & ${1\over 2}^-$ 
   & 1
   & (1,0) 
   & $0^+$
   & $1^-$ 
   & $2792.0\pm0.5$ 
   & $8.9\pm 1.0$ 
   & $\Xi'_c\pi$
   \\
 \hline
 $\Xi_c(2790)^0$ 
   & ${1\over 2}^-$ 
   & 1
   & (1,0) 
   & $0^+$
   & $1^-$  
   & $2792.8\pm1.2$ 
   & $10.0\pm1.1$ 
   & $\Xi'_c\pi$ 
   \\
 \hline
 $\Xi_c(2815)^+$ 
   & ${3\over 2}^-$ 
   & 1
   & (1,0) 
   & $0^+$
   & $1^-$   
   & $2816.67\pm0.31$ 
   & $2.43\pm0.26$ 
   & $\Xi^*_c\pi,\Xi_c\pi\pi,\Xi_c'\pi$ 
   \\
 \hline
 $\Xi_c(2815)^0$ 
   & ${3\over 2}^-$ 
   & 1
   & (1,0) 
   & $0^+$
   & $1^-$   
   & $2820.22\pm0.32$ 
   & $2.54\pm0.25$ 
   & $\Xi^*_c\pi,\Xi_c\pi\pi,\Xi_c'\pi$ 
   \\
 \hline
$\Xi_c(2930)^0$ 
  & $?^?$ 
  & ?
  & ? 
  & ? 
  & $?$ 
  & $2931\pm6$ 
  & $36\pm13$
  & $\Lambda_c \ov K$ 
  \\
\hline
 $\Xi_c(2970)^+$ 
   & $?^?$ 
   & ?
   & ? 
   & ? 
   & $?$ 
   & $2969.4\pm0.8$ 
   & $20.9^{+2.4}_{-3.5}$
   & $\Sigma_c \ov K,\Lambda_c \ov K\pi,\Xi_c\pi\pi$  
   \\
 \hline
 $\Xi_c(2970)^0$ 
   & $?^?$ 
   & ?
   & ? 
   & ? 
   & $?$ 
   & $2967.8\pm0.8$ 
   & $28.1^{+3.4}_{-4.0}$
   & $\Sigma_c \ov K,\Lambda_c \ov K\pi,\Xi_c\pi\pi$
   \\
 \hline
 $\Xi_c(3055)^+$ 
   & $?^?$ 
   & ?
   & ? 
   & ? 
   & $?$ 
   & $3055.9\pm0.4$
   & $7.8\pm1.9$
   & $\Sigma_c \ov K,\Lambda_c \ov K\pi,D\Lambda$  
   \\
 \hline
 $\Xi_c(3080)^+$ 
   & $?^?$ 
   & ?
   & ? 
   & ? 
   & $?$ 
   & $3077.2\pm0.4$ 
   & $3.6\pm1.1$ 
   & $\Sigma_c \ov K,\Lambda_c \ov K\pi,D\Lambda$  
   \\
\hline
 $\Xi_c(3080)^0$ 
   & $?^?$ 
   & ?
   & ? 
   & ? 
   & $?$ 
   & $3079.9\pm1.4$ 
   & $5.6\pm2.2$
   & $\Sigma_c \ov K,\Lambda_c \ov K\pi,D\Lambda$ 
   \\
\hline
$\Xi_c(3123)^+$ 
   & $?^?$ 
   & ?
   & ? 
   & ? 
   & $?$ 
   & $3122.9\pm1.3$ 
   & $4\pm4$
   & $\Sigma_c^* \ov K,\Lambda_c \ov K\pi$ 
   \\
 \hline
 $\Omega_c^0$ 
   & ${1\over 2}^+$ 
   & 1
   & (0,0) 
   & $1^+$
   & $1^+$ 
   & $2695.2\pm1.7$ 
   & 
   & weak 
   \\
 \hline
 $\Omega_c(2770)^0$ 
   & ${3\over 2}^+$ 
   & 1
   & (0,0) 
   & $1^+$
   & $1^+$  
   & $2765.9\pm2.0$ 
   & 
   & $\Omega_c\gamma$ 
   \\
\hline 
 $\Omega_c(3000)^0$ 
   & $?^?$
   & ?
   & ?
   & ?
   & ?
   & $3000.4\pm0.4$ 
   & $4.5\pm0.7$
   & $\Xi_c\bar K$ 
   \\
\hline
 $\Omega_c(3050)^0$ 
   & $?^?$ 
   & ?
   & ?
   & ?
   & ? 
   & $3050.2\pm0.33$ 
   & $<1.2$
   & $\Xi_c\bar K$ 
   \\
\hline
 $\Omega_c(3065)^0$ 
   & $?^?$ 
   & ?
   & ?
   & ?
   & ?
   & $3065.6\pm0.4$ 
   & $3.5\pm0.4$
   & $\Xi_c\bar K$ 
   \\
\hline
 $\Omega_c(3090)^0$ 
   & $?^?$ 
   & ?
   & ? 
   & ? 
   & ?
   & $3090.2\pm0.7$ 
   & $8.7\pm1.3$
   & $\Xi^{(\prime)}_c\bar K$ 
   \\
\hline
 $\Omega_c(3119)^0$ 
   & $?^?$ 
   & ?
   & ? 
   & ? 
   & ?
   & $3119.1\pm1.0$ 
   & $<2.6$
   & $\Xi^{(\prime)}_c\bar K$
   \\
\hline
\hline
\end{tabular}
}
\end{center}
\end{table}

\begin{table}[t!]
\caption{Allowed configurations with $L_k+L_K=0,1,2$ are shown. 
The angular momenta are defined as
$\vec S_{[qq]}\equiv\vec L_k+\vec S_{qq}$, $\vec J_l\equiv\vec S_{[qq]}+\vec L_K$ and $\vec J\equiv\vec J_l+\vec S_Q$,
which are the angular momenta of the diquark system, the light-degree of freedom and the whole baryon, respectively. The quantum number assignments are from Tables~\ref{tab:quantum numbers} and \ref{tab:spectrumC}, while those with $(\dagger)$ are taken from \cite{Cheng:2017ove}. There are different assignments of the quantum number of $\Lambda_c(2940)$, see text for more details.
There are plenty of states to be discovered.}
 \label{tab:quantum number C}
\scriptsize
{
 \begin{center}
\begin{tabular}{| l c c  c c c c c c|}
\hline
$n$
    & $L_K$
    & $L_k$
    & ${\rm flavor}$
    & $S_{qq}$
    & $S_{[qq]}^P$
    & $J_l^P$
    & $J^P$
    & $\B_c$
    \\
    \hline
1
    & 0   
    & 0
    & ${\bf \bar 3_f}$
    &  $0$
    & $0^+$
    & $0^+$
    & $\frac{1}{2}^+$
    & $\Lambda^+_c$, $\Xi_c^{+,0}$
    \\    
2
    & 0
    & 0
    & ${\bf \bar 3_f}$
    &  $0$
    & $0^+$
    & $0^+$
    & $\frac{1}{2}^+$
    & $\Lambda_c(2765)^+ (\dagger)$ 
    \\        
1
    &0
    & 0
    & ${\bf 6_f}$
    & $1$
    & $1^+$
    & $1^+$
    & $\frac{1}{2}^+$
    & $\Sigma_c(2455)^{++,+,0}$, $\Xi_c^{\prime +,0}$, $\Omega^0_c$
    \\
2
    & 0
    & 0
    & ${\bf 6_f}$
    & $1$
    & $1^+$
    & $1^+$
    & $\frac{1}{2}^+$
    & $\Xi'_c(2970)^{+,0}(\dagger)$, $\Omega_c(3090)^0 (\dagger)$ 
    \\ 
1
    & 0
    & 0
    & ${\bf 6_f}$
    & $1$
    & $1^+$
    & $1^+$
    & $\frac{3}{2}^+$
    & $\Sigma_c(2520)^{++,+,0}$, $\Xi_c(2645)^{+,0}$, $\Omega_c(2770)^0$
    \\
2
    & 0
    & 0
    & ${\bf 6_f}$
    & $1$
    & $1^+$
    & $1^+$
    & $\frac{3}{2}^+$
    & $\Omega_c(3119)^0(\dagger)$
    \\
1
    &2    
    & 0
    & ${\bf \bar 3_f}$
    &  $0$
    & $0^+$
    & $2^+$
    & $\frac{3}{2}^+$
    & $\Lambda_c(2860)^+$, $\Xi_c(3055)^{+,0} (\dagger)$
    \\    
1
    &2    
    & 0
    & ${\bf \bar 3_f}$
    &  $0$
    & $0^+$
    & $2^+$
    & $\frac{5}{2}^+$
    & $\Lambda_c(2880)^+$, $\Xi_c(3080)^{+,0}(\dagger)$
    \\
$n$
    & 2
    & 0
    & ${\bf 6_f}$
    & $1$
    & $1^+$
    & $1^+$
    & $\frac{1}{2}^+,\frac{3}{2}^+$
    &
    \\
$n$
    & 2
    & 0
    & ${\bf 6_f}$
    & $1$
    & $1^+$ 
    & $2^+$
    & $\frac{3}{2}^+,\frac{5}{2}^+$
    &
    \\            
$n$
    & 2
    & 0
    & ${\bf 6_f}$
    & $1$
    & $1^+$
    & $3^+$
    & $\frac{5}{2}^+,\frac{7}{2}^+$
    &
    \\                                          
$n$
    &0    
    & 2
    & ${\bf \bar 3_f}$
    &  $0$
    & $2^+$
    & $2^+$
    & $\frac{3}{2}^+,\frac{5}{2}^+$
    &
    \\    
$n$
    &0
    & 2
    & ${\bf 6_f}$
    & $1$
    & $1^+$
    & $1^+$
    & $\frac{1}{2}^+,\frac{3}{2}^+$
    &
    \\
$n$
    &0
    & 2
    & ${\bf 6_f}$
    & $1$
    & $2^+$
    & $2^+$
    & $\frac{3}{2}^+,\frac{5}{2}^+$
    &
    \\
$n$
    &0
    & 2
    & ${\bf 6_f}$
    & $1$
    & $3^+$
    & $3^+$
    & $\frac{5}{2}^+,\frac{7}{2}^+$
    &
    \\          
$n$
    &1
    & 1
    & ${\bf \bar 3_f}$
    & $1$
    & $0^-$
    & $1^+$
    & $\frac{1}{2}^+,\frac{3}{2}^+$
    &
    \\  
$n$
    & 1
    & 1
    & ${\bf \bar 3_f}$
    & $1$
    & $1^-$
    & $0^+$
    & $\frac{1}{2}^+$
    &
    \\   
$n$
    & 1
    & 1
    & ${\bf \bar 3_f}$
    & $1$
    & $1^-$ 
    & $1^+$
    & $\frac{1}{2}^+,\frac{3}{2}^+$
    &
    \\   
$n$
    & 1
    & 1
    & ${\bf \bar 3_f}$
    & $1$
    & $1^-$
    & $2^+$
    & $\frac{3}{2}^+,\frac{5}{2}^+$
    &
    \\            
$n$
    & 1
    & 1
    & ${\bf \bar 3_f}$
    & $1$
    & $2^-$
    & $1^+$
    & $\frac{1}{2}^+,\frac{3}{2}^+$
    &
    \\ 
$n$
    & 1
    & 1
    & ${\bf \bar 3_f}$
    & $1$
    & $2^-$
    & $2^+$
    & $\frac{3}{2}^+,\frac{5}{2}^+$
    &
    \\   
$n$
    & 1
    & 1
    & ${\bf \bar 3_f}$
    & $1$
    & $2^-$
    & $3^+$
    & $\frac{5}{2}^+,\frac{7}{2}^+$
    &
    \\                   
$n$
    & 1
    & 1
    & ${\bf 6_f}$
    &  $0$
    & $1^-$
    & $0^+$
    & $\frac{1}{2}^+$
    &
    \\ 
$n$
    & 1
    & 1
    & ${\bf 6_f}$
    &  $0$
    & $1^-$
    & $1^+$
    & $\frac{1}{2}^+,\frac{3}{2}^+$
    &
    \\ 
$n$
    & 1
    & 1
    & ${\bf 6_f}$
    &  $0$
    & $1^-$ 
    & $2^+$
    & $\frac{3}{2}^+,\frac{5}{2}^+$
    &
    \\         
\hline
$n$
    & $L_K$
    & $L_k$
    & ${\rm flavor}$
    & $S_{qq}$
    & $S_{[qq]}^P$
    & $J_l^P$
    & $J^P$
    & $\B_c$
    \\
    \hline 
1
    & 1   
    & 0
    & ${\bf \bar 3_f}$
    &  $0$
    & $0^+$
    & $1^-$
    & $\frac{1}{2}^-$
    & $\Lambda_c(2595)^+$, $\Xi_c(2790)^{+,0}$
    \\
2
    & 1   
    & 0
    & ${\bf \bar 3_f}$
    &  $0$
    & $0^+$
    & $1^-$
    & $\frac{1}{2}^-$
    & $\Lambda_c(2940)^+ (\dagger)$
    \\            
1
    & 1   
    & 0
    & ${\bf \bar 3_f}$
    &  $0$
    & $0^+$
    & $1^-$
    & $\frac{3}{2}^-$
    & $\Lambda_c(2625)^+$, $\Xi_c(2815)^{+,0}$
    \\
2
    & 1   
    & 0
    & ${\bf \bar 3_f}$
    &  $0$
    & $0^+$
    & $1^-$
    & $\frac{3}{2}^-$
    & $\Lambda_c(2940)^+$
    \\
1
    & 1
    & 0
    & ${\bf 6_f}$
    & $1$
    & $1^+$
    & $0^-$
    & $\frac{1}{2}^-$
    & 
    \\
1
    & 1
    & 0
    & ${\bf 6_f}$
    & $1$
    & $1^+$
    & $1^-$
    & $\frac{1}{2}^-,\frac{3}{2}^-$
    & 
    \\            
1
    & 1
    & 0
    & ${\bf 6_f}$
    & $1$
    & $1^+$
    & $2^-$
    & $\frac{3}{2}^-$
    & $\Sigma_c(2800)^{++,+,0}(\dagger)$, $\Xi'_c(2930)^{ +,0}(\dagger)$, $\Omega_c(3050)^0(\dagger)$
    \\
1
    & 1
    & 0
    & ${\bf 6_f}$
    & $1$
    & $1^+$
    & $2^-$
    & $\frac{5}{2}^-$
    & $\Omega_c(3066)^0(\dagger)$  
    \\                  
$n$
    & 0
    & 1
    & ${\bf \bar 3_f}$
    & $1$
    & $0^-$
    & $0^-$
    & $\frac{1}{2}^-$
    &
    \\  
$n$
    & 0
    & 1
    & ${\bf \bar 3_f}$
    & $1$
    & $1^-$
    & $1^-$
    & $\frac{1}{2}^-,\frac{3}{2}^-$
    &
    \\  
$n$
    & 0
    & 1
    & ${\bf \bar 3_f}$
    & $1$
    & $2^-$
    & $2^-$
    & $\frac{3}{2}^-,\frac{5}{2}^-$
    &
    \\            
$n$
    & 0
    & 1
    & ${\bf 6_f}$
    &  $0$
    & $1^-$
    & $1^-$
    & $\frac{1}{2}^-,\frac{3}{2}^-$
    &
    \\
    \hline
\end{tabular}
\end{center}
}
\end{table}

\subsection{Charmed Baryons}

The observed mass spectra and decay widths of charmed baryons are
summarized in Table~\ref{tab:spectrumC}.
The $J^P$ quantum numbers of
$\Lambda_c^+$, $\Lambda_c(2595)^+$, $\Lambda_c(2860)^+$, $\Lambda_c(2880)^+$, $\Lambda_c(2940)^+$ and
$\Sigma_c(2455)$,  
are determined up to different levels of certainty, while 
the $J^P$ quantum numbers given in
Table~\ref{tab:spectrumC} other states are either from quark model predictions or totally undetermined. 
In fact, there are 16 states out of 40 states in Table \ref{tab:spectrumC} having unknown quantum numbers.

In Table~\ref{tab:quantum number C}
configurations with $L_k+L_K=0,1,2$ for charmed baryons are shown. 
The quantum number assignments are from Tables~\ref{tab:quantum numbers} and \ref{tab:spectrumC}, while those with $(\dagger)$ are taken from ref.~\cite{Cheng:2017ove}. 
Only several multiplets are well established.
These include the $J^P={1\over 2}^+$ ${\bf \bar 3_f}$ states: ($\Lambda_c^+$, $\Xi_c^+,\Xi_c^0)$, 
$J^P={1\over 2}^-$ ${\bf \bar 3_f}$ states:
($\Lambda_c(2595)^+$, $\Xi_c(2790)^+,\Xi_c(2790)^0)$; 
$J^P={3\over 2}^-$ ${\bf \bar 3_f}$ states:
$(\Lambda_c(2625)^+$, $\Xi_c(2815)^+,\Xi_c(2815)^0)$;
$J^P={1\over 2}^+$ and ${3\over 2}^+$ ${\bf 6_f}$ states:
($\Omega_c,\Sigma_c,\Xi'_c$) and ($\Omega_c^*,\Sigma_c^*,\Xi'^*_c$), respectively.
Ref.~\cite{Cheng:2017ove} makes further suggestions on the classification on some other states.
As noted previously PDG and LHCb assign $\Lambda_c(2940)^+$ as a $\frac{3}{2}^-$ state~\cite{PDG,Aaij:2017vbw}, while the authors of ref.~\cite{Cheng:2017ove} take it as a $\frac{1}{2}^-$ state.
We follow the suggestions of ref.~\cite{Cheng:2017ove} on the quantum numbers of $\Lambda_c(2940)^+$ and some other states.
Note that other quantum number assignments on the newly observed $\Omega^{(*,**)}_c$ states, such as those avocated in refs.~\cite{Chen:2017sci,Karliner:2017kfm,Wang:2017hej,Padmanath:2017lng,Wang:2017zjw,Chen:2017gnu}, are not shown in the table.

From Table~\ref{tab:quantum number C} we see that there are plenty of states in the $L_k+L_K=0,1,2$ sector to be discovered.

\begin{table}[!]
\caption{Mass spectra and widths (in units of MeV) of
bottom baryons. Experimental values are taken from the Particle
Data Group~\cite{PDG}, except those of $\Xi_b(6227)^-$, which are from \cite{Aaij:2018yqz}.
}
\label{tab:spectrumB}
\begin{center}
\footnotesize{
\begin{tabular}{|c|c cc c cc c|c|} \hline \hline
State 
   & $J^P$
   & $n$
   & $(L_K,L_k)$ 
   & $S_{[qq]}^P$
   & $J_\ell^{P_\ell}$ 
   & Mass 
   & Width 
   & Decay modes
   \\
\hline
$\Lambda_b^0$ 
   & ${1\over 2}^+$ 
   & 1
   & (0,0) 
   & $0^+$ 
   & $0^+$
   & $5619.60\pm0.17$ 
   & 
   & weak  
   \\
\hline  
 $\Lambda_b(5912)^0$ 
   & ${1\over 2}^-$ 
   & 1
   & (1,0)
   & $0^+$
   & $1^-$ 
   & $5912.20\pm0.21$ 
   & $<0.66$ 
   & $\Lambda_b\pi\pi$ 
   \\      
 \hline 
  $\Lambda_b(5920)^0$ 
   & ${3\over 2}^-$ 
   & 1
   & (1,0)
   & $0^+$ 
   & $1^-$ 
   & $5919.92\pm0.19$ 
   &$<0.63$ 
   & $\Lambda_b^0\pi\pi$ 
   \\     
 \hline
 $\Sigma_b^{+}$ 
   & ${1\over 2}^+$ 
   & 1
   & (0,0) 
   & $1^+$ 
   & $1^+$ 
   & $5811.3\pm1.9$ 
   & $9.7^{+4.0}_{-3.0}$ 
   & $\Lambda_b\pi$ 
   \\
 \hline
 $\Sigma_b^{-}$ 
   & ${1\over 2}^+$ 
   & 1
   & (0,0)
   & $1^+$ 
   & $1^+$ 
   & $5815.5\pm1.8$ 
   & $4.9^{+3.3}_{-2.4}$ 
   & $\Lambda_b\pi$
   \\   
\hline  
 $\Sigma^{*+}_b$ 
   & ${3\over 2}^+$ 
   & 1
   & (0,0) 
   & $1^+$ 
   & $1^+$ 
   & $5832.1\pm1.9$
   & $11.5\pm2.8$ 
   & $\Lambda_b\pi$
   \\
 \hline
 $\Sigma^{*-}_b$ 
   & ${3\over 2}^+$ 
   & 1
   & (0,0) 
   & $1^+$ 
   & $1^+$ 
   & $5835.1\pm1.9$
   & $7.5\pm2.3$ 
   & $\Lambda_b\pi$ 
   \\
\hline 
 $\Xi_b^0$ 
   & ${1\over 2}^+$ 
   & 1
   & (0,0) 
   & $0^+$ 
   & $0^+$ 
   & $5791.9\pm0.5$ 
   & 
   & weak 
   \\ 
 \hline
 $\Xi_b^-$ 
   & ${1\over 2}^+$ 
   & 1
   & (0,0) 
   & $0^+$ 
   & $0^+$ 
   & $5794.5\pm1.4$ 
   & 
   & weak 
   \\  
   \hline
 $\Xi'_b(5935)^-$ 
   & ${1\over 2}^+$ 
   & 1
   & (0,0) 
   & $1^+$ 
   & $1^+$ 
   & $5935.02\pm0.05$ 
   & $<0.08$
   & $\Xi^0_b\pi^-$ 
   \\ 
 \hline
 $\Xi_b(5945)^0$ 
   & ${3\over 2}^+$ 
   & 1
   & (0,0)
   & $1^+$ 
   & $1^+$ 
   & $5949.8\pm 1.4$ 
   & $0.90\pm0.18$ 
   & $\Xi_b\pi$ 
   \\
\hline
 $\Xi_b(5955)^-$ 
   & ${3\over 2}^+$ 
   & 1
   & (0,0)
   & $1^+$ 
   & $1^+$ 
   & $5955.33\pm0.13$ 
   & $1.65\pm0.33$ 
   & $\Xi_b\pi$ 
   \\
\hline
 $\Xi_b(6227)^-$ 
   & $?^?$ 
   & ?
   & ?
   & ?
   & ? 
   & $6226.9\pm2.0\pm0.3\pm0.2$ 
   & $18.1\pm5.4\pm1.8$ 
   & $\Lambda_b K^-$, $\Xi_b\pi^-$ 
   \\   
 \hline
  $\Omega_b^0$ 
   & ${1\over 2}^+$ 
   & 1
   & (0,0) 
   & $1^+$ 
   & $1^+$ 
   & $6046.1\pm1.7$ 
   & 
   & weak 
   \\
 \hline        
\hline
\end{tabular}
}
\end{center}
\end{table}

\begin{table}[t]
\caption{Same as Table~\ref{tab:quantum number C} but for bottom baryons. The quantum number assignments are basically taken from Tables~\ref{tab:quantum numbers} and \ref{tab:spectrumB}. 
There are plenty of states to be discovered.}
 \label{tab:quantum number B}
\footnotesize
{
 \begin{center}
\begin{tabular}{| l c c  c c c c c c|}
\hline
$n$
    & $L_K$
    & $L_k$
    & ${\rm flavor}$
    & $S_{qq}$
    & $S_{[qq]}^P$
    & $J_l^P$
    & $J^P$
    & $\B_b$
    \\
    \hline
1
    & 0   
    & 0
    & ${\bf \bar 3_f}$
    &  $0$
    & $0^+$
    & $0^+$
    & $\frac{1}{2}^+$
    & $\Lambda^0_b$, $\Xi_b^{0,-}$
    \\    
1
    &0
    & 0
    & ${\bf 6_f}$
    & $1$
    & $1^+$
    & $1^+$
    & $\frac{1}{2}^+$
    & $\Sigma_b^{+,-}$, $\Xi_b^{\prime}(5935)^-$, $\Omega^0_b$
    \\
1
    &0
    & 0
    & ${\bf 6_f}$
    & $1$
    & $1^+$
    & $1^+$
    & $\frac{3}{2}^+$
    & $\Sigma_b^{*+,-}$, $\Xi_b(5945)^+,\Xi_b(5955)^-$
    \\
$n$
    &2    
    & 0
    & ${\bf \bar 3_f}$
    &  $0$
    & $0^+$
    & $2^+$
    & $\frac{3}{2}^+,\frac{5}{2}^+$
    & 
    \\      
$n$
    &2
    & 0
    & ${\bf 6_f}$
    & $1$
    & $1^+$
    & $1^+$
    & $\frac{1}{2}^+,\frac{3}{2}^+$
    &
    \\
$n$
    &2
    & 0
    & ${\bf 6_f}$
    & $1$
    & $1^+$
    & $2^+$
    & $\frac{3}{2}^+,\frac{5}{2}^+$
    &
    \\            
$n$
    &2
    & 0
    & ${\bf 6_f}$
    & $1$
    & $1^+$
    & $3^+$
    & $\frac{5}{2}^+,\frac{7}{2}^+$
    &
    \\            
$n$
    &0    
    & 2
    & ${\bf \bar 3_f}$
    &  $0$
    & $2^+$
    & $2^+$
    & $\frac{3}{2}^+,\frac{5}{2}^+$
    &
    \\    
$n$
    &0
    & 2
    & ${\bf 6_f}$
    & $1$
    & $1^+$
    & $1^+$
    & $\frac{1}{2}^+,\frac{3}{2}^+$
    &
    \\
$n$
    &0
    & 2
    & ${\bf 6_f}$
    & $1$
    & $2^+$
    & $2^+$
    & $\frac{3}{2}^+,\frac{5}{2}^+$
    &
    \\
$n$
    &0
    & 2
    & ${\bf 6_f}$
    & $1$
    & $3^+$
    & $3^+$
    & $\frac{5}{2}^+,\frac{7}{2}^+$
    &
    \\                        
$n$
    &1
    & 1
    & ${\bf \bar 3_f}$
    & $1$
    & $0^-$
    & $1^+$
    & $\frac{1}{2}^+,\frac{3}{2}^+$
    &
    \\  
$n$
    & 1
    & 1
    & ${\bf \bar 3_f}$
    & $1$
    & $1^-$
    & $0^+$
    & $\frac{1}{2}^+$
    &
    \\   
$n$
    & 1
    & 1
    & ${\bf \bar 3_f}$
    & $1$
    & $1^-$
    & $1^+$
    & $\frac{1}{2}^+,\frac{3}{2}^+$
    &
    \\   
$n$
    & 1
    & 1
    & ${\bf \bar 3_f}$
    & $1$
    & $1^-$
    & $2^+$
    & $\frac{3}{2}^+,\frac{5}{2}^+$
    &
    \\            
$n$
    & 1
    & 1
    & ${\bf \bar 3_f}$
    & $1$
    & $2^-$
    & $1^+$
    & $\frac{1}{2}^+,\frac{3}{2}^+$
    &
    \\ 
$n$
    & 1
    & 1
    & ${\bf \bar 3_f}$
    & $1$
    & $2^-$
    & $2^+$
    & $\frac{3}{2}^+,\frac{5}{2}^+$
    &
    \\   
$n$
    & 1
    & 1
    & ${\bf \bar 3_f}$
    & $1$
    & $2^-$
    & $3^+$
    & $\frac{5}{2}^+,\frac{7}{2}^+$
    &
    \\                   
$n$
    & 1
    & 1
    & ${\bf 6_f}$
    &  $0$
    & $1^-$
    & $0^+$
    & $\frac{1}{2}^+$
    &
    \\ 
$n$
    & 1
    & 1
    & ${\bf 6_f}$
    &  $0$
    & $1^-$
    & $1^+$
    & $\frac{1}{2}^+,\frac{3}{2}^+$
    &
    \\ 
$n$
    & 1
    & 1
    & ${\bf 6_f}$
    &  $0$
    & $1^-$
    & $2^+$
    & $\frac{3}{2}^+,\frac{5}{2}^+$
    &
    \\         
\hline
$n$
    & $L_K$
    & $L_k$
    & ${\rm flavor}$
    & $S_{qq}$
    & $S_{[qq]}^P$
    & $J_l^P$
    & $J^P$
    & $\B_b$
    \\
    \hline 
1
    & 1   
    & 0
    & ${\bf \bar 3_f}$
    &  $0$
    & $0^+$
    & $1^-$
    & $\frac{1}{2}^-$
    & $\Lambda_b(5912)^0$
    \\
1
    & 1   
    & 0
    & ${\bf \bar 3_f}$
    &  $0$
    & $0^+$
    & $1^-$
    & $\frac{3}{2}^-$
    & $\Lambda_b(5920)^0$
    \\              
$n$
    & 0
    & 1
    & ${\bf \bar 3_f}$
    & $1$
    & $0^-$
    & $0^-$
    & $\frac{1}{2}^-$
    &
    \\  
$n$
    & 0
    & 1
    & ${\bf \bar 3_f}$
    & $1$
    & $1^-$
    & $1^-$
    & $\frac{1}{2}^-,\frac{3}{2}^-$
    &
    \\  
$n$
    & 0
    & 1
    & ${\bf \bar 3_f}$
    & $1$
    & $2^-$
    & $2^-$
    & $\frac{3}{2}^-,\frac{5}{2}^-$
    &
    \\            
$n$
    & 0
    & 1
    & ${\bf 6_f}$
    &  $0$
    & $1^-$
    & $1^-$
    & $\frac{1}{2}^-,\frac{3}{2}^-$
    &
    \\
$n$
    & 1
    & 0
    & ${\bf 6_f}$
    & $1$
    & $1^+$
    & $0^-$
    & $\frac{1}{2}^-$
    & 
    \\    
$n$
    & 1
    & 0
    & ${\bf 6_f}$
    & $1$
    & $1^+$
    & $1^-$
    & $\frac{1}{2}^-,\frac{3}{2}^-$
    & 
    \\   
$n$
    & 1
    & 0
    & ${\bf 6_f}$
    & $1$
    & $1^+$
    & $2^-$
    & $\frac{3}{2}^-,\frac{5}{2}^-$
    & 
    \\            
    \hline
\end{tabular}
\end{center}
}
\end{table}

\subsection{Bottom Baryons}

The observed mass spectra and decay widths of bottom baryons are
summarized in Table \ref{tab:spectrumB}.
Note that except
$\Xi^\prime_b(5935)^-$ and $\Xi_b(5955)^-$  
other $J^P$ quantum numbers given in
Table \ref{tab:spectrumB} are unmeasured. One has to rely on the
quark model to determine the $J^P$ assignments.

In Table~\ref{tab:quantum number B}
configurations with $L_k+L_K=0,1,2$ for charmed baryons are shown. 
The quantum number assignments are from Tables~\ref{tab:quantum numbers} and \ref{tab:spectrumB}. 
Only the $J^P={1\over 2}^+$ ${\bf \bar 3_f}$ multiplet with states: ($\Lambda_b^0$, $\Xi_b^0,\Xi_b^-)$, is established.
Several miltiplets are to be completed with the yet to be discovered states, such as $\Sigma_b^0$, $\Sigma_b^{*0}$, $\Xi_b^{\prime}(5935)^0$ and so on.
From Table~\ref{tab:quantum number B} we see that there are plenty of states in the $L_k+L_K=0,1,2$ sector to be discovered.

As shown in Table~\ref{tab:spectrumB}, $\Lambda_b$, $\Xi_b^{0,-}$ and $\Omega_b$ are the few singly bottom baryons that decay weakly.
We will study their decay modes in this work. In particular, $\Lambda^0_b\to \Lambda^{(*,**)}_c M$, $\Xi_b\to\Xi_c^{(*,**)} M$ and $\Omega_b\to\Omega^{(*,**)}_c M$ decays with $M=\pi, K,\rho, K^*$ will be explored.
In Table~\ref{tab:transition}, we summery the transitions we are about to study. 
There are altogether 8 different $\B_b\to\B_c$ transitions,
which can be classified into 3 types according to the quantum numbers of the initial and final state baryons.
These three types of transitions are
${\cal B}_b({\bf \bar 3_f},1/2^+)$ to ${\cal B}_c({\bf \bar 3_f},1/2^+)$,
${\cal B}_b({\bf 6_f},1/2^+)$ to ${\cal B}_c({\bf 6_f},1/2^+)$
and
${\cal B}_b({\bf \bar 3_f},1/2^+)$ to ${\cal B}_c({\bf \bar 3_f},1/2^-)$ transitions.
The type (i) and (iii) transitions have ${\bf \bar 3_f}$ scalar as spectators, while the type (ii) transition has ${\bf 6_f}$ axial-vector spectator diquarks.
Among the final states three charmed baryons are denoted with $(\dagger)$, they are states with unspecified or ambiguous  quantum numbers as noted in Tables~\ref{tab:spectrumC} and \ref{tab:quantum number C}.
As a working assumption we shall use the suggestion from ref.~\cite{Cheng:2017ove} for their quantum numbers.
Accordingly, we take $\Lambda_c(2765)$ as a radial excited $s$-wave state, 
$\Lambda_c(2940)$ a radial excited $p$-wave state
and $\Omega_c(3090)$ a radial excited $s$-wave state.
The study on these $\B_b\to \B_c$ transitions may shed light on the quantum numbers of these charmed baryons.

\begin{table}[t!]
\caption{Bottom baryon to charmed baryon transitions studied in this work are summarized in this table. 
There are three basic transition types.
Type (i) is the $\B_b({\bf \bar 3_f}, 1/2^+)\to \B_c({\bf \bar 3_f}, 1/2^+)$ transition,
type (ii) is the $\B_b({\bf 6_f}, 1/2^+)\to \B_c({\bf 6_f}, 1/2^+)$ transition
and type (iii) is the $\B_b({\bf \bar 3_f}, 1/2^+)\to \B_c({\bf \bar 3_f}, 1/2^-)$ transition.
Note that type (i) and (iii) transitions involve scalar diquarks, while type (ii) transitions involve axial-vector diquarks.
Type (iii) has odd parity baryons in the final states.
The quantum number assignments are from Tables~\ref{tab:quantum number C} and \ref{tab:quantum number B}, 
while those with $(\dagger)$ are taken from ref.~\cite{Cheng:2017ove}.
The asterisks indicate that the baryons in the final states are radial excited.}
 \label{tab:transition}
{
 \begin{center}
\begin{tabular}{| l c c |}
\hline
Type
    &~~~$(n=1, L_K, S_{[qq]}^P, J_l^P,J^P)_b\to
    (n, L_K, S_{[qq]}^P,J_l^P, J^P)_c$
    & $\B_b\to\B_c$
    \\
    \hline
(i) 
    & $ (1, 0, 0^+,0^+, \frac{1}{2}^+)\to(1, 0, 0^+,0^+, \frac{1}{2}^+)$ 
    & $\Lambda_b^0\to \Lambda^+_c$, $\Xi_b^{0(-)}\to \Xi_c^{+(0)}$
    \\ 
(i)$^*$      
    & $(1, 0, 0^+,0^+, \frac{1}{2}^+)\to (2, 0, 0^+,0^+, \frac{1}{2}^+)$ 
    & $\Lambda_b^0\to\Lambda_c(2765)^+ (\dagger)$ 
    \\  
(ii) 
    & $(1, 0, 1^+,1^+, \frac{1}{2}^+)\to (1, 0, 1^+,1^+, \frac{1}{2}^+)$
    & $\Omega^-_b\to\Omega^0_c$
    \\
(ii)$^*$
    & $(1, 0, 1^+,1^+, \frac{1}{2}^+)\to (2, 0, 1^+,1^+, \frac{1}{2}^+)$
    & $\Omega^-_b\to\Omega_c(3090)^0 (\dagger)$  
    \\           
(iii) 
    & $(1, 0, 0^+,0^+, \frac{1}{2}^+) \to (1, 1, 0^+,1^-, \frac{1}{2}^-)$   
    & $\Lambda_b^0\to\Lambda_c(2595)^+$, $\Xi^{0(-)}_b\to\Xi_c(2790)^{+(0)}$
    \\
(iii)$^*$  
    & $(1, 0, 0^+,0^+, \frac{1}{2}^+)\to(2, 1, 0^+,1^-, \frac{1}{2}^-)$   
    & $\Lambda_b^0\to\Lambda_c(2940)^+ (\dagger)$
    \\            
    \hline
\end{tabular}
\end{center}
}
\end{table}

\section{Form factors in the light-front approach}

We consider a heavy baryon consisting a heavy quark $Q$ and a scalar isosinglet diquark $[qq]$ or an axial-vector isovector diquark $[qq]$.
In the light-front approach, the baryon bound state with the
total momentum $P$ and spin $J$ can be written as (see, for example \cite{Cheng97,CCH})
\begin{eqnarray}
        |\B_Q(P,J,J_z)\rangle
                =\int &&\{d^3p_1\}\{d^3p_2\} 2(2\pi)^3 \delta^3(
                \tilde P -\tilde p_1-\tilde p_2)~\nonumber\\
        &&\times \sum_{\lambda_1,m,\alpha-\epsilon,b-e}
                \Psi^{JJ_z}_{nL_KS_{[qq]}J_l}(\tilde p_1,\tilde p_2,\lambda_1,\lambda_2)~
                C_{\alpha\beta\gamma} F^{bc}
        \non\\
        &&\times ~
             \Big|Q^\alpha(p_1,\lambda_1) [q_b^\beta q_c^\gamma](p_2,\lambda_2) \Big\rangle,
 \label{lfmbs}
\end{eqnarray}
where $S_{[qq]}$ is the spin of the diquark, $L_K$ is the orbital angular momentum of the $Q-[qq]$ system,
$J_l$ is the total angular momentum of the light degree of freedom,
$n$ is the quantum number of the wave-function (see later),
$\alpha,\beta,\gamma$ and $b,c$ are color and
flavor indices, respectively, $\lambda_i$ denotes helicity, $p_1$ and $p_2$ are the on-mass-shell light-front momenta,
\begin{equation}
        \tilde p=(p^+, \vec p_\bot)~, \quad \vec p_\bot = (p^1, p^2)~,
                \quad p^- = {m^2+p_\bot^2\over p^+},
\end{equation}
and
\begin{eqnarray}
        &&\{d^3p\} \equiv {dp^+d^2p_\bot\over 2(2\pi)^3},
        \quad \delta^3(\tilde p)=\delta(p^+)\delta^2(\vec p_\bot),
        \nonumber \\
        &&\Big|Q(p_1,\lambda_1) [q_b q_c](p_2,\lambda_2)\Big\rangle
        = b^\dagger_{\lambda_1}(p_1) a^\dagger_{\lambda_2}(p_2) |0\rangle,\\
        &&[a_{\lambda'}(p'),a^\dagger_\lambda(p)] =2(2\pi)^3~\delta^3(\tilde p'-\tilde
        p)\,\delta_{\lambda',\lambda},\,
        \non\\
        &&\{b_{\lambda'}(p'),b_{\lambda}^\dagger(p)\} =
        2(2\pi)^3~\delta^3(\tilde p'-\tilde p)~\delta_{\lambda'\lambda},
                \nonumber
\end{eqnarray}
with $\lambda_2=S_2=0$ for scalar diquark and $\lambda_2=0,\pm1$ and $S_2=1$ for axial vector diquark . 
The coefficient $C_{\alpha\beta\gamma}$ is a
normalized color factor and $F^{bc}$ is a normalized flavor
coefficient, obeying the relation
 \be
 &&C_{\alpha'\beta'\gamma'} F^{b'c'}
 C_{\alpha\beta\gamma} F^{bc}
         \Big \la Q^{\alpha'}(p'_1,\lambda'_1)
                 [q_{b'}^{\beta'} q_{c'}^{\gamma'}](p'_2,\lambda'_2)
             \Big|Q^\alpha(p_1,\lambda_1) [q_a^\beta q_b^\gamma](p_2,\lambda_2) 
             \Big\rangle
 \non\\
&&=2^2(2\pi)^6~\delta^3(\tilde p'_1-\tilde p_1)
 \delta^3(\tilde p'_2-\tilde p_2)
 \delta_{\lambda'_1\lambda_1}\delta_{\lambda'_2\lambda_2}.
 \label{eq:norm}
 \en

Tthe momenta can be defined 
in terms of the light-front relative momentum variables, 
$(x_i, \vec k_{i\bot})$ for $i=1,2$, 
\begin{eqnarray}
        && p^+_i=x_i P^{+}, \quad \sum_{i=1}^2 x_i=1, \nonumber \\
        &&\vec  p_{i\bot}=x_i \vec P_\bot+\vec k_{i\bot}, \quad \sum_{i=1}^2 \vec k_{i\bot}=0.
\end{eqnarray}
The momentum-space wave-function $\Psi^{JJ_z}_{nL_KS_{[qq]} J_l}$ can be expressed
as
\be
\Psi^{JJ_z}_{nL_KS_{[qq]}J_l}(\tilde p_1,\tilde p_2,\lambda_1,\lambda_2)
&=& \langle \lambda_1|{\cal R}_M^\dagger(p_1^+,\vec p_{1\bot}, m_1)|s_1\rangle
         \langle \lambda_2|{\cal R}_M^\dagger(p_2^+,\vec p_{2\bot}, m_2)|s_2\rangle
\non\\ 
          && \la S_1 J_l; s_1 J_{lz}|S_1 J_l; J J_z\ra
                \la L_K S_{[qq]}; L_z s_2|L_k S_{[qq]};J_l J_{l z}\ra
\non\\
&&          
                  ~\phi_{n L_K L_z}(x_1,x_2,k_{1\bot},k_{2\bot}),
\label{eq: Psi}
\en
where $\phi_{n L_K L_z}(x_1,x_2,k_{1\bot},k_{2\bot})$
describes the momentum distribution of the constituents in the
bound state,
$\la J' J''; m' m''|J' J'';J m\ra$ is the
Clebsch-Gordan coefficients and $\langle
\lambda_i|{\cal R}_M^\dagger(p^+_1,\vec p_{1\bot}, m_i)|s_i\rangle$ is
the well normalized Melosh transform matrix element. 
We will return to these quantities later.

We normalize the state as
\begin{equation}
        \langle \B_Q(P',J',J'_z)|\B_Q(P,J,J_z)\rangle = 2(2\pi)^3 P^+
        \delta^3(\tilde P'- \tilde P)\delta_{J'J}\delta_{J'_z J_z}~,
\label{wavenor1}
\end{equation}
consequently, $\phi_{nLL_z}(x,p_\bot)$ satisfies the following orthonormal condition,
\begin{equation}
        \int {dx\,d^2p_{\bot}\over 2(2\pi)^3}~\phi^{\prime*}_{n'L^\prime L^\prime_z}(x,p_\bot)
                                                   \phi_{nLL_z}(x,p_\bot)
        =\delta_{n',n}~\delta_{L^\prime,L}~\delta_{L^\prime_z,L_z}.
\label{momnor}
\end{equation}

The wave function is defined as 
 \be
   \phi_{nLm}(\{x\},\{k_\bot\})
   &=&
   \sqrt{\frac{d k_{2z}}{d x_2}}
  ~\varphi_{nLm}
  \left(\frac{\vec k_1-\vec k_2}{2},\beta\right), 
\en
with
\be  
  \varphi_{n00}(\vec k,\beta)
  &=&\varphi_{ns}(\vec k,\beta),
\non\\  
  \varphi_{n1m}(\vec k,\beta)
  &=&k_{m} \varphi_{np}(\vec k,\beta)
  =-\varepsilon(k_1+k_2,m)\cdot k\varphi_{np}(\vec k,\beta),
  \label{eq: varphi}
 \en
where $k_m\equiv\vec\varepsilon(m)\cdot\vec k$ (or, explicitly
$k_{L_z=\pm1}\equiv\mp(k^x\pm i k^y)/\sqrt2$,
$k_{L_z=0}\equiv k^z$) are proportional to the spherical harmonics
$Y_{1L_z}$ in momentum space, 
and $\varphi_{ns}$ and $\varphi_{np}$ 
are the
distribution amplitudes of $s$-wave and $p$-wave 
states,
respectively. 
 For a Gaussian-like wave function, one has
(the first two are from refs. \cite{Cheng97,CCH})
\begin{eqnarray} 
\varphi_{n=1, L_K=s}(\vec k,\beta)
    &=&4 \left({\pi\over{\beta^{2}}}\right)^{3\over{4}}
               ~{\rm exp}
               \left(-{k^2_z+k^2_\bot\over{2 \beta^2}}\right),
\non\\               
\varphi_{n=1,L_K=p}(\vec k,\beta)
    &=&\sqrt{2\over{\beta^2}}~\varphi_{n=1}(\vec k,\beta),
\non\\
\varphi_{n=2,L_K=s}(\vec k,\beta)
   &=& \sqrt{\frac{3}{2}}\bigg(1-\frac{2}{3}\frac{\vec k^2}{\beta^2}\bigg) \varphi_{n=1}(\vec k,\beta), 
\non\\   
\varphi_{n=2,L_K=p}(\vec k,\beta)
   &=&\sqrt{\frac{5}{2}}\bigg(1-\frac{2}{5}\frac{\vec k^2}{\beta^2}\bigg) \varphi_{n=1,L_K=p}(\vec k,\beta).
 \label{eq:wavefn}
\end{eqnarray}
The kinematics are given by
 \be
 M_0^{(\prime) 2}&=&\sum_{i=1}^2\frac{m_i^{(\prime)2}+k^{(\prime)2}_{i\bot}}{x_i},\quad
 k^{(\prime)}_i
 =(\frac{m_i^{(\prime)2}+k^{(\prime)2}_{i\bot}}{x^{(\prime)}_i M^{(\prime)}_0},x^{(\prime)}_i M^{(\prime)}_0,\,\vec k^{(\prime)}_{i\bot})
 =(e^{(\prime)}_i-k^{(\prime)}_{iz},e^{(\prime)}_i+k^{(\prime)}_{iz},\vec k^{(\prime)}_{i\bot}),
 \non\\
 M^{(\prime)}_0&=&e^{(\prime)}_1+e^{((\prime)}_2,\quad
 e^{(\prime)}_i =\sqrt{m^{(\prime)2}_i+k^{(\prime)2}_{i\bot}+k^{(\prime)2}_{iz}}
 =\frac{x^{(\prime)}_i M^{(\prime)}_0}{2}+\frac{m_i^{(\prime)2}+k^{(\prime)2}_{i\bot}}{2 x^{(\prime)}_i M^{(\prime)}_0},
 \non\\
 k^{(\prime)}_{iz}&=&\frac{x^{(\prime)}_i M^{(\prime)}_0}{2}-\frac{m_i^{(\prime)2}+k^{(\prime)2}_{i\bot}}{2 x^{(\prime)}_i M^{(\prime)}_0},
 \qquad
 2M^{(\prime)}_0 (e^{(\prime)}_{1(2)}+m^{(\prime)}_{1(2)})=(M^{(\prime)}_0+m^{(\prime)}_{1(2)})^2-m_{2(1)}^{(\prime)2}.
 \label{eq: kinematics}
 \en
 Under the constraint of $1-\sum_{i=1}^2 x_i=\sum_{i=1}^2
(k_i)_{x,y,z}=0$, we have 
 \be
 \frac{d k_{2z}}{d x_2}=\frac{e_1 e_2 }{x_1 x_2 M_0}=\frac{d k_{1z}}{dx_1}.
 \en

Now we turn to the Melosh transform.
For the heavy quark part, we have
\cite{Jaus90,deAraujo:1999ugw},
 \be
        \la \lambda_1|{\cal R}^\dagger_M (p^+_1,\vec p_{1\bot},m_1)|s_1\ra
        &=&\frac{\bar
        u(p_1,\lambda_1) u_D(p_1,s_1)}{2 m_1}
\en 
with $u_{(D)}$, a Dirac spinor in the light-front (instant)
form.
For the diquark part, if it is a scalar diquark the Melosh transform is a trivial one, i.e.
\be
\la \lambda_2|{\cal R}^\dagger_M (p^+_2,\vec p_{2\bot},m_2)|s_2\ra
        &=&1,
\en
but if it is a axial vector diquark, the Melosh transform is more interesting,
\be
\la \lambda_2|{\cal R}^\dagger_M (p^+_2,\vec p_{2\bot},m_2)|s_2\ra
        &=&-\varepsilon_{LF}^*(p_2,\lambda_2)\cdot \varepsilon_{I}(p_2,s_2),
\en
where $\varepsilon_{LF}$ and $\varepsilon_{I}$ are polarization vectors in light-front and instant forms, respectively.
Note that we have
 $u_D(k,s)=u(k,\lambda) \la \lambda|{\cal R}^\dagger_M|s\ra$
and
$\varepsilon_I(k,s)=\varepsilon_{LF}(k,\lambda) \la \lambda|{\cal R}^\dagger_M|s\ra$.
Consequently, the state 
$|Q(k,\lambda)\ra$ $\la \lambda|{\cal R}^\dagger_M|s\ra$ and 
$|[qq](k,\lambda)\ra$ $\la \lambda|{\cal R}^\dagger_M|s\ra$
transforms like $|Q(k,s)\ra$ and $|[qq](k,s)\ra$, respectively, under rotation,
i.e. their transformation do not depend on their momentum. A crucial
feature of the light-front formulation of a bound state, such as
the one shown in Eq.~(\ref{lfmbs}), is the frame-independence of
the light-front wave function~\cite{Brodsky:1997de,Jaus90}.
Namely, the hadron can be boosted to any (physical) ($P^+$,
$P_\bot$) without affecting the internal variables ($x_i$,
$\vec k_{\bot i}$) of the wave function, which is certainly not the
case in the instant-form formulation.

In practice it is more convenient to use the covariant form for
$\Psi^{1/2 J_z}_{nL_K S_{[qq]} J_l}$: 
\be
\Psi^{1/2 J_z}_{nL_K S_{[qq]} J_l}(\tilde p_1,\tilde p_2,\lambda_1,\lambda_2)
&=&  
      \frac{1}{\sqrt{(M_0+m_1)^2-m_2^2}}
        ~\bar u(p_1,\lambda_1)\Gamma_{L_K S_{[qq]} J_l} u(\bar P,J_z)
\non\\
&&
         ~\phi_{n L_K}(x_1,x_2,k_{1\bot},k_{2\bot}),     
\en
with
\begin{eqnarray}
&&
       \Gamma_{s00}=1,
\non\\
&&
       \Gamma_{s11}
       =\frac{\gamma_5}{\sqrt3}
          \bigg(\not\!\varepsilon_{LF}^*(p_2,\lambda_2)
        -
        \frac{M_0+m_1+m_2}{\bar P\cdot p_2+m_2 M_0}\varepsilon_{LF}^*(p_2,\lambda_2)\cdot \bar P\bigg),
\non\\
&&
     \Gamma_{p01}=\frac{\gamma_5}{2\sqrt3}
     \bigg(\not\! p_1-\not \! p_2-\frac{m_1^2-m_2^2}{M_0} \bigg),
\label{eq: Gamma}
\end{eqnarray}
for baryon states with a $S_2=0$ or $S_2=1$ diquark. 
The derivation of the above results can be found in Appendix \ref{appendix: vertex}.
Note that $\Gamma_{s00}$ agrees with the one in ref.~\cite{Ke:2007tg},
while $\Gamma_{s11}$ and $\Gamma_{p01}$ are new results and $\Gamma_{s11}$ is different from those in ref.~\cite{Ke:2012wa,Zhao:2018zcb,Wang:2017mqp}, which have $\Gamma_{s11}$ proportional to 
$\gamma_5\not\!\varepsilon_{LF}^*(p_2,\lambda_2)$, instead.

It should be remarked that in the conventional LF approach $\bar
P=p_1+p_2$ is not equal to the baryon's four-momentum  as all
constituents are on-shell and consequently $u(\bar P,S_z)$ is not
equal to $u(P,S_z)$; they satisfy different equations of motions
$(\not\!\!\bar P-M_0)u(\bar P,S_z)=0$ and $(\not\!\!P-M)u(
P,S_z)=0$. This is similar to the case of a vector meson bound
state where the polarization vectors $\varepsilon(\bar P,S_z)$ and
$\varepsilon(P,S_z)$ are different and satisfy different equations
$\varepsilon(\bar P,S_z)\cdot\bar P=0$ and
$\varepsilon(P,S_z)\cdot P=0$~\cite{Jaus91}. Although $u(\bar
P,S_z)$ is different than $u(P,S_z)$, they satisfy the relation
 \be
 \gamma^+ u(\bar P,S_z)=\gamma^+ u(P, S_z),
  \en
followed from $\gamma^+\gamma^+=0$, $\bar P^+=P^+$, $\bar
P_\bot=P_\bot$. This is again in analogy with the case of
$\varepsilon(\bar P,\pm 1)=\varepsilon(P,\pm 1)$.

Note that the normalization of state, Eq.~(\ref{wavenor1}), implies
\be
\delta_{J'_z,J_z}
&=&  
             \int \frac{dx_2d^2 k_{2\bot}}{2(2\pi)^3}
              \frac{ \phi^*_{nL_K}(\{x_j\};\{\vec k_{j\bot}\})\,
                \phi_{n L_K}(\{x_i\}; \{\vec k_{i\bot}\})}{2\sqrt{\bar P\cdot p_1+m_1 M'_0}\sqrt{\bar P\cdot p_1+m_1 M_0}}
\non\\                
   &&\qquad
               \times             
                \bar u(\bar P,J'_z)\bar\Gamma_{L_K S_{[qq]} J_l} 
                (\not \! p_1+m_1)
                \Gamma_{L_K S_{[qq]} J_l} u(\bar P,J_z),
\label{eq: J'zJz}                
\en
with $\bar\Gamma_{L_K S_{[qq]} J_l} \equiv\gamma_0\Gamma^\dagger_{L_K S_{[qq]} J_l} \gamma_0$.
To verify it we note that the right-hand-side of Eq.~(\ref{eq: J'zJz}) is a matrix element of a $2\times2$ hermitian matrix. Hence, it's value can be extracted by taking traces with unit and sigma matrices, giving
\be
1 &=&  \frac{1}{2}
             \int \frac{dx_2d^2 k_{2\bot}}{2(2\pi)^3}
              \frac{ \phi^*_{nL_K}(\{x_j\};\{\vec k_{j\bot}\})\,
                \phi_{n L_K}(\{x_i\}; \{\vec k_{i\bot}\})}{2\sqrt{\bar P\cdot p_1+m_1 M'_0}\sqrt{\bar P\cdot p_1+m_1 M_0}}
\non\\                
   &&\qquad
               \times             
                Tr[(\not\!\bar P+M_0)\bar\Gamma_{L_K S_{[qq]} J_l} 
                (\not \! p_1+m_1)
                \Gamma_{L_K S_{[qq]} J_l}],
\label{eq: J'zJz0}                                    
\end{eqnarray}
and
\be
0
&=&  
             \int \frac{dx_2d^2 k_{2\bot}}{2(2\pi)^3}
              \frac{ \phi^*_{nL_K}(\{x_j\};\{\vec k_{j\bot}\})\,
                \phi_{n L_K}(\{x_i\}; \{\vec k_{i\bot}\})}{8 P^+\sqrt{\bar P\cdot p_1+m_1 M'_0}\sqrt{\bar P\cdot p_1+m_1 M_0}}
\non\\                
   &&\qquad
               \times  
               Tr[ (\not\!\bar P+M_0)\gamma^+\gamma_5(\not\!\bar P+M_0)       
               \bar\Gamma_{L_K S_{[qq]} J_l} 
                (\not \! p_1+m_1)
                \Gamma_{L_K S_{[qq]} J_l}],   
\non\\
0
&=&  
             \int \frac{dx_2d^2 k_{2\bot}}{2(2\pi)^3}
              \frac{ \phi^*_{nL_K}(\{x_j\};\{\vec k_{j\bot}\})\,
                \phi_{n L_K}(\{x_i\}; \{\vec k_{i\bot}\})}{8 P^+\sqrt{\bar P\cdot p_1+m_1 M'_0}\sqrt{\bar P\cdot p_1+m_1 M_0}}
\non\\                
   &&\qquad
               \times  
               Tr[ (\not\!\bar P+M_0)\sigma^{i+}\gamma_5(\not\!\bar P+M_0)       
               \bar\Gamma_{L_K S_{[qq]} J_l} 
                (\not \! p_1+m_1)
                \Gamma_{L_K S_{[qq]} J_l}],                              
\label{eq: J'zJz1}                
\en 
where we have made use of the following identities in the above equations,
  \be
 \frac{1}{2}\sum_{J_z,J'_z} u(\bar P, J_z)(\sigma^3)_{J_z J'_z}\bar u(\bar P,J'_z)
  &=&\frac{1}{4P^+}(\not\!\bar P+M_0)\gamma^+\gamma_5(\not\!\bar P+M_0),
 \non\\
 \frac{1}{2}\sum_{J_z,J'_z} u(\bar P, J_z)(\sigma_\bot^i)_{J_z J'_z}\bar u(\bar P,J'_z)
  &=&\frac{i}{4P^+}(\not\!\bar P+M_0)\sigma^{i+}\gamma_5(\not\!\bar P+M'_0).
  \label{eq:spinorprojectionbar0}
 \en
Eqs. (\ref{eq: J'zJz0}) and (\ref{eq: J'zJz1}) are non-trivial requirements and we check that using $\Gamma_{s00}$, $\Gamma_{s11}$ and $\Gamma_{p01}$ in Eq.~(\ref{eq: Gamma}) and $\phi_{nL_K}$ in Eqs.~(\ref{eq: varphi}) and (\ref{eq:wavefn}), the above relations are indeed satisfied.~\footnote{
Note that some authors used vertex functions that do not satisfy Eq.~(\ref{eq: J'zJz0}),
while some authors employed some ad hoc additional normalization factors to the vertex functions in order to satisfy Eq.~(\ref{eq: J'zJz0}).
In this work, Eqs.~(\ref{eq: J'zJz0}) and (\ref{eq: J'zJz1}) are satisfied automatically. 
} 

\subsection{$\B_b(1/2)\to \B_c(1/2)$ weak transitions, a general discussion}

\begin{figure}[t!]
\centerline{
          \includegraphics[width=0.6\textwidth]  {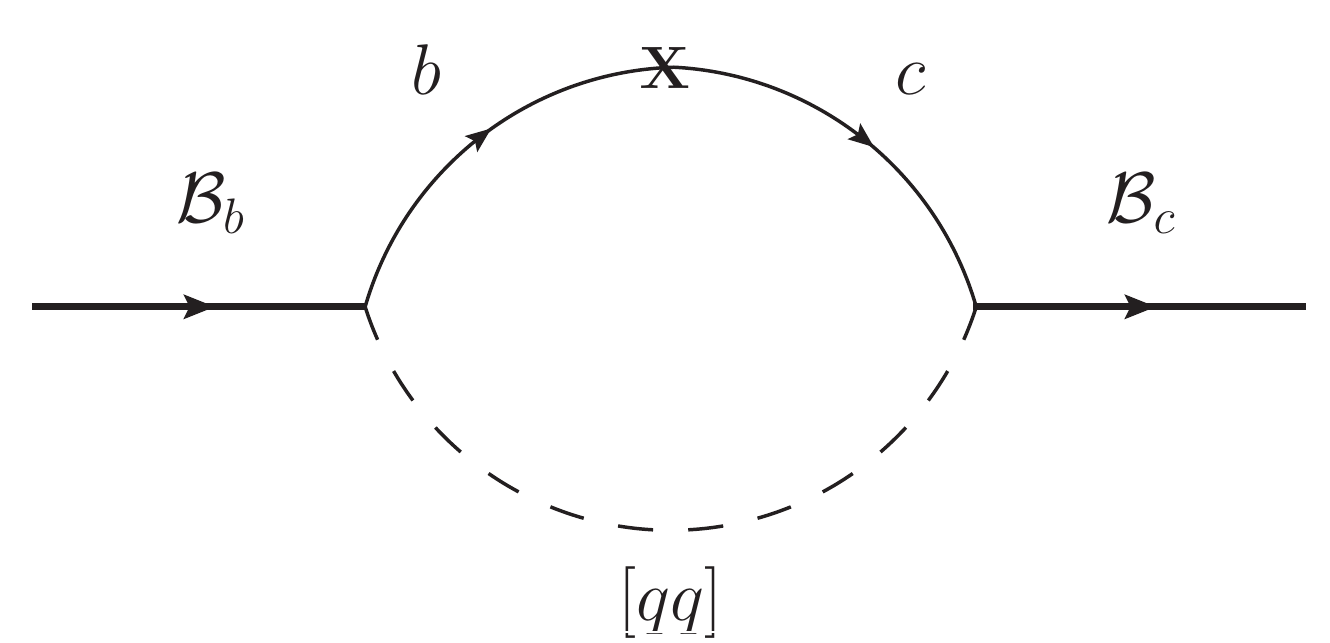}}
\caption{Feynman diagram for a typical $\B_b\to\B_c$
transition, where the scalar or axial-vector diquark 
is denoted by a dashed line and the corresponding $V-A$ current
vertex by X.} \label{fig: BbBc} 
\end{figure}

The Feynman diagram for a typical $\B_b\to\B_c$
transition, is shown in Fig.~\ref{fig: BbBc}. 
For the $\B_{b}(1/2^+)\to \B_{c}(1/2^+)$ transition, 
the matrix element can be parameterized as
 \be
 &&\la \B_{c}(P',J'_z)|\bar c\gamma_\mu  b|\B_b(P,J_z)\ra
 \non\\
 &&\qquad=\bar u(P',J'_z)\Big[f^V_1(q^2)\gamma_\mu+i{f^V_2(q^2)\over M+M'} \sigma_{\mu\nu}q^\nu
    +{f^V_3(q^2)\over M+M'}q_\mu\Big] u(P,J_z),
 \non\\
 &&\la \B_{c}(P',J'_z)|\bar c\gamma_\mu \gamma_5 b|\B_b(P,J_z)\ra
 \non\\
 &&\qquad=\bar u(P',J'_z)\Big[g^A_1(q^2)\gamma_\mu+i{g^A_2(q^2)\over M+M'} \sigma_{\mu\nu}q^\nu
     +{g^A_3(q^2)\over M+M'}q_\mu\Big]\gamma_5 u(P,J_z),
 \label{eq:figi}
 \en
 with $q=P-P'$. 
For the $\B_{b}(1/2^+)\to \B_{c}(1/2^-)$ transition, we have
  \be
 &&\la \B_{c}(P',J'_z)|\bar c\gamma_\mu  b|\B_b(P,J_z)\ra
 \non\\
 &&\qquad=\bar u(P',J'_z)\Big[g^V_1(q^2)\gamma_\mu+i{g^V_2(q^2)\over M+M'} \sigma_{\mu\nu}q^\nu
    +{g^V_3(q^2)\over M+M'}q_\mu\Big] \gamma_5u(P,J_z),
 \non\\
 &&\la \B_{c}(P',J'_z)|\bar c\gamma_\mu \gamma_5 b|\B_b(P,J_z)\ra
 \non\\
 &&\qquad=\bar u(P',J'_z)\Big[f^A_1(q^2)\gamma_\mu+i{f^A_2(q^2)\over M+M'} \sigma_{\mu\nu}q^\nu
     +{f^A_3(q^2)\over M+M'}q_\mu\Big] u(P,J_z).
 \label{eq:figi1}
 \en

Armed with the light-front quark model description of
$|\B_b(P,J_z)\ra$ in the previous subsection, we are ready to
calculate the weak transition matrix element of heavy baryons.
For a $\B_b(1/2)\to\B_c(1/2)$ transition, we have the
general expressions  
\be
&&
     \la \B_{c}(P',J'_z)|\bar c\gamma^\mu (1-\gamma_5) b|\B_{b}(P,J_z)\ra
\non\\
&&\qquad
     =\int \{d^3p_2\}~\frac{\phi^{\prime*}_{nL'_K}(\{x'\},\{k'_{\bot}\})\phi_{1 L_K}(\{x\},\{k_{\bot}\})}
                                       {2\sqrt{p_1^+ p_1^{\prime +}(p_1\cdot \bar P+m_1 M_0)(p'_1\cdot\bar P'+m'_1 M'_0)}}
\non\\
&&
     \qquad\qquad\qquad\times~
     \bar u(\bar P',J'_z)\bar \Gamma_{L'_K S_{[qq]} J'_l} 
     (\not\! p'_1+m'_1)\gamma^\mu(1-\gamma_5)(\not\! p_1+m_1)\Gamma_{L_K S_{[qq]} J_l}  
     u(\bar P,J_z),
\label{eq: B->B'}
\en
where the diquark acts as an spectator and
\begin{eqnarray}
        && p^{(\prime)+}_i=x^{(\prime)}_i P^{(\prime)+},\qquad
           p^{(\prime)}_{i\bot}=x^{(\prime)}_i
           \vec P^{(\prime)}_\bot+\vec k^{(\prime)}_{i\bot},\qquad 1-\sum_{i=1}^2
           x^{(\prime)}_i=\sum_{i=1}^2 \vec k^{(\prime)}_{i\bot}=0,
        \non\\
        && \tilde p_1-\tilde p_1^{\prime}=\tilde q,\qquad
        \tilde p_2=\tilde p_2^{\prime},
\end{eqnarray}
with $\Gamma_{L_K S_{[qq]} J_l}$ given in Eq.
(\ref{eq: Gamma}). 
As in
\cite{CCH,Schlumpf, Cheng:2004cc}, we consider the $q^+=0$, $\vec q_\bot\not=\vec 0$ case.
We follow~\cite{Schlumpf, Cheng:2004cc} to project out various form
factors from the above transition matrix elements (see Appendix B for details). 
The results are given below.

\subsection{Form factors for $\B_b({\bf \bar 3_f}, 1/2^+)\to \B_c({\bf \bar 3_f}, 1/2^+)$ transition [type (i)]}

The $\B_b({\bf \bar 3_f},{1}/{2}^+)\to\B_c({\bf \bar 3_f},{1}/{2}^+)$ transitions involve initial states in
$(n, L_K, S_{[qq]}^P, J_l^P,J^P)_b
 =(1, 0, 0^+,0^+, \frac{1}{2}^+)$ configuration and final states in
$(n, L_K, S_{[qq]}^P, J_l^P,J^P)_c=(n, 0, 0^+,0^+, \frac{1}{2}^+)$ configurations (with $n$=1,2). 
Explicitly, we have 
$\Lambda_b^0\to \Lambda^+_c$, $\Xi_b^{0(-)}\to \Xi_c^{+(0)}$ and $\Lambda_b^0\to\Lambda_c(2765)^+$
transitions, where we follow ref.~\cite{Cheng:2017ove} to take $\Lambda_c(2765)^+$ as a radial excited $s$-wave state. 
In these transitions the scalar diquarks are spectators.

We obtain the following transition form factors for type (i) transition:
 \be
 f^V_1(q^2)&=&\int \frac{dx_2 d^2 k_{2\bot}}{2 (2\pi)^3}~
 \frac{\phi^{\prime*}_{ns}(\{x'\},\{k'_{\bot}\})~\phi_{1s}(\{x\},\{k_{\bot}\})}
 {\sqrt{[(m_1+x_1 M_0)^2+k_{1\bot}^2][(m'_1+x_1 M'_0)^2+k_{1\bot}^{\prime 2}]}}
 \non\\
 &&\qquad\qquad\qquad\qquad\qquad\times~
   [k_{1\bot}\cdot k'_{1\bot}+(m_1+x_1 M_0)(m'_1+x'_1 M'_0)],
 \non\\
 \frac{f^V_2(q^2)}{M+M'}&=&\frac{1}{\vec q^2_\bot}\int \frac{dx_2 d^2 k_{2\bot}}{2 (2\pi)^3}~
 \frac{\phi^{\prime*}_{ns}(\{x'\},\{k'_{\bot}\})~\phi_{1s}(\{x\},\{k_{\bot}\})}
 {\sqrt{[(m_1+x_1 M_0)^2+k_{1\bot}^2][(m'_1+x_1 M'_0)^2+k_{1\bot}^{\prime 2}]}}
 \non\\
 &&\qquad\qquad\qquad\qquad\qquad\times~
   [(m_1+x_1 M_0)~\vec k^{\prime}_{1\bot}\cdot \vec q_\bot-(m'_1+x'_1 M'_0)~\vec k_{1\bot}\cdot \vec q_\bot],
 \non\\
  g^A_1(q^2)&=&\int \frac{dx_2 d^2 k_{2\bot}}{2 (2\pi)^3}~
 \frac{\phi^{\prime*}_{ns}(\{x'\},\{k'_{\bot}\})~\phi_{1s}(\{x\},\{k_{\bot}\})}
 {\sqrt{[(m_1+x_1 M_0)^2+k_{1\bot}^2][(m'_1+x_1 M'_0)^2+k_{1\bot}^{\prime 2}]}}
 \non\\
 &&\qquad\qquad\qquad\qquad\qquad\times~
   [-k_{1\bot}\cdot k'_{1\bot}+(m_1+x_1 M_0)(m'_1+x'_1 M'_0)],
 \non\\
 \frac{g^A_2(q^2)}{M+M'}&=&\frac{1}{\vec q^2_\bot}\int \frac{dx_2 d^2 k_{2\bot}}{2 (2\pi)^3}~
 \frac{\phi^{\prime*}_{ns}(\{x'\},\{k'_{\bot}\})~\phi_{1s}(\{x\},\{k_{\bot}\})}
 {\sqrt{[(m_1+x_1 M_0)^2+k_{1\bot}^2][(m'_1+x_1 M'_0)^2+k_{1\bot}^{\prime 2}]}}
 \non\\
 &&\qquad\qquad\qquad\qquad\qquad\times~
   [(m'_1+x'_1 M'_0)~\vec k_{1\bot}\cdot\vec q_\bot+(m_1+x_1 M_0)~\vec k^{\prime}_{1\bot}\cdot \vec q_\bot].
 \label{eq:ff type i}
 \en
Note that we have $\vec k_{1\bot}-\vec k'_{1\bot}=x_2 \vec q_\bot$ and $q^2=-q^2_\bot$.
For the transition with low laying final state ($n=1$),
the above equations are similar to those obtained in ref.~\cite{Cheng:2004cc} 
and are identical to those in ref.~\cite{Ke:2007tg}.

\subsection{Form factors for $\B_b({\bf 6_f}, 1/2^+)\to \B_c({\bf 6_f}, 1/2^+)$ transition [type (ii)]}

The $\B_b({\bf 6_f}, 1/2^+)\to \B_c({\bf 6_f}, 1/2^+)$ transitions involve initial states in
$(n, L_K, S_{[qq]}^P, J_l^P,J^P)_b
 =(1, 0, 1^+,1^+, \frac{1}{2}^+)$ configuration and final states in
$(n, L_K, S_{[qq]}^P, J_l^P,J^P)_c=(n, 0, 1^+,1^+, \frac{1}{2}^+)$ configurations (with $n$=1,2). 
Explicitly, we have 
$\Omega^-_b\to\Omega^0_c$ and $\Omega^-_b\to\Omega_c(3090)^0$ 
transitions, where we follow ref.~\cite{Cheng:2017ove} to consider $\Omega_c(3090)^0$ as a radial excited $s$-wave state. 
In these transitions the axial-vector diquarks are spectators. 

We obtain the following transition form factors for type (ii) transition:
 \be
 f^V_1(q^2)&=&\int \frac{dx_2 d^2 k_{2\bot}}{2 (2\pi)^3}~
 \frac{\phi^{\prime*}_{ns}(\{x'\},\{k'_{\bot}\})~\phi_{1s}(\{x\},\{k_{\bot}\})}
 {\sqrt{[(m_1+x_1 M_0)^2+k_{1\bot}^2][(m'_1+x_1 M'_0)^2+k_{1\bot}^{\prime 2}]}}
 \non\\
 &&\qquad\qquad\qquad\qquad\qquad\times~
   (A_++B_++C_++D_+),
 \non\\
 \frac{f^V_2(q^2)}{M+M'}&=&\frac{1}{\vec q^2_\bot}\int \frac{dx_2 d^2 k_{2\bot}}{2 (2\pi)^3}~
 \frac{\phi^{\prime*}_{ns}(\{x'\},\{k'_{\bot}\})~\phi_{1s}(\{x\},\{k_{\bot}\})}
 {\sqrt{[(m_1+x_1 M_0)^2+k_{1\bot}^2][(m'_1+x_1 M'_0)^2+k_{1\bot}^{\prime 2}]}}
 \non\\
 &&\qquad\qquad\qquad\qquad\qquad\times~
   (H_++I_++J_++K_+),
 \non\\
  g^A_1(q^2)&=&\int \frac{dx_2 d^2 k_{2\bot}}{2 (2\pi)^3}~
 \frac{\phi^{\prime*}_{ns}(\{x'\},\{k'_{\bot}\})~\phi_{1s}(\{x\},\{k_{\bot}\})}
 {\sqrt{[(m_1+x_1 M_0)^2+k_{1\bot}^2][(m'_1+x_1 M'_0)^2+k_{1\bot}^{\prime 2}]}}
 \non\\
 &&\qquad\qquad\qquad\qquad\qquad\times~
    (A_-+B_-+C_-+D_-),
 \non\\
 \frac{g^A_2(q^2)}{M+M'}&=&\frac{1}{\vec q^2_\bot}\int \frac{dx_2 d^2 k_{2\bot}}{2 (2\pi)^3}~
 \frac{\phi^{\prime*}_{ns}(\{x'\},\{k'_{\bot}\})~\phi_{1s}(\{x\},\{k_{\bot}\})}
 {\sqrt{[(m_1+x_1 M_0)^2+k_{1\bot}^2][(m'_1+x_1 M'_0)^2+k_{1\bot}^{\prime 2}]}}
 \non\\
 &&\qquad\qquad\qquad\qquad\qquad\times~
   (H_-+I_-+J_-+K_-),
 \label{eq:ff type ii}
 \en
where we have
\be
A_+&=&\frac{2 e_2 M_0}{6 m_2^2}
\bigg[
 \left[4 {e'_2} {M'_0}+x_2 ({M'_0}+{m'_1}) (M_0-2 {M'_0}+m_1)+2 m_2^2
   (x_2-1)\right]
\non\\
&&   
   +2 x_2 
   \bigg({e'_2} {M'_0} \left[(M_0+m_1) (-2 M_0+{M'_0}+{m'_1})+2
   m_2^2\right]
\non\\
&&   
   +m_2^2 \left[M_0^2-M_0 (4 {M'_0}+m_1+2 {m'_1})+{M'_0} ({M'_0}-2
   m_1-{m'_1})+2 q^2\right]\bigg)
\non\\
&&   
   +2 m_2^2 \left[-2 {e'_2} {M'_0}+{m'_1} (M_0+2
   {M'_0}+m_1)+M_0 {M'_0}+2 M_0 m_1+{M'_0} m_1-m_2^2-2 q^2\right]
\non\\
&&   
   -x_2^2
   \left[(M_0-{M'_0})^2+q^2\right] \left((M_0+m_1) ({M'_0}+{m'_1})+2 m_2^2\right)
   \bigg],
\en
\be
B_+&=&
\frac{1}{6 M_0 m_2^2 (e_2+m_2)}
\bigg[
-2 e_2^2 M_0^2 \left[4 {e'_2} {M'_0}+x_2 ({M'_0}+{m'_1}) (M_0-2 {M'_0}+m_1+m_2)\right]
\non\\
&&
+e_2
   M_0 \bigg(-2 x_2 \Big\{{e'_2} {M'_0} (-2 M_0+{M'_0}+{m'_1}) (M_0+m_1+m_2)
\non\\
&&  
   +m_2
   \Big[M_0^2 m_2-M_0 {M'_0} ({M'_0}+{m'_1}+2 m_2)
\non\\
&&   
   +{M'_0} ({M'_0} (m_1+2
   m_2)+(m_1+m_2) ({m'_1}-2 m_2))+m_2 q^2\Big]\Big\}
\non\\
&&   
   +4 m_2 \left({e'_2} {M'_0}
   (-M_0+m_1+2 m_2)+m_2 \left(M_0^2-{M'_0} {m'_1}+q^2\right)\right)
\non\\
&&   
   +x_2^2 ({M'_0}+{m'_1})
   \left((M_0-{M'_0})^2+q^2\right) (M_0+m_1+m_2)\bigg)
\non\\
&&   
   +m_2^2 
   \bigg(2 {e'_2} {M'_0}
         \{M_0^2 (-x_2)
         +M_0 [{M'_0}-x_2 (m_1+m_2)+{m'_1}+m_2]
\non\\
&&         
         +(m_1+m_2)
               ({M'_0}+{m'_1}-m_2)
          \}         
          -2 m_2 
          [-M_0^3+M_0^2 (m_1+m_2)
\non\\
&&           
          +M_0 {M'_0}
          ({m'_1}+m_2)-M_0 q^2+(m_1+m_2) (-{M'_0} {m'_1}+{M'_0} m_2+q^2)
          ]
\non\\
&&          
          -x_2
          \{2 M_0^4-M_0^3 ({M'_0}-2 m_1+{m'_1}-3 m_2)
          -M_0^2 [{M'_0} (m_1+2{m'_1}+m_2)
\non\\
&&                  
                 +(m_1+m_2) ({m'_1}+m_2)-2 q^2]               
                 +M_0 {M'_0} 
                 [{M'_0}^2+{M'_0}
                        ({m'_1}+m_2)-2 {m'_1} (m_1+m_2)
                 ]
\non\\
&&                 
                 +M_0 q^2 ({M'_0}+2 m_1+{m'_1}+3 m_2)+
                 \left({M'_0}^2+q^2\right) (m_1+m_2) ({M'_0}+{m'_1}-m_2)
           \}
\non\\
&&           
           +M_0 x_2^2
           [(M_0-{M'_0})^2+q^2] (M_0+m_1+m_2)
           \bigg)
   \bigg],
\en
\be
C_+&=&
\frac{1}{6 {M'_0} m_2^2 ({e'_2}+m_2)}
\bigg[
2 e_2 M_0 
\bigg(-4 {e'_2}^2 {M'_0}^2+{e'_2} {M'_0} (2 m_2 (-{M'_0}+{m'_1}+2 m_2)
\non\\
&&
-x_2
   (M_0-2 {M'_0}+m_1) ({M'_0}+{m'_1}+m_2))
\non\\
&&   
   +m_2^2         
        \{M_0 ({M'_0}+{m'_1}+m_2)
        +{M'_0}^2(-x_2)
        +{M'_0} (m_1-x_2 ({m'_1}+m_2)+m_2)
\non\\
&&
        +(m_1-m_2)
   ({m'_1}+m_2)
        \}
\bigg)
+x_2 
\bigg(-2 {e'_2}^2 {M'_0}^2 (M_0+m_1) (-2
   M_0+{M'_0}+{m'_1}+m_2)
\non\\
&&   
   -2 {e'_2} {M'_0} m_2 
   [M_0^2 (-{M'_0}+{m'_1}+2 m_2)-M_0{M'_0} (m_1+2 m_2)
\non\\
&&   
   +M_0 (m_1-2 m_2) ({m'_1}+m_2)+m_2 ({M'_0}^2+q^2)]
\non\\
&&    
    +m_2^2 
    \{M_0^3 (-{M'_0}-{m'_1}-m_2)-M_0^2 ({M'_0}
   (m_1+m_2)+(m_1-m_2) ({m'_1}+m_2))
\non\\
&&   
   +M_0 
           ({M'_0}^2+2 {M'_0} m_1-q^2)
   ({M'_0}+{m'_1}+m_2)
\non\\
&&   
   -q^2 
           [2 {M'_0}^2+{M'_0} (m_1+2 {m'_1}+3 m_2)
           +(m_1-m_2)({m'_1}+m_2)]
\non\\
&&           
           +{M'_0}^2 
           \left(-2 {M'_0}^2+{M'_0} (m_1-2 {m'_1}-3 m_2)+(m_1+m_2)
   ({m'_1}+m_2)
           \right)
   \}
\bigg)
\non\\
&&
 -2 m_2^2 
\bigg(m_2 
      \{M_0 {M'_0} (m_1+m_2)
      +({m'_1}+m_2) (-M_0 m_1+M_0 m_2+q^2)
      -{M'_0}^3
\non\\
&&      
      +{M'_0}^2({m'_1}+m_2)-{M'_0} q^2
      \}   
      -2 {e'_2} {M'_0} 
      \left(-M_0
   m_1+{M'_0}^2+q^2
      \right)
\bigg)
\non\\
&&
+{M'_0} x_2^2 \left((M_0-{M'_0})^2+q^2\right)
   ({M'_0}+{m'_1}+m_2) \left({e'_2} (M_0+m_1)+m_2^2\right)
\bigg],
\en
\be
D_+
&=&\frac{   \left(m_2^2 \left(M_0^2+{M'_0}^2+q^2\right)-2 e_2 {e'_2} M_0 {M'_0}\right)}{12 M_0 {M'_0} m_2^2 (e_2+m_2)
   ({e'_2}+m_2)}
\non\\
&&
\times   
\bigg[   
-2 x_2
          \{e_2 M_0 ({M'_0}+{m'_1}+m_2) (M_0-2 {M'_0}+m_1+m_2)
\non\\
&&          
          +m_2^2 \left({M'_0}
   ({e'_2}+{M'_0}+m_1+{m'_1})+M_0^2+M_0 (-2 {M'_0}+m_1+{m'_1})
                 \right)
\non\\
&&                 
   +m_2 [{e'_2} {M'_0}(-M_0+{M'_0}+m_1+{m'_1})
   +m_1 {m'_1} (M_0+{M'_0})
\non\\
&&   
   +{M'_0} m_1 ({M'_0}-M_0)
   +M_0{m'_1} (M_0-{M'_0})
   -M_0 {M'_0} (M_0+{M'_0})]
\non\\
&&   
   +{e'_2} {M'_0} (M_0+m_1) (-2
   M_0+{M'_0}+{m'_1})+m_2^3 (M_0+{M'_0})
   \}
\non\\
&&   
   +4 (e_2 M_0 (m_2
   (-{M'_0}+{m'_1}+m_2)-2 {e'_2} {M'_0})+m_2 ({e'_2} {M'_0} (-M_0+m_1+m_2)
\non\\
&&   
   +M_0
   m_2 ({m'_1}+m_2)+{M'_0} m_2 (m_1+m_2)))
\non\\
&&   
   +x_2^2 \left((M_0-{M'_0})^2+q^2\right)
   (M_0+m_1+m_2) ({M'_0}+{m'_1}+m_2)\bigg],
\en
\be
D_-&=&D_++\frac{(2 e_2 M_0 e_2' M_0'-m_2^2 \left(M_0^2+\left(M_0'\right)^2+q^2\right))}
{6 M_0 m_2^2(e_2+m_2) M_0' (e_2'+m_2)}
 (M_0+m_1+m_2) \left(M_0'+m_1'+m_2\right)
\non\\
 &&\qquad\qquad  \times\left\{x_2 \left[-2 e_2 M_0-2 e'_2 M'_0+x_2
  \left(M_0^2+M_0^{\prime 2}+q^2\right)\right]+2 m_2^2\right\},
\en
\be
H_+&=&\frac{1}{3 m_2^2}
\bigg[-2 e_2 M_0 ({M'_0}+{m'_1}) (\vec k_{2\bot}\cdot \vec q_\bot-q^2 x_2)+2 {e'_2} \vec k_{2\bot}\cdot \vec q_\bot {M'_0}
   (M_0+m_1)
\non\\
&&   
   +x_2 \left(\vec k_{2\bot}\cdot \vec q_\bot (M_0-{M'_0}) \left((M_0+m_1) ({M'_0}+{m'_1})+2
   m_2^2\right)+m_2^2 q^2 (4 M_0+{M'_0}+2 m_1+{m'_1})\right)
\non\\
&&   
   +m_2^2 (\vec k_{2\bot}\cdot \vec q_\bot (-3 M_0+3
   {M'_0}-m_1+{m'_1})-2 q^2 (M_0+{M'_0}+m_1+{m'_1}))
\non\\
&&   
   -M_0 q^2 x_2^2
   \left((M_0+m_1) ({M'_0}+{m'_1})+2 m_2^2\right)
\bigg],   
\en
\be
I_+&=&\frac{1}{3 M_0 m_2^2 (e_2+m_2)}
\bigg[\vec k_{2\bot}\cdot \vec q_\bot 
   \bigg(2 e_2^2 M_0^2 ({M'_0}+{m'_1})  
   +e_2 M_0 
         \{-2 {e'_2} {M'_0}
   (M_0+m_1+m_2)
\non\\
&&   
   +m_2 
               \left(m_2 (M_0-3 {M'_0}+m_1-{m'_1})+(M_0-m_1)
   ({M'_0}+{m'_1})+m_2^2
               \right)
\non\\
&&               
               -x_2 (M_0-{M'_0}) ({M'_0}+{m'_1})
   (M_0+m_1+m_2)
          \}
\non\\
&&          
 +m_2^2 
          \{M_0^3 (1-x_2)
          +M_0^2 [{M'_0} x_2-x_2(m_1+m_2)+m_1+m_2]
\non\\
&&
          +M_0 [{M'_0} x_2 (m_1+m_2)-{M'_0}({m'_1}+m_2)+q^2]
          +(m_1+m_2) ({M'_0} (m_2-{m'_1})+q^2)
          \}
     \bigg)
\non\\
&&     
     +q^2 
     \bigg(-M_0 x_2    
              \{2 e_2^2 M_0 ({M'_0}+{m'_1})
              +e_2 m_2 (M_0 ({M'_0}+{m'_1}+2m_2)
   -(m_1+m_2) ({M'_0}+{m'_1}-2 m_2))
\non\\
&&   
   +m_2^2 ({M'_0}+{m'_1})
   (M_0+m_1+m_2)
            \}
\non\\
&&            
            +M_0^2 x_2^2 (M_0+m_1+m_2) 
            \left(e_2
   ({M'_0}+{m'_1})+m_2^2
            \right)
\non\\
&&            
            +m_2^2 (2 e_2 M_0 ({M'_0}+{m'_1})+M_0 m_2
   ({M'_0}+{m'_1}+m_2)-m_2 (m_1+m_2) ({M'_0}+{m'_1}-m_2))
   \bigg)
   \bigg], 
\en
\be
J_+&=&\frac{1  }{3 {M'_0} m_2^2 ({e'_2}+m_2)}
\bigg[
\vec k_{2\bot}\cdot \vec q_\bot 
\bigg(2 e_2 {e'_2} M_0 {M'_0} ({M'_0}+{m'_1}+m_2)-2 {e'_2}^2 {M'_0}^2
   (M_0+m_1)
\non\\
&&   
   +{e'_2} {M'_0} \{x_2 ({M'_0}-M_0) (M_0+m_1) ({M'_0}+{m'_1}+m_2)
\non\\
&&   
   +m_2
   [M_0 (-{M'_0}+{m'_1}+3 m_2)-{M'_0} (m_1+m_2)+(m_1-m_2) ({m'_1}+m_2)]\}
\non\\
&&   
   -m_2^2
   \{{M'_0}^2 [M_0 x_2-x_2 ({m'_1}+m_2)+{m'_1}+m_2]
   +{M'_0} [-M_0(m_1+m_2)
\non\\
&&
   +M_0 x_2 ({m'_1}+m_2)+q^2]
   +({m'_1}+m_2) [M_0 (m_2-m_1)+q^2]
   +{M'_0}^3 (1-x_2)\}
\bigg)
\non\\
&&   
   +q^2 
   \bigg({e'_2} {M'_0} 
            \{-x_2 
                  [2
   e_2 M_0 ({M'_0}+{m'_1}+m_2)+m_2^2 (2 M_0+{M'_0}+{m'_1}+m_2)
                  ]
\non\\
&&                  
                  +M_0 x_2^2
   (M_0+m_1) ({M'_0}+{m'_1}+m_2)+2 m_2^2 (M_0+m_1)
            \}
\non\\
&&            
            +m_2^2 
            \{M_0 x_2
                   [m_2 (-2 {M'_0}-m_1+{m'_1})-({M'_0}+m_1) ({M'_0}+{m'_1})+m_2^2
                   ]
\non\\
&&                   
                   +m_2^2
   (-M_0+{M'_0}-m_1+{m'_1})+m_2 (M_0+m_1) ({M'_0}-{m'_1})+M_0 {M'_0} x_2^2
   ({M'_0}+{m'_1}+m_2)
\non\\
&&   
   +x_2 
                   \left({M'_0}^2-{M'_0} m_1+q^2\right)
   ({M'_0}+{m'_1}+m_2)+m_2^3
           \}
    \bigg)
\bigg] ,
\en
\be
K_+&=&\frac{\left(m_2^2 \left(M_0^2+{M'_0}^2+q^2\right)-2 e_2 {e'_2} M_0 {M'_0}\right) }
                     {6 M_0 {M'_0} m_2^2 (e_2+m_2)({e'_2}+m_2)}
\non\\
&&
\times                     
\bigg[q^2 x_2
   ({M'_0}+{m'_1}+m_2) 
   [-2 e_2 M_0+M_0 x_2 (M_0+m_1+m_2)+m_2(-M_0+m_1+m_2)]
\non\\
&&   
   -\vec k_{2\bot}\cdot \vec q_\bot \bigg(-2 e_2 M_0 ({M'_0}+{m'_1}+m_2)+2 {e'_2} {M'_0}
   (M_0+m_1+m_2)
\non\\
&&   
   +x_2 (M_0-{M'_0}) (M_0+m_1+m_2) ({M'_0}+{m'_1}+m_2)
\non\\
&&   
   -2 M_0
   m_2 ({m'_1}+m_2)+2 {M'_0} m_2 (m_1+m_2)\bigg)
\bigg],   
\en
\be
K_-&=&K_+-\frac{\vec k_{2\bot}\cdot \vec q_\bot (M_0+m_1+m_2)}
{3 M_0 M_0' m_2^2(e_2+m_2) (e_2'+m_2)}
\non\\
&&\times
 \left(m_2^2 \left(M_0^2+\left(M_0'\right)^2+q^2\right)-2 e_2 e_2' M_0 M_0'\right) 
\non\\
&&\times   
[-2 e'_2 M'_0+M'_0 x_2 (M'_0+m'_1+m_2)+m_2 (-M'_0+m'_1+m_2)],
\en
and
$A_-$ is equal to $A_+$, but with $M'_0\to -M'_0$ and $m'_1\to -m'_1$;
$(e_2+m_2) B_-$ is equal to $(e_2+m_2) B_+$, 
but with $M'_0\to -M'_0$, $m'_1\to -m'_1$ and $e'_2\to -e'_2$ ;
$(e'_2+m_2) C_-$ is equal to $-(e'_2+m_2) C_+$, 
but with $M'_0\to -M'_0$, $m'_1\to -m'_1$, $e'_2\to -e'_2$ and $m_2\to -m_2$;
$H_-$ is equal to $-H_+$, but with $M'_0\to -M'_0$ and $m'_1\to -m'_1$;
$(e_2+m_2) I_-$ is equal to $-(e_2+m_2) I_+$, 
but with $M'_0\to -M'_0$, $m'_1\to -m'_1$ and $e'_2\to -e'_2$ ;
$(e'_2+m_2) J_-$ is equal to $(e'_2+m_2) J_+$, 
but with $M'_0\to -M'_0$, $m'_1\to -m'_1$, $e'_2\to -e'_2$ and $m_2\to -m_2$.
Note that we have $\vec k_{1\bot}-\vec k'_{1\bot}=x_2 \vec q_\bot$ and $q^2=-q^2_\bot$.
The above formulas of the form factors are new results.

\subsection{Form factors for $\B_b({\bf \bar 3_f}, 1/2^+)\to \B_c({\bf \bar 3_f}, 1/2^-)$ transition [type (iii)]}

The $\B_b({\bf \bar 3_f},{1}/{2}^+)\to\B_c({\bf \bar 3_f},{1}/{2}^-)$ transitions
involve initial states in
$(n, L_K, S_{[qq]}^P, J_l^P,J^P)_b
 =(1, 0, 0^+,0^+, \frac{1}{2}^+)$ configuration and final states in
$(n, L_K, S_{[qq]}^P, J_l^P,J^P)_c=(n, 1, 0^+,1^-, \frac{1}{2}^-)$ configurations (with $n$=1,2). 
Explicitly, we have 
$\Lambda_b^0\to\Lambda_c(2595)^+$, $\Xi^{0(-)}_b\to\Xi_c(2790)^{+(0)}$ and $\Lambda_b^0\to\Lambda_c(2940)^+$ 
transitions, where we follow ref.~\cite{Cheng:2017ove} to consider $\Lambda_c(2940)^+$ as a radial excited $p$-wave state. 
In these transitions, the scalar diquarks are spectators.

We obtain the following transition form factors for type (iii) transition:
 \be
 f^A_1(q^2)&=&\int \frac{dx_2 d^2 k_{2\bot}}{2 (2\pi)^3}~
 \frac{\phi^{\prime*}_{np}(\{x'\},\{k'_{\bot}\})~\phi_{1s}(\{x\},\{k_{\bot}\})R_+}
 {\sqrt{[(m_1+x_1 M_0)^2+k_{1\bot}^2][(m'_1+x_1 M'_0)^2+k_{1\bot}^{\prime 2}]}} ,
 \non\\
 \frac{f^A_2(q^2)}{M+M'}&=&\frac{1}{\vec q^2_\bot}\int \frac{dx_2 d^2 k_{2\bot}}{2 (2\pi)^3}~
 \frac{\phi^{\prime*}_{np}(\{x'\},\{k'_{\bot}\})~\phi_{1s}(\{x\},\{k_{\bot}\})S_+}
 {\sqrt{[(m_1+x_1 M_0)^2+k_{1\bot}^2][(m'_1+x_1 M'_0)^2+k_{1\bot}^{\prime 2}]}},
 \non\\
  g^V_1(q^2)&=&\int \frac{dx_2 d^2 k_{2\bot}}{2 (2\pi)^3}~
 \frac{\phi^{\prime*}_{np}(\{x'\},\{k'_{\bot}\})~\phi_{1s}(\{x\},\{k_{\bot}\}) R_-}
 {\sqrt{[(m_1+x_1 M_0)^2+k_{1\bot}^2][(m'_1+x_1 M'_0)^2+k_{1\bot}^{\prime 2}]}},
 \non\\
 \frac{g^V_2(q^2)}{M+M'}&=&\frac{1}{\vec q^2_\bot}\int \frac{dx_2 d^2 k_{2\bot}}{2 (2\pi)^3}~
 \frac{\phi^{\prime*}_{np}(\{x'\},\{k'_{\bot}\})~\phi_{1s}(\{x\},\{k_{\bot}\}) S_-}
 {\sqrt{[(m_1+x_1 M_0)^2+k_{1\bot}^2][(m'_1+x_1 M'_0)^2+k_{1\bot}^{\prime 2}]}}, 
 \label{eq:ff type iii} 
 \en
where we have
\be
R_+
&=&
\frac{1}{4 \sqrt{3}{M'_0}}
\bigg[
2 x_2 \bigg(({M'_0}+{m'_1}) \left(e_2 M_0 ({M'_0}+{m'_1})-{M'_0} {m'_1} (3
   M_0+m_1)+M_0 {m'_1}^2+{M'_0}^2 (-m_1)\right)
\non\\
&&   
   +m_2^2 (-e_2 M_0+4 M_0 {M'_0}-M_0
   {m'_1}+{M'_0} m_1)\bigg)
\non\\
&&   
   +2 {e'_2} {M'_0} \left(4 {M'_0} (M_0+m_1)+x_2 \left(-4 M_0
   {M'_0}+({M'_0}+{m'_1})^2-m_2^2\right)\right)
\non\\
&&   
   -2 m_2^2 \left(-{m'_1} (M_0-2 {M'_0}+m_1)+{M'_0} (3
   M_0+{M'_0}+3 m_1)+{m'_1}^2\right)
\non\\
&&   
   -2 (M_0+m_1) ({M'_0}-{m'_1})^2 ({M'_0}+{m'_1})
\non\\
&&   
   -x_2^2
   \left((M_0-{M'_0})^2+q^2\right) 
   ({M'_0}+{m'_1}-m_2) ({M'_0}+{m'_1}+m_2)+2 m_2^4
   \bigg],
\en
\be
S_+
&=&
\frac{1}{2 \sqrt{3} {M'_0}}
\bigg[
4 {e'_2} \vec k_{2\bot}\cdot \vec q_\bot {M'_0}^2
+\vec k_{2\bot}\cdot \vec q_\bot 
\bigg(M_0 (x_2-1) \left({M'_0}^2+2 {M'_0}
   {m'_1}+{m'_1}^2-m_2^2\right)
\non\\
&&   
   +{M'_0}^3 (-x_2-1)+{M'_0}^2 (-m_1-2 {m'_1} x_2+{m'_1})+{M'_0}
   \left(-2 m_1 {m'_1}+{m'_1}^2 (1-x_2)+m_2^2 (x_2-3)\right)
\non\\
&&   
   -(m_1+{m'_1})
   \left({m'_1}^2-m_2^2\right)\bigg)   
   +q^2 x_2 (M_0 (-x_2)+M_0+m_1) \left({M'_0}^2+2 {M'_0}
   {m'_1}+{m'_1}^2-m_2^2\right)
\bigg]   ,
\non\\
\en
\be
R_-&=&
\frac{1}{4 \sqrt{3} {M'_0}}
\bigg[
-2 x_2 ({M'_0}+{m'_1}) \left(e_2 M_0 ({M'_0}+{m'_1})+{M'_0} {m'_1} (3
   M_0+m_1)-M_0 {m'_1}^2+{M'_0}^2 m_1\right)
\non\\
&&   
   +2 m_2^2 x_2 (e_2 M_0+4 M_0
   {M'_0}-M_0 {m'_1}+{M'_0} m_1)
\non\\
&&   
   +2 {e'_2} {M'_0} \left(4 {M'_0} (M_0+m_1)-x_2 \left(4
   M_0 {M'_0}+({M'_0}+{m'_1})^2-m_2^2\right)\right)
\non\\
&&   
   +2 m_2^2 \left({m'_1} (M_0+2
   {M'_0}+m_1)+{M'_0} (-3 M_0+{M'_0}-3 m_1)+{m'_1}^2\right)
\non\\
&&   
   - 
   2 (M_0+m_1) ({M'_0}-{m'_1})^2
   ({M'_0}+{m'_1})
\non\\
&&   
   +x_2^2 \left((M_0+{M'_0})^2+q^2\right) ({M'_0}+{m'_1}-m_2)
   ({M'_0}+{m'_1}+m_2)-2 m_2^4
   \bigg],
\en
and
\be
S_-&=&\frac{1}{2 \sqrt{3} {M'_0}}
\bigg[
-4 {e'_2} \vec k_{2\bot}\cdot \vec q_\bot {M'_0}^2
+\vec k_{2\bot}\cdot \vec q_\bot 
   \bigg(M_0 (x_2-1) \left({M'_0}^2+2 {M'_0}
   {m'_1}+{m'_1}^2-m_2^2\right)
\non\\
&&   
   +{M'_0}^3 (x_2+1)-{M'_0}^2 (m_1-2 {m'_1} x_2+{m'_1})-2 {M'_0}
   m_1 {m'_1}+{M'_0} {m'_1}^2 (x_2-1)
\non\\
&&   
   -{M'_0} m_2^2 (x_2-3)-(m_1-{m'_1})
   \left({m'_1}^2-m_2^2\right)\bigg)
\non\\
&&  
   +q^2 x_2 (M_0 (-x_2)+M_0+m_1) \left({M'_0}^2+2 {M'_0}
   {m'_1}+{m'_1}^2-m_2^2\right)
\bigg]. 
\en
Note that we have $\vec k_{1\bot}-\vec k'_{1\bot}=x_2 \vec q_\bot$ and $q^2=-q^2_\bot$.
The above formulas of the form factors are new results.

\section{Numerical results}

In this section we will show the numerical results of
various $\B_b\to \B_c$ transition form factors using formulas obtained in Sec. III. 
We then proceed to estimate
the decay rates and up-down asymmetries of
$\Lambda_b\to \Lambda^{(*,**)}_c M^-$, $\Xi_b\to\Xi_c^{(**)} M^-$ and $\Omega_b\to\Omega^{(*)}_c M^-$ decays
using na\"{i}ve factorization.

\subsection{$\B_b\to \B_c$ form factors}

\begin{table}[t!]
\caption{\label{tab:input} The input
parameters $m^{S,A}_{[qq']}$, $m_q$ and $\beta$'s (in units of GeV)
appearing in the Gaussian-type wave function (\ref{eq:wavefn}).
(The superscript $S$ and $A$ mean scalar and axial vector, repestively.)
The constituent quark and diquark masses are taken from ref.~\cite{Ebert:2010af}.}
\begin{ruledtabular}
\begin{tabular}{ccccccc}
           $m^S_{[ud]}$
          & $m^S_{[us]}$
          & $m^A_{[ss]}$
          & $m_b$
          & $m_c$
          & $\beta_{b}$
          & $\beta_{c}$
          \\
\hline    
           $0.710$
          & $0.948$
          & $1.203$
          & $4.88$
          & $1.55$
          & $0.72$
          & $0.37$
\end{tabular}
\end{ruledtabular}
\end{table}

\begin{table}[t!]
\caption{\label{tab:fg type i} The transition form factors for various
$\B_b({\bf \bar 3_f},1/2^+)\to\B_c({\bf \bar 3_f},1/2^+)$ transitions [type (i)]. 
We use a three parameter form for these form factors, see Eq.~(\ref{eq:FFpara1}).
}
\begin{ruledtabular}
\begin{tabular}{ccccccccc}
 $\B_b\to\B_c$
          & $F$
          & $F(0)$
          & $a$
          & $b$
          & $F$
          & $F(0)$
          & $a$
          & $b$
          \\
\hline     
$\Lambda_b\to\Lambda_c$          
          & $f_1^V$
          & 0.48
          & 0.42
          & 0.31
          & $g_1^A$
          & 0.47
          & 0.40
          & 0.32
          \\
          &$f_2^V$
          & $-0.05$
          & 1.02
          & 0.64
          & $g_2^A$
          & $-0.14$
          & 0.77
          & 0.50
          \\
\hline     
$\Lambda_b\to\Lambda_c(2765)$          
          & $f_1^V$
          & 0.34
          & 0.57
          & 0.58
          & $g_1^A$
          & 0.33
          & 0.55
          & 0.58
          \\          
          &$f_2^V$
          & $-0.07$
          & 1.07
          & 0.84
          & $g_2^A$
          & $-0.10$
          & 0.57
          & 0.87
          \\  
\hline     
$\Xi_b\to\Xi_c$          
          & $f_1^V$
          & 0.40
          & 1.02
          & 0.84
          & $g_1^A$
          & 0.39
          & 0.99
          & 0.82
          \\          
          &$f_2^V$
          & $-0.05$
          & 1.58
          & 1.67
          & $g_2^A$
          & $-0.14$
          & 1.36
          & 1.34
          \\                                     
\end{tabular}
\end{ruledtabular}
\end{table}

\begin{table}[t!]
\caption{\label{tab:fg type ii} The transition form factors for various
$\B_b({\bf 6_f},1/2^+)\to\B_c({\bf 6_f},1/2^+)$ transitions [type (ii)]. 
We use a three parameter form for these form factors, see Eq.~(\ref{eq:FFpara1}).
}
\begin{ruledtabular}
\begin{tabular}{ccccccccc}
 $\B_b\to\B_c$
          & $F$
          & $F(0)$
          & $a$
          & $b$
          & $F$
          & $F(0)$
          & $a$
          & $b$
          \\
\hline     
$\Omega_b\to\Omega_c$          
          & $f_1^V$
          & 0.32
          & 0.35
          & 1.36
          & $g_1^A$
          & $-0.11$
          & 1.76
          & $-0.07$
          \\  
          &$f_2^V$
          & $0.43$
          & 1.30
          & 2.14
          & $g_2^A$
          & $-0.013$
          & $-5.91$
          & $10.55$
          \\ 
\hline     
$\Omega_b\to\Omega_c(3090)$          
          & $f_1^V$
          & 0.20
          & 0.58
          & 2.79
          & $g_1^A$
          & $-0.07$
          & 2.47
          & 1.27
          \\ 
          &$f_2^V$
          & $0.29$
          & 1.55
          & 5.06
          & $g_2^A$
          & $-0.018$
          & $-6.18$
          & $15.50$
          \\                                                 
\end{tabular}
\end{ruledtabular}
\end{table}

\begin{table}[t!]
\caption{\label{tab:fg type iii} The transition form factors for various
$\B_b({\bf 3_f},1/2^+)\to\B_c({\bf 3_f},1/2^-)$ transitions [type (iii)]. 
We use a three parameter form for these form factors, see Eq.~(\ref{eq:FFpara1}).
}
\begin{ruledtabular}
\begin{tabular}{ccccccccc}
 $\B_b\to\B_c$
          & $F$
          & $F(0)$
          & $a$
          & $b$
          & $F$
          & $F(0)$
          & $a$
          & $b$
          \\
\hline     
$\Lambda_b\to\Lambda_c(2595)$          
          & $f_1^A$
          & 0.31
          & $-0.94$
          & 1.08
          & $g_1^V$
          & 0.27
          & 0.40
          & $-0.04$
          \\     
          &$f_2^A$
          & $-0.34$
          & 0.46
          & 0.43
          & $g_2^V$
          & $-0.27$
          & 0.79
          & 0.56
          \\ 
\hline     
$\Lambda_b\to\Lambda_c(2940)$          
          & $f_1^A$
          & 0.29
          & $-1.14$
          & 1.36
          & $g_1^V$
          & 0.25
          & 0.69
          & $-0.04$
          \\       
          &$f_2^A$
          & $-0.33$
          & 0.41
          & 0.58
          & $g_2^V$
          & $-0.25$
          & 0.79
          & 0.77
          \\
\hline     
$\Xi_b\to\Xi_c(2790)$          
          & $f_1^A$
          & 0.30
          & $-0.55$
          & 1.19
          & $g_1^V$
          & 0.26
          & 1.09
          & 0.30
          \\ 
          &$f_2^A$
          & $-0.37$
          & 1.06
          & 1.03
          & $g_2^V$
          & $-0.28$
          & 1.41
          & 1.51
          \\                                                                      
\end{tabular}
\end{ruledtabular}
\end{table}

\begin{figure}[t!]
\centering
\subfigure[]{
 \includegraphics[width=0.44\textwidth]  {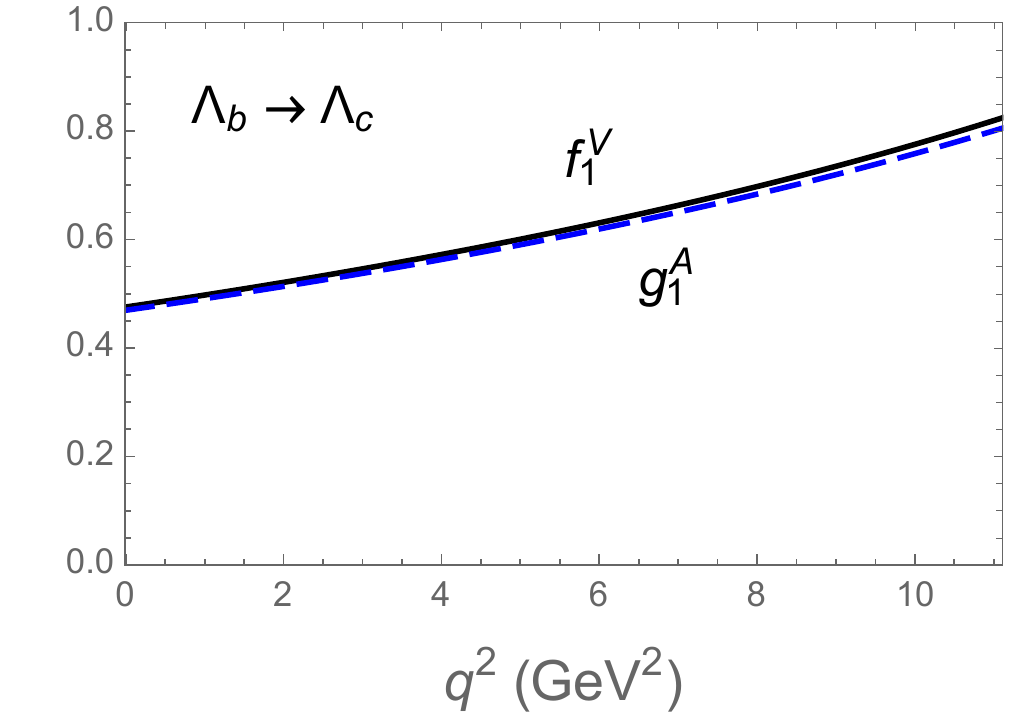}
}
\subfigure[]{
  \includegraphics[width=0.44\textwidth]  {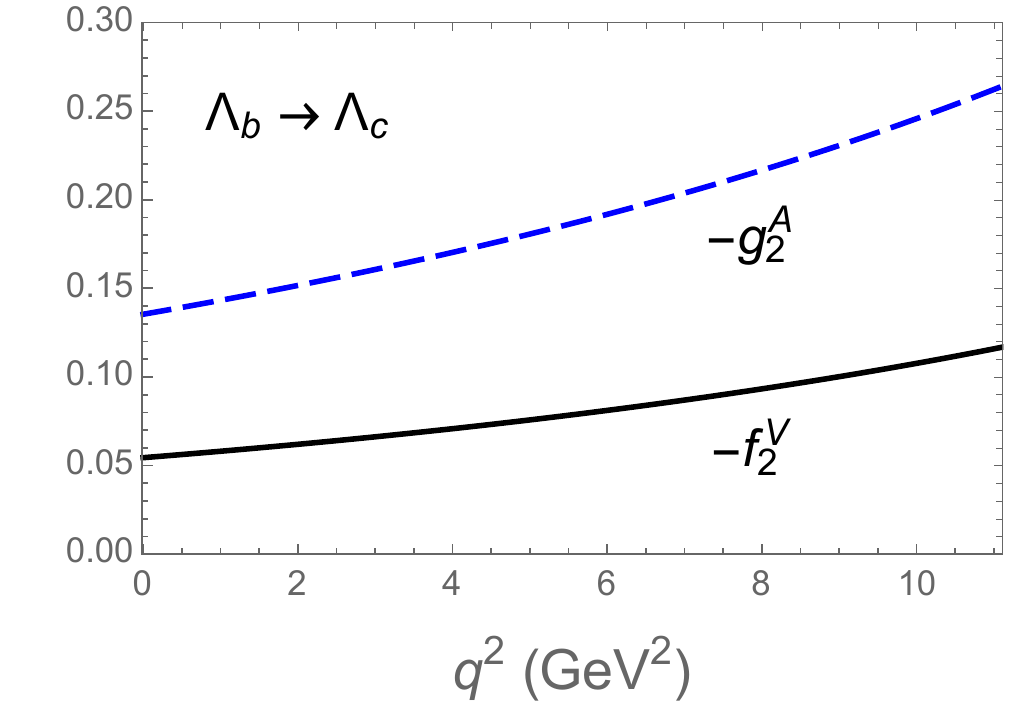}
}
\subfigure[]{
 \includegraphics[width=0.44\textwidth]  {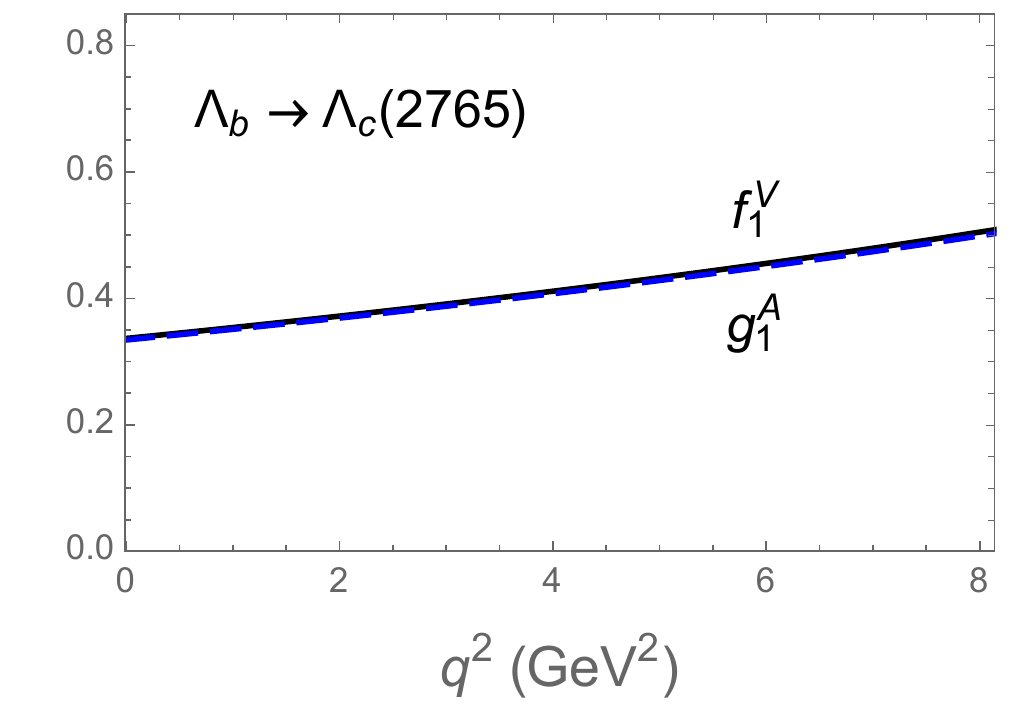}
}
\subfigure[]{
  \includegraphics[width=0.44\textwidth]  {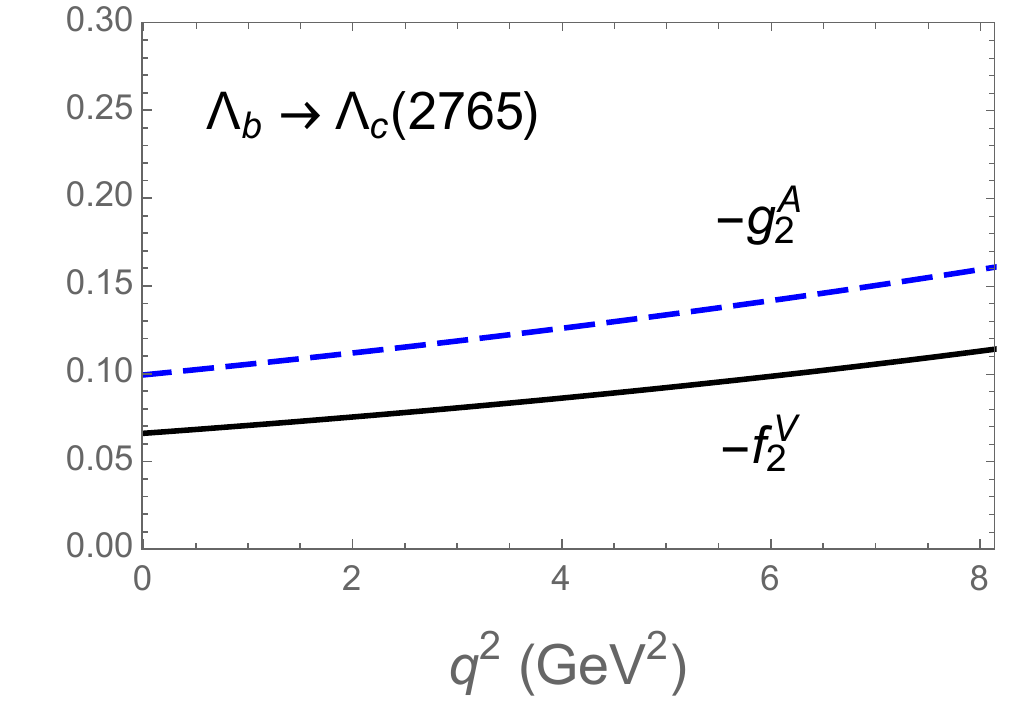}
}
\subfigure[]{
 \includegraphics[width=0.44\textwidth]  {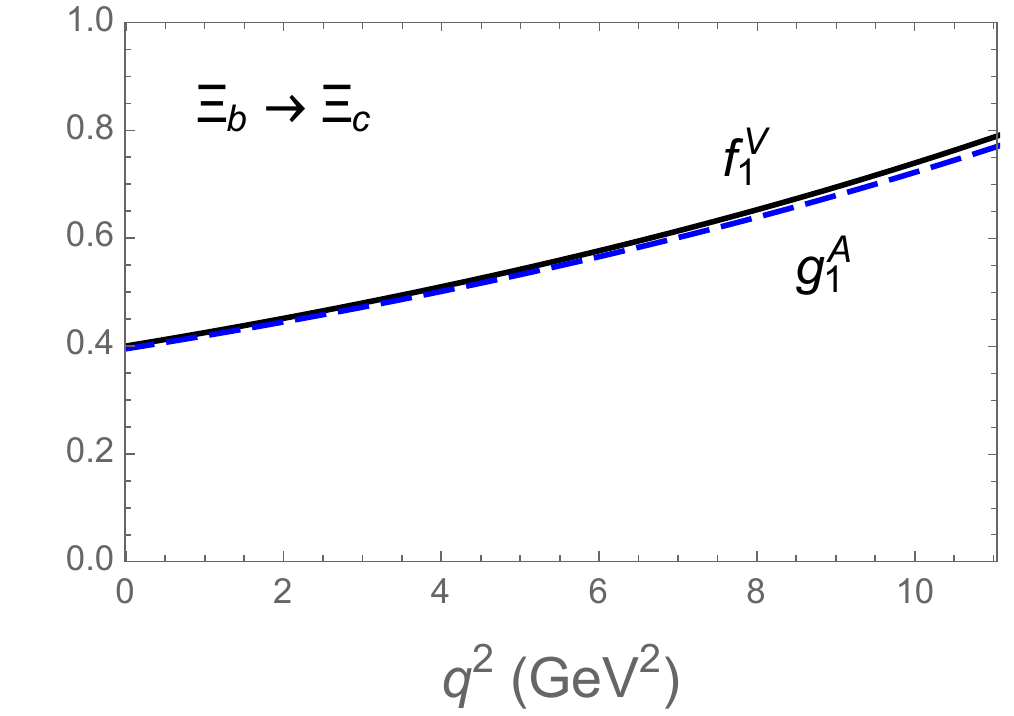}
}
\subfigure[]{
  \includegraphics[width=0.44\textwidth]  {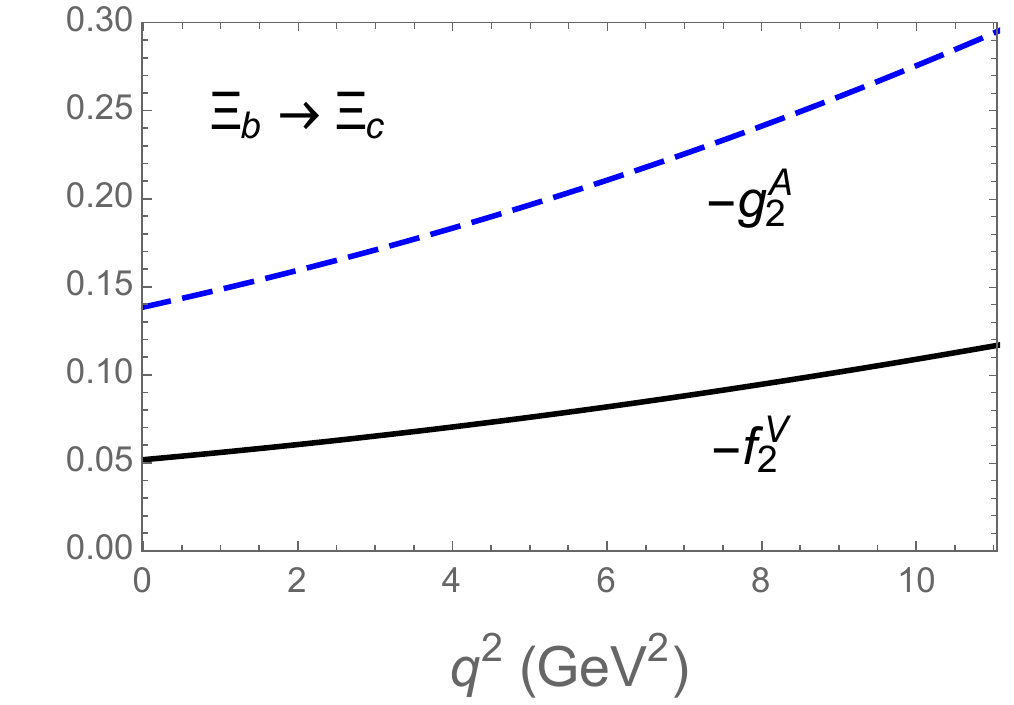}
}
\caption{Form factors $f_{1,2}(q^2)$ and $g_{1,2}(q^2)$ for 
$\Lambda_b\to\Lambda_c, \Lambda_c(2765)$ and $\Xi_b\to\Xi_c$ transitions.
The transitions are $\B_b({\bf\bar 3_f},1/2^+)\to\B_c({\bf \bar 3_f},1/2^+)$ transitions [type (i)].}
\label{fig:fg type i}
\end{figure}

\begin{figure}[t!]
\centering
\subfigure[]{
 \includegraphics[width=0.44\textwidth]  {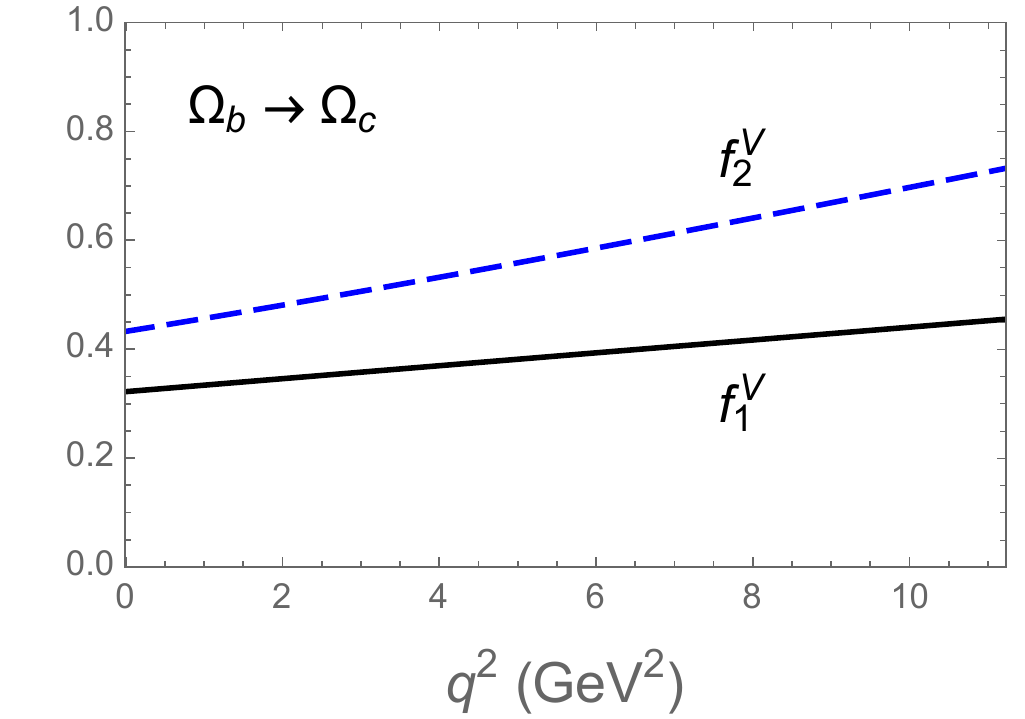}
}
\subfigure[]{
  \includegraphics[width=0.44\textwidth]  {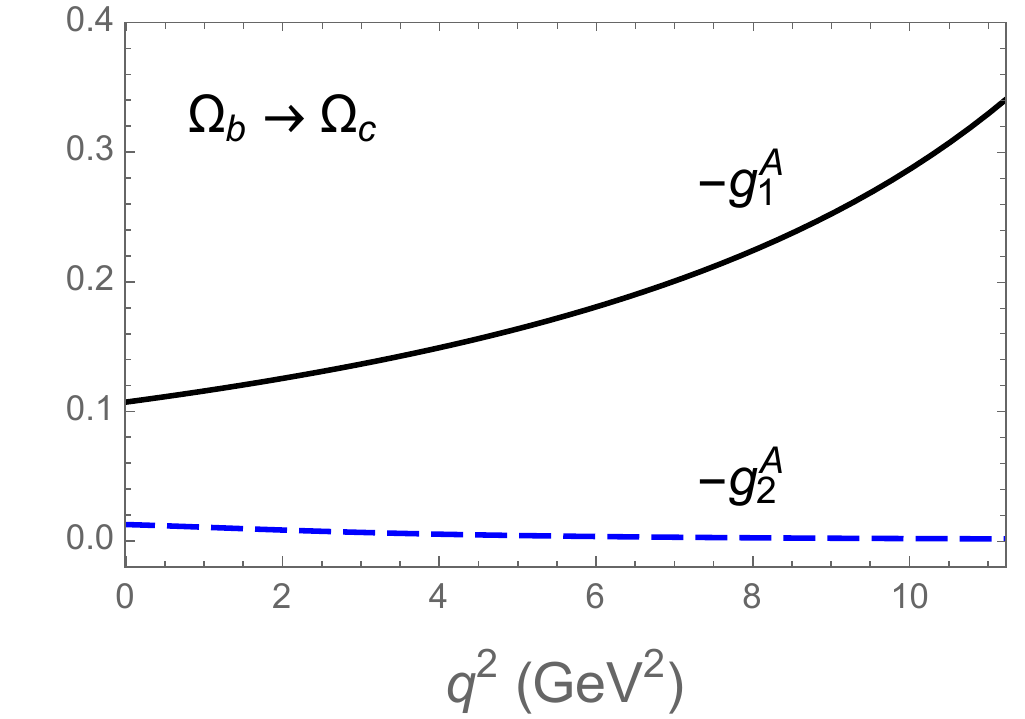}
}
\subfigure[]{
 \includegraphics[width=0.44\textwidth]  {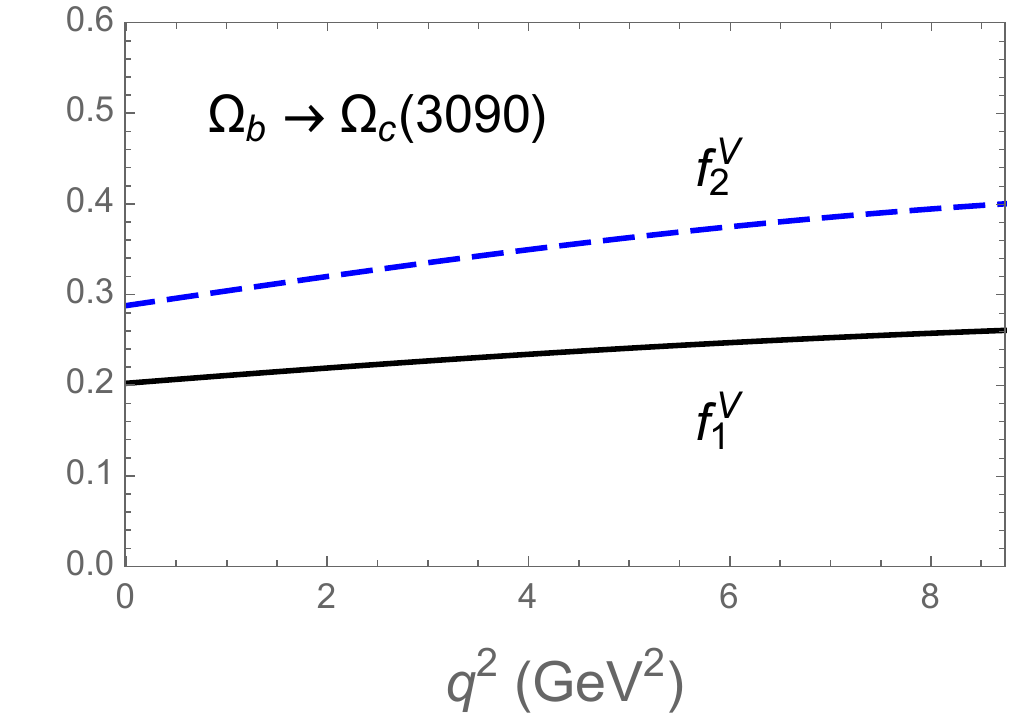}
}
\subfigure[]{
  \includegraphics[width=0.44\textwidth]  {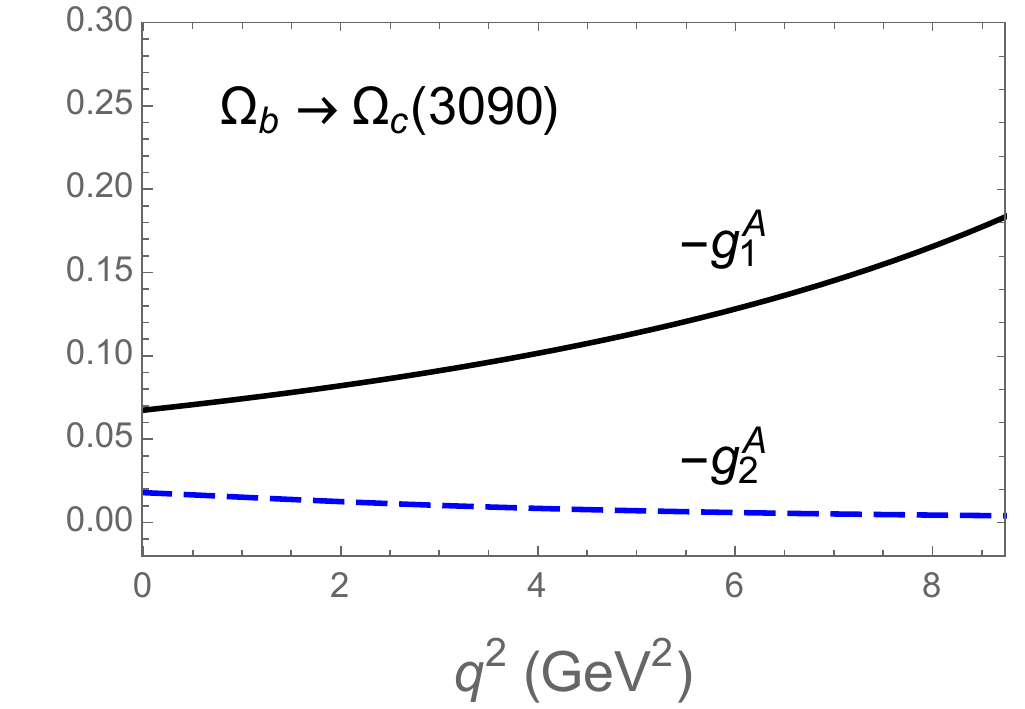}
}
\caption{Form factors $f_{1,2}(q^2)$ and $g_{1,2}(q^2)$ for
$\Omega_b\to\Omega_c$ and $\Omega_c(3090)$
transitions. The transitions are $\B_b({\bf 6_f},1/2^+)\to\B_c({\bf 6_f},1/2^+)$ transitions [type (ii)].}
\label{fig:fg type ii}
\end{figure}

\begin{figure}[t!]
\centering
\subfigure[]{
 \includegraphics[width=0.44\textwidth]  {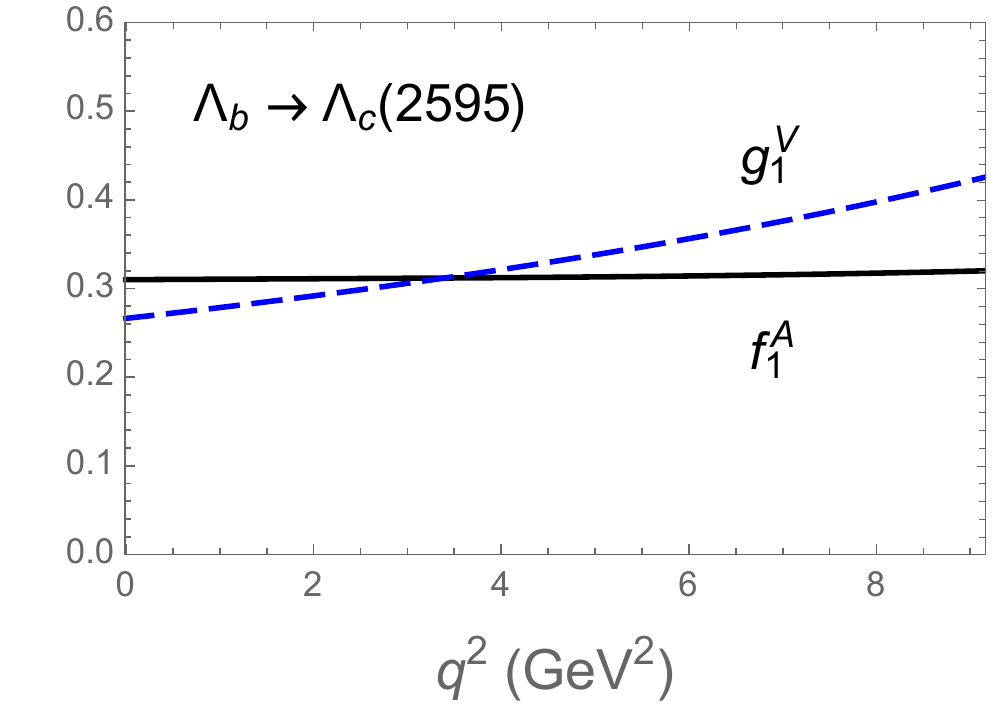}
}
\subfigure[]{
  \includegraphics[width=0.44\textwidth]  {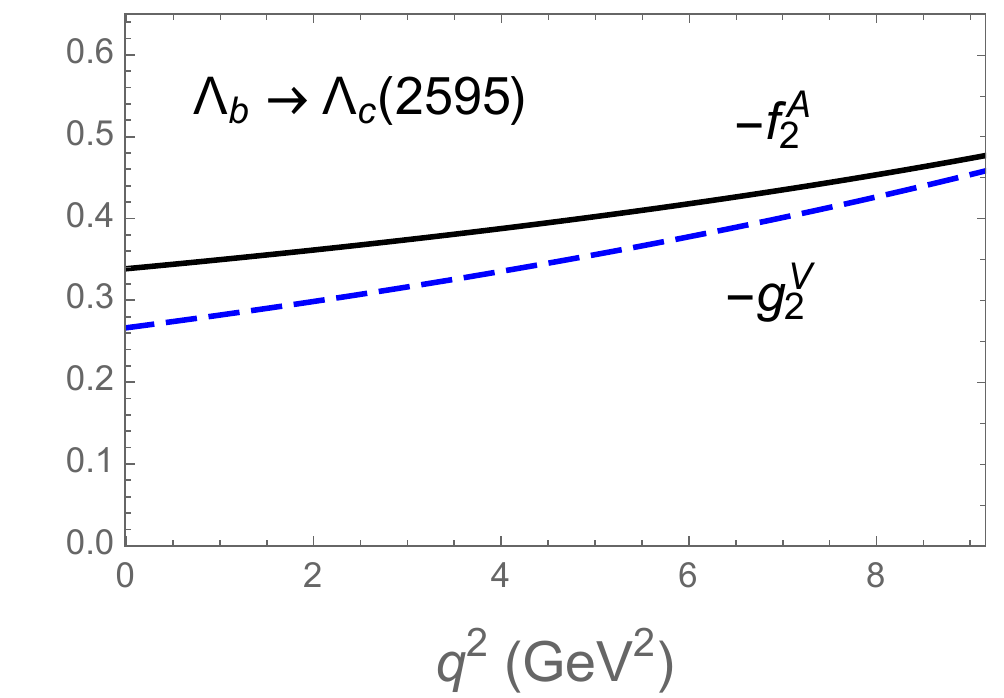}
}
\subfigure[]{
 \includegraphics[width=0.44\textwidth]  {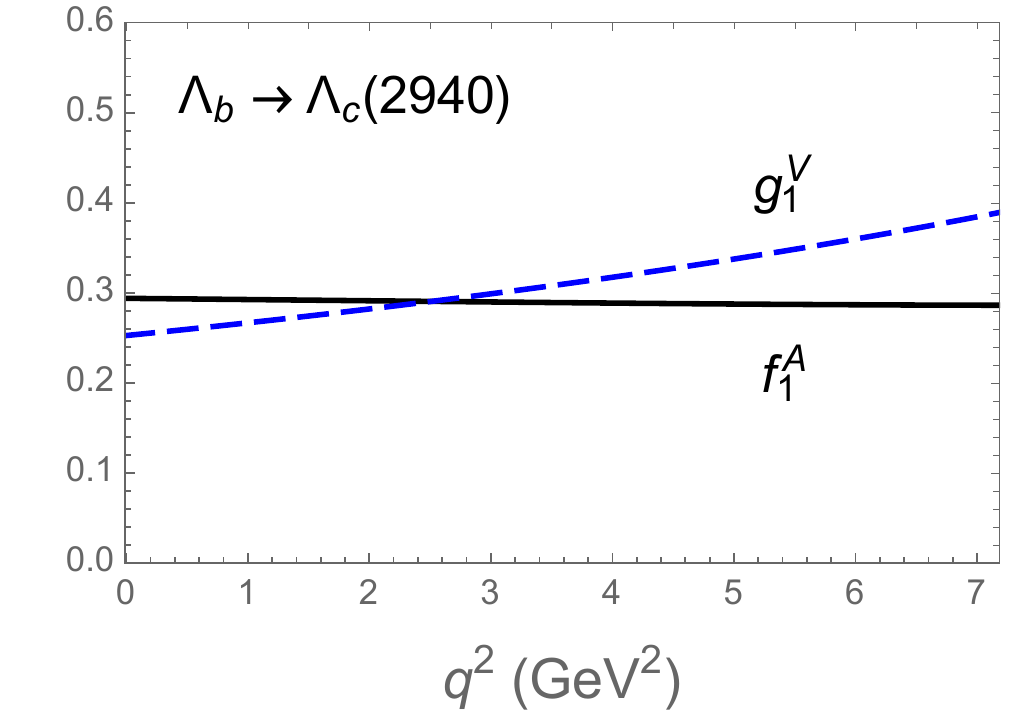}
}
\subfigure[]{
  \includegraphics[width=0.44\textwidth]  {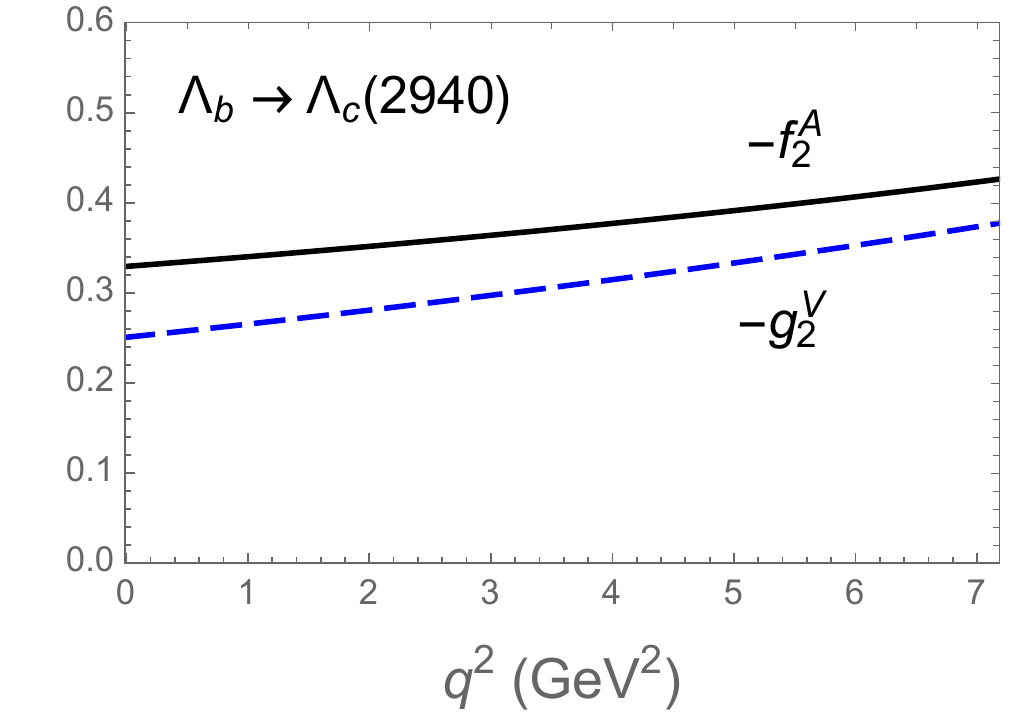}
}
\subfigure[]{
 \includegraphics[width=0.44\textwidth]  {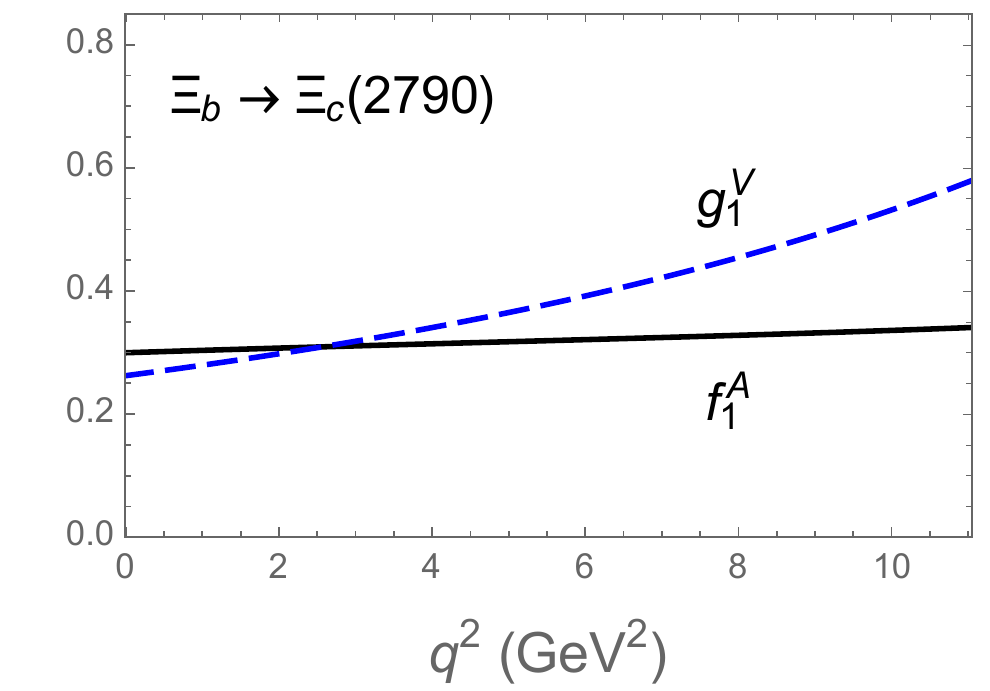}
}
\subfigure[]{
  \includegraphics[width=0.44\textwidth]  {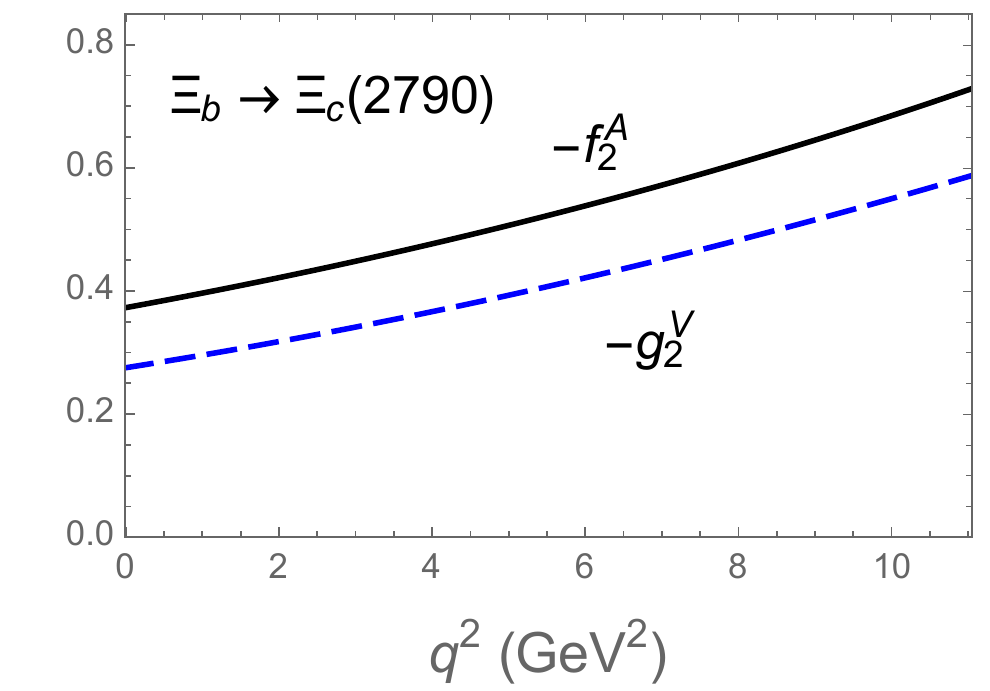}
}
\caption{Form factors $f_{1,2}(q^2)$ and $g_{1,2}(q^2)$ for
$\Lambda_b\to\Lambda_c(2595)$, $\Lambda_c(2940)$ and $\Xi_b\to\Xi_c(2790)$
transitions. The transitions are $\B_b({\bf\bar 3_f},1/2^+)\to\B_c({\bf\bar 3_f},1/2^-)$ transitions [type (iii)].}
\label{fig:fg type iii}
\end{figure}

The input parameters $m_{[qq']}$, $m_q$, $\beta$ are summarized in Table~\ref{tab:input}. 
The constituent quark and diquark masses are taken from ref.~\cite{Ebert:2010af}.
For the diquark masses, we use $m^S_{[ud]}$ for $\Lambda_b$ and $\Lambda^{(*,**)}_c$, 
$m^S_{[us]}$ for $\Xi_b$ and $\Xi^{(**)}_c$, 
and $m^A_{[ss]}$ for $\Omega_b$ and $\Omega_c^{(*)}$.
The $\beta$s are chosen to reproduce the $Br(\Lambda_b\to\Lambda_c P)$ data (see later discussion).

The form factors of various $\B_b\to\B_c$ transitions can be obtained using formulas in the previous section.
There are three types of transitions.
For the type (i) transition, the $\B_b({\bf \bar 3_f},1/2^+)\to \B_c({\bf \bar 3_f},1/2^+)$ transition, the form factors can be obtained by using Eq. (\ref{eq:ff type i}),
for the type (ii) transition, $\B_b({\bf 6_f},1/2^+)\to \B_c({\bf 6_f},1/2^+)$ transition, we use Eq. (\ref{eq:ff type ii}) and
for the type (iii) transition, the $\B_b({\bf \bar 3_f},1/2^+)\to \B_c({\bf \bar 3_f},1/2^-)$ transition, Eq. (\ref{eq:ff type iii}) should be employed.

As our calculation of form factors is done in the $q^+=0$ frame,
where $q^2\leq 0$, we shall follow~\cite{Jaus96,CCH,CC2004} to analytically
continue the form factors to the timelike region.
We find that the momentum dependence of the form
factors in the spacelike region can be well parameterized and
reproduced in the three-parameter form:
 \be 
 F(q^2)&=&\,{F(0)\over (1-q^2/M^2)
 [1-a(q^2/M^2)+b(q^2/M^2)^2]} 
 \label{eq:FFpara1}
 \en
for $\B_b\to \B_c$ transitions. The parameters $a$, $b$ and
$F(0)$ are first determined in the spacelike region. We then
employ this parametrization to determine the physical form factors
at $q^2\geq 0$. The parameters $a,b$ are expected to be of order
${\cal O}(1)$. As we shall see this is usually true in our numerical results. 
Occasionally some $a$s and $b$s are larger than ${\cal O}(1)$, 
but in most of these cases the corresponding form factors are small 
and do not have much impact on decay rates.

The $\B_b({\bf\bar 3_f},1/2^+)\to\B_c({\bf \bar 3_f},1/2^+)$ transition form factors $f^V_{1,2}(q^2)$ and $g^A_{1,2}(q^2)$
are given in Table~\ref{tab:fg type i} and are plotted in Fig.~\ref{fig:fg type i}.
These include the form factors for 
$\Lambda_b\to\Lambda_c, \Lambda_c(2765)$ and $\Xi_b\to\Xi_c$ transitions.
In this case, we have $f_1^V$, $g_1^A>0$ and $f_2^V$, $g_2^A<0$. 
We see that $|f_1^V|$ and $|g_1^A|$ are larger than $|f_2^V|$ and $|g_2^A|$ in these transitions.
Note that 
except $|f_2^V|$, the $\Lambda_b\to \Lambda_c(2765)$ transition form factors have smaller sizes comparing to those in the other two transitions.
This is reasonable, since $\Lambda_c(2765)$ is a radial excited state.  
The configurations of the final states in excited state differ 
from those in the low lying states
and larger mis-match between initial and final state configurations, usually lead to smaller form factors.

The $\B_b({\bf 6_f},1/2^+)\to\B_c({\bf 6_f},1/2^+)$ transition form factors $f^V_{1,2}(q^2)$ and $g^A_{1,2}(q^2)$
are given in Table~\ref{tab:fg type ii} and are plotted in Fig.~\ref{fig:fg type ii}.
These includes the form factors for 
$\Omega_b\to\Omega_c$ and $\Omega_c(3090)$ transitions.
In this case, we have $f_1^V$, $f_2^V>0$, $g_1^A$ and $g_2^A<0$. 
We see that $|f_1^V|$ and $|f_2^V|$ are larger than $|g_1^A|$ and $|g_2^A|$ in these transitions.
Note that 
except $g^A_2$, the $\Omega_b\to \Omega_c(2940)$ transition form factors have smaller sizes comparing to those in the $\Omega_b\to\Omega_c$ transition.
This is reasonable, since we take $\Omega_c(2940)$ as a radial excited state. 
Larger mis-match between initial and final state configurations, usually lead to smaller form factors.

The $\B_b({\bf\bar 3_f},1/2^+)\to\B_c({\bf \bar 3_f},1/2^-)$ transition form factors $f^A_{1,2}(q^2)$ and $g^V_{1,2}(q^2)$
are given in Table~\ref{tab:fg type iii} and are plotted in Fig.~\ref{fig:fg type iii}.
These includes the form factors for 
$\Lambda_b\to\Lambda_c(2595)$, $\Lambda_c(2940)$ and $\Xi_b\to\Xi_c(2790)$ transitions.
In this case, we have $f_1^A$, $g_1^V>0$ and $f_2^A$, $g_2^V<0$.
The signs of the form factors are identical to those in the $\B_b({\bf \bar 3_f},1/2^+)\to\B_c({\bf \bar 3_f},1/2^+)$ case. 
The transitions in this case have $p$-wave final state baryons. 
In the previous two cases, the initial and final state baryons belong to the same categories [$\B_b({\bf \bar 3_f},1/2^+)$ or $\B_c({\bf 6_f},1/2^+)$], while in this case they are in different categories, the initial state is a $s$-wave baryon, but the final state is a $p$-wave baryon. 
We see that some of these form factors behavior rather differently from the previous ones.
For example, as shown in Fig.~\ref{fig:fg type iii}, $f_1^A(q^2)$ are almost independent of $q^2$, which are different from the $f_1^V(q^2)$ in the previous cases. 
Furthermore, all four form factors are of similar sizes in this case, while in the previous cases either one or two form factors are much smaller than the others.
Note that the transition form factors of $\Lambda_b\to\Lambda_c(2940)$ are similar to those in $\Lambda_b\to\Lambda_c(2595)$, even though $\Lambda_c(2940)$ is a radial excited $p$-wave state. 
This feature is also different from the two previous cases, where form factors involving radial excited states are usually smaller in sizes.

\subsection{$\B_b\to \B_c M$ decay rates and up-down asymmetries}

Under the factorization approximation, the decay amplitudes for
color-allowed $\B_b\to\B_c M^-$ decays are given by
\be
{\cal A}(\B_b\to\B_c M^-)=\frac{G_F}{\sqrt2} V_{cb} V^*_{ij} a_1
\langle \B_c |V_\mu-A_\mu|\B_b\rangle \langle M^-(\bar q_i q_j) |V^\mu-A^\mu|0\rangle,
\en
where $V_{cb, ij}$ are the Cabibbo-Kobayashi-Maskawa (CKM) matrix elements
and
$a_1$ is the effective color-allowed Wilson coefficient. 
In na\"ive factorization $a_1$ is given by $c_1+c_2/N_c$ with $c_1= 1.081$ and $c_2=-0.190$ at the scale of $\mu=4.2$ GeV~\cite{BBNS}.
The matrix element $\langle \B_c |V_\mu-A_\mu|\B_b\rangle$ is given by Eqs.~(\ref{eq:figi}) and (\ref{eq:figi1}),
while $\langle M^-(\bar q_i q_j) |V^\mu-A^\mu|0\rangle$ for $M=P, V, A$ (with $P$, $V$ and $A$ stand for pseudoscalar, vector and axial vector mesons, respectively) are given by
\be
\langle P |V^\mu-A^\mu|0\rangle=i q^\mu f_P,
\quad
\langle V |V^\mu-A^\mu|0\rangle=m_V f_V \varepsilon^*_V,
\quad
\langle A |V^\mu-A^\mu|0\rangle=-m_A f_A \varepsilon^*_A,
\en
where $f_{P,V,A}$ are the corresponding decay constants. 

In type (i) and (ii) transitions 
[$\B_b({\bf\bar 3_f},1/2^+)\to \B_c({\bf\bar 3_f},1/2^+)$ and $\B_b({\bf 6_f},1/2^+)\to \B_c({\bf 6_f},1/2^+)$ transitions],  the decay amplitudes are given by~\cite{Cheng97a}
 \be
 {\cal A}(\B_b\to\B_c P)&=&i\bar u'(A+B\gamma_5) u,
 \non\\
  {\cal A}(\B_b\to\B_c V)&=&\bar u'\varepsilon^{*\mu}(A_1\gamma_\mu\gamma_5+A_2 P'_\mu\gamma_5
 +B_1\gamma_\mu+B_2 P'_\mu) u,
 \non\\
 {\cal A}(\B_b\to\B_c A)&=&\bar u'\varepsilon^{*\mu}(A'_1\gamma_\mu\gamma_5+A'_2 P'_\mu\gamma_5
 +B'_1\gamma_\mu+B'_2 P'_\mu) u,
 \en
with
\be
A&=&\frac{G_f}{\sqrt2} V_{cb} V^*_{q_1q_2}\, a_1 f_P \left((M-M') f^V_1(m_P^2)+\frac{m_P^2}{M+M'} f^V_3(m_P^2)\right),
\non\\
B&=&\frac{G_f}{\sqrt2} V_{cb} V^*_{q_1 q_2}\, a_1 f_P \left((M+M') g^A_1(m_P^2)+\frac{m_P^2}{M+M'} g^A_3(m_P^2)\right),
\non\\
A_1&=&-\frac{G_f}{\sqrt2} V_{cb} V^*_{q_1q_2}\, a_1 f_V m_V
 \left[g^A_1(m_V^2)+g^A_2(m_V^2)\frac{M-M'}{M+M'}\right], 
\non\\
A_2&=&-2\frac{G_f}{\sqrt2} V_{cb} V^*_{q_1q_2}\, a_1 f_V m_V
\frac{g^A_2(m_V^2)}{M+M'},
\non\\
B_1&=&\frac{G_f}{\sqrt2} V_{cb} V^*_{q_1q_2}\, a_1 f_V m_V
\left[f^V_1(m_V^2)-f^V_2(m_V^2)\right],
\non\\
B_2&=&2\frac{G_f}{\sqrt2} V_{cb} V^*_{q_1q_2}\, a_1 f_V m_V\frac{f^V_2(m_V^2)}{M+M'}, 
\non\\
A'_1&=&\frac{G_f}{\sqrt2} V_{cb} V^*_{q_1q_2}\, a_1 f_A m_A
 \left[g^A_1(m_A^2)+g^A_2(m_A^2)\frac{M-M'}{M+M'}\right], 
\non\\
A'_2&=&2\frac{G_f}{\sqrt2} V_{cb} V^*_{q_1q_2}\, a_1 f_A m_A
\frac{g^A_2(m_V^2)}{M+M'},
\non\\
B'_1&=&-\frac{G_f}{\sqrt2} V_{cb} V^*_{q_1q_2}\, a_1 f_A m_A
\left[f^V_1(m_A^2)-f^V_2(m_A^2)\right],
\non\\
B'_2&=&-2\frac{G_f}{\sqrt2} V_{cb} V^*_{q_1q_2}\, a_1 f_A m_A\frac{f^V_2(m_A^2)}{M+M'}.
\label{eq: AB}
\en

For the type (iii) transition [$\B_b({\bf\bar 3_f},1/2^+)\to \B_c({\bf\bar 3_f},1/2^-)$ transition], one simply replaces $f_i^V$ and $g_i^A$ in the above equations by $-f^A_i$ and $-g^V_i$, respectively.

The decay rates and asymmetries read \cite{Cheng97a,Cheng:2018hwl}
 \be
 \Gamma(\B_b\to \B_c P)&=&\frac{p_c}{8\pi}\left[\frac{(M+M')^2-m_P^2}
 {M^2}|A|^2+\frac{(M-M')^2-m_P^2}{M^2}|B|^2\right],
 \non\\
 \Gamma[\B_b\to \B_c V(A)]&=&\frac{p_c}{4\pi}\frac{E'+M'}{M}\left[2(|S^{(\prime)}|^2+|P^{(\prime)}_2|^2)
 +\frac{E_{V(A)}^2}{m_{V(A)}^2}(|S^{(\prime)}+D^{(\prime)}|^2+|P^{(\prime)}_1|^2)\right],
 \en
\be
\alpha(\B_b\to \B_c P)&=&-\frac{2\kappa {\rm Re}(A^* B)}{|A|^2+\kappa^2|B|^2},
\non\\
\alpha[\B_b\to \B_c V(A)]&=&
\frac{4 m_{V(A)}^2{\rm Re}(S^{(\prime)*} P_2)+2 E_V^2 {\rm Re}(S^{(\prime)}+D^{(\prime)})^* P^{(\prime)}_1}
{2m_{V(A)}^2(|S^{(\prime)}|^2+|P^{(\prime)}_2|^2)+E_{V(A)}^2(|S^{(\prime)}+D^{(\prime)}|^2+|P^{(\prime)}_1|^2)},
\en 
with $\kappa\equiv p_c/(E'+M')$,
 \be
 S^{(\prime)}&=&-A^{(\prime)}_1,
 \qquad 
 P^{(\prime)}_1=-\frac{p_c}{E_{V(A)}}\left(\frac{M+M'}{E'+M'} B^{(\prime)}_1+M B^{(\prime)}_2\right),
 \non\\
 P^{(\prime)}_2&=&\frac{p_c}{E'+M'} B^{(\prime)}_1,
 \qquad
 D^{(\prime)}=-\frac{p_c^2}{E_{V(A)}(E'+M')}(A^{(\prime)}_1-M A^{(\prime)}_2),
 \en
where $p_c$ is the momentum in the center of mass frame.

All hadron masses and life-times are taken from PDG~\cite{PDG}.
The CKM matrix elements are taken from the latest results of the CKM fitter group~\cite{ckmfitter}.
The values of decay constants of pseudoscalars are 
$f_\pi=130.2$~MeV, 
$f_K=155.6$~MeV,
$f_D=211.9$~MeV and
$f_{D_s}=249.0$~MeV, 
which are the center values of the averaged values given in the review by Rosner, Stone and Van de Water in ref.~\cite{PDG},
while those of vectors and the axial-vector particles are 
$f_\rho=216$~MeV, 
$f_{K^*}=210$~MeV,
$f_{D^*}=220$~MeV and
$f_{D^*_s}=230$~MeV
and
$f_{K^*}=-203$~MeV,
which are taken from ref.~\cite{CCH}.
In this work the decay rates are estimated using the na\"{i}ve factorization approach.
Note that in ref.~\cite{Beneke:2000ry} using QCD factorization the authors obtained $|a_1(\bar B\to DP)|=1.055^{+0.019}_{-0.017}-(0.013^{+0.011}_{-0.006})\alpha_1^P$ with $\alpha_1^\pi=0$ and $|\alpha_1^K|<1$ [see Eq.~(230) in \cite{Beneke:2000ry}].
The $|a_1(DP)|$ agrees with the na\"{i}ve factorization value (ref.~\cite{Beneke:2000ry} used $a^{\rm LO}_1=1.025$) within few \%.
For estimations, we assign 10\% uncertainty in the effective Wilson coefficient $a_1$ and 10\% uncertainty in form factors.
Note that in $\B_b\to\B_c P$ decays, in principle one needs $f_3$ and $g_3$ contributions, see Eq. (\ref{eq: AB}). Since these contributions are suppressed by a $m_P^2/(M+M')^2$ factor compared to the $f_1$ and $g_1$ terms and $f_3, g_3$ are expected to be vanishing in the heavy quark limit~\cite{Ke:2007tg}, we shall neglect them, but enlarge the form factor uncertainties to $15\%$ in $\B_b\to\B_c D$ and $\B_b\to \B_c D_s$ decays. 

\begin{table}[t!]
\caption{\label{tab:rate P} Branching ratios of $\B_b\to \B_c P$ decays. The branching ratios are given in the unit of $10^{-3}$. The asterisks in the first column indicate that the baryons in the final states are radial excited.}
\begin{ruledtabular}
\begin{tabular}{lccccc}
Type          
          & Mode
          & $P=\pi^-$ 
          & $P=K^-$ 
          & $P=D^-$ 
          & $P=D_s^-$ 
          \\
\hline  
(i)  
          & $Br(\Lambda_b\to \Lambda_c P)$
          & $4.19^{+1.94}_{-1.44}$
          & $0.32^{+0.15}_{-0.11}$
          & $0.53^{+0.32}_{-0.22}$
          & $13.58^{+8.15}_{-5.63}$
          \\    
(i)      
          & $Br^{\rm expt.}(\Lambda_b\to \Lambda_c P)$
          & $4.9\pm 0.4$
          & $0.359\pm0.030$
          & $0.46\pm0.06$
          & $11.0\pm1.0$
          \\
(i)
          & $Br(\Xi^0_b\to \Xi^+_c P)$
          & $3.08^{+1.43}_{-1.06}$
          & $0.23^{+0.11}_{-0.08}$
          & $0.43^{+0.26}_{-0.18}$
          & $11.20^{+6.72}_{-4.65}$
          \\  
(i)
          & $Br(\Xi^-_b\to \Xi^0_c P)$
          & $3.27^{+1.52}_{-1.12}$%
          & $0.25^{+0.12}_{-0.09}$%
          & $0.46^{+0.28}_{-0.19}$%
          & $11.90^{+7.14}_{-4.94}$%
          \\                     
(i)$^*$
          & $Br[\Lambda_b\to \Lambda_c(2765) P]$
          & $1.58^{+0.73}_{-0.54}$%
          & $0.12^{+0.06}_{-0.04}$%
          & $0.18^{+0.11}_{-0.07}$%
          & $4.39^{+2.63}_{-1.82}$%
          \\  
(ii)
          & $Br(\Omega_b\to \Omega_c P)$
          & $1.33^{+0.62}_{-0.46}$%
          & $0.10^{+0.05}_{-0.03}$%
          & $0.18^{+0.11}_{-0.08}$%
          & $4.75^{+2.85}_{-1.97}$%
          \\  
(ii)$^*$
          & $Br[\Omega_b\to \Omega_c(3090) P]$
          & $0.41^{+0.19}_{-0.14}$%
          & $0.031^{+0.015}_{-0.011}$%
          & $0.054^{+0.033}_{-0.023}$%
          & $1.38^{+0.83}_{-0.57}$%
          \\  
(iii)
          & $Br[\Lambda_b\to \Lambda_c(2595) P]$
          & $1.31^{+0.61}_{-0.45}$%
          & $0.10^{+0.05}_{-0.03}$%
          & $0.13^{+0.08}_{-0.05}$%
          & $3.19^{+1.92}_{-1.32}$%
          \\ 
(iii)
          & $Br[\Xi^0_b\to \Xi^-_c(2790) P]$
          & $1.27^{+0.59}_{-0.44}$
          & $0.10^{+0.04}_{-0.03}$
          & $0.14^{+0.08}_{-0.06}$
          & $3.41^{+2.05}_{-1.41}$
          \\  
(iii)
          & $Br[\Xi^-_b\to \Xi^0_c(2790) P]$
          & $1.36^{+0.63}_{-0.47}$%
          & $0.10^{+0.05}_{-0.04}$%
          & $0.15^{+0.09}_{-0.06}$%
          & $3.63^{+2.18}_{-1.51}$%
          \\               
(iii)$^*$
          & $Br[\Lambda_b\to \Lambda_c(2940) P]$
          & $0.93^{+0.43}_{-0.32}$%
          & $0.069^{+0.032}_{-0.024}$%
          & $0.078^{+0.047}_{-0.032}$%
          & $1.86^{+1.12}_{-0.77}$%
          \\
\end{tabular}
\end{ruledtabular}
\end{table}

\begin{table}[t!]
\caption{\label{tab:rate V A} The predicted branching ratios of $\B_b\to \B_c V$ and $B_b\to \B_c A$ decays. The branching ratios are given in unit of $10^{-3}$. The asterisks in the first column indicate that the baryons in the final states are radial excited.}
\begin{ruledtabular}
\begin{tabular}{lcccccc}
Type
          & Mode
          & $M=\rho^-$ 
          & $M=K^{*-}$ 
          & $M=D^{*-}$ 
          & $M=D_s^{*-}$ 
          & $M=a^-_1$ 
          \\
\hline   
(i)
          & $Br(\Lambda_b\to \Lambda_c M)$
          & $12.39^{+5.79}_{-4.28}$
          & $0.63^{+0.30}_{-0.22}$
          & $0.79^{+0.38}_{-0.28}$
          & $16.33^{+7.94}_{-5.81}$
          & $11.53^{+5.44}_{-4.01}$
          \\ 
(i)
          & $Br(\Xi^0_b\to \Xi^+_c M)$
          & $9.28^{+4.34}_{-3.21}$
          & $0.47^{+0.22}_{-0.16}$
          & $0.65^{+0.32}_{-0.23}$
          & $13.50^{+6.62}_{-4.83}$
          & $8.84^{+4.19}_{-3.09}$
          \\   
(i)
          & $Br(\Xi^-_b\to \Xi^0_c M)$
          & $9.86^{+4.61}_{-3.41}$
          & $0.50^{+0.24}_{-0.17}$
          & $0.69^{+0.33}_{-0.24}$
          & $14.34^{+7.03}_{-5.13}$
          & $9.39^{+4.45}_{-3.28}$
          \\                            
(i)$^*$
          & $Br[\Lambda_b\to \Lambda_c(2765) M]$
          & $4.75^{+2.22}_{-1.65}$
          & $0.24^{+0.11}_{-0.08}$
          & $0.28^{+0.14}_{-0.10}$
          & $5.77^{+2.83}_{-2.07}$
          & $4.41^{+2.09}_{-1.54}$
          \\
(ii)
          & $Br(\Omega_b\to \Omega_c M)$
          & $1.63^{+1.17}_{-0.75}$
          & $0.082^{+0.058}_{-0.037}$
          & $0.088^{+0.055}_{-0.035}$
          & $1.84^{+1.11}_{-0.71}$
          & $1.41^{+0.99}_{-0.63}$
          \\  
(ii)$^*$
          & $Br[\Omega_b\to \Omega_c(3090) M]$
          & $0.54^{+0.37}_{-0.24}$
          & $0.027^{+0.018}_{-0.012}$
          & $0.030^{+0.017}_{-0.011}$
          & $0.62^{+0.34}_{-0.23}$
          & $0.47^{+0.31}_{-0.20}$
          \\ 
(iii)
          & $Br[\Lambda_b\to \Lambda_c(2595) M]$
          & $4.95^{+2.33}_{-1.72}$
          & $0.25^{+0.12}_{-0.09}$
          & $0.22^{+0.12}_{-0.08}$
          & $4.38^{+2.33}_{-1.66}$
          & $4.23^{+2.05}_{-1.50}$
          \\                         
(iii)
          & $Br[\Xi^0_b\to \Xi^+_c(2790) M]$
          & $5.03^{+2.37}_{-1.75}$
          & $0.25^{+0.12}_{-0.09}$
          & $0.25^{+0.13}_{-0.10}$
          & $5.08^{+2.70}_{-1.93}$
          & $4.43^{+2.14}_{-1.57}$
          \\
(iii)
          & $Br[\Xi^-_b\to \Xi^0_c(2790) M]$
          & $5.35^{+2.52}_{-1.86}$
          & $0.27^{+0.13}_{-0.09}$
          & $0.27^{+0.14}_{-0.10}$
          & $5.41^{+2.88}_{-2.05}$
          & $4.72^{+2.28}_{-1.67}$
          \\          
(iii)$^*$
          & $Br[\Lambda_b\to \Lambda_c(2940) M]$
          & $3.35^{+1.58}_{-1.17}$
          & $0.17^{+0.08}_{-0.06}$
          & $0.14^{+0.07}_{-0.05}$
          & $2.60^{+1.42}_{-1.01}$
          & $2.80^{+1.36}_{-1.00}$
          \\                                      
\end{tabular}
\end{ruledtabular}
\end{table}

Note that as shown in refs.~\cite{Ivanov:1997ra, Ivanov:1997hi} non-factorizable contributions to $\B_b\to\B_c P$ non-leptonic decay amplitudes can contribute as large as 30\% comparing to the factorized ones.
A precise estimation of non-factorization contributions is beyond the scope of the present work.~\footnote{One is referred to
\cite{Zhu:2018jet} for a recent attempt on applying QCD factorization to $\Lambda_b$ decays.}  
If needed, one can scale up the uncertainties of our numerical results on rates.

The branching ratios for $\B_b\to\B_c P$, $\B_c V$ and $\B_c A$ decays, with $P=\pi, K, D, D_s$, and 
are summarized in Tables~\ref{tab:rate P} and \ref{tab:rate V A}.
As shown in Table~\ref{tab:rate P}
the $\Lambda_b\to\Lambda_c P$ rates can reasonably reproduce the data within errors.
We see that the $\Lambda_b\to\Lambda_c \pi$ and $\Lambda_b\to\Lambda_c K$ rates prefer lower values,
while the $\Lambda_b\to\Lambda_c D$ and $\Lambda_b\to\Lambda_c D_s$ rates prefer higher values.
Branching ratios for other modes are predictions.
We find that for $\Lambda_b$ decays, we have the following pattern in the decay rates: 
\be
Br(\Lambda_b\to\Lambda_c P)
>
Br(\Lambda_b\to \Lambda_c(2765) P)
>
Br(\Lambda_b\to\Lambda_c(2595) P)
>
Br(\Lambda_b\to\Lambda_c(2940) P).
\en
The first two decays are of type (i) transitions [$\B_b({\bf \bar 3_f},1/2^+)\to\B_c({\bf \bar 3_f},1/2^+)$ transitions], while $\Lambda_c(2765)$ is a radial excited state,
and the last two decays are of type (iii) transitions [$\B_b({\bf \bar 3_f},1/2^+)\to\B_c({\bf \bar 3_f},1/2^-)$ transitions], while $\Lambda_c(2940)$ is a radial excited state.
We see that rates of type (i) transitions are greater than those of type (iii) transitions, and decay rates involving excited states are smaller within the same type. 
These are reasonable as the configurations of the final states in excited $s$-wave $\B_c({\bf \bar 3_f},1/2^+)$ state and low lying or excited $p$-wave $\B_c({\bf \bar 3_f},1/2^-)$ states differ 
from those in the low lying $s$-wave $\B_{b,c}({\bf \bar 3_f},1/2^+)$ states.
Larger mis-match between initial and final state configurations, usually lead to smaller form factors, and, consequently, smaller rates.

The $\Xi_b\to\Xi_c P$ modes are of type (i) decays, 
while $\Xi_b\to\Xi_c(2790) P$ decays are of type (iii) decays, 
where $\Xi_c(2790)$ is a $p$-wave baryon.
From Table~\ref{tab:rate P} we have
\be
Br(\Xi_b\to\Xi_c P)
>Br(\Xi_b\to\Xi_c(2790) P).
\en
We see again that rates of type (i) transitions are greater than those of type (iii) transitions.
Note that the $\Xi_b\to\Xi_c P$ rate is slightly smaller than the $\Lambda_b\to\Lambda_c P$ rate.

For $\Omega_b$ decays, we have
\be
Br(\Omega_b\to\Omega_c P)>Br(\Omega_b\to\Omega_c(3090) P).
\en
These decays are type (ii) decays [$\B_b({\bf 6_f},1/2^+)\to\B_c({\bf 6_f},1/2^+)$ transitions] and $\Omega_c(3090)$ is a radial excited state. Again the decay rate involving an excite state is smaller.
Note that in $\B_b\to \B_cP$ decays, rates in type~(ii) transition are smaller than those in type (i) transition, but similar to those in type (iii) transition.

The branching ratios for the weak
decays $\B_b\to\B_c V(A)$, with $V=\rho^-, K^{*-}, D^{*-}, D^{*-}_s$ and $A=a^-_1$
are summarized in Table~\ref{tab:rate V A}.
We find that for $\Lambda_b$ decays, except for $V=\rho^-$, we have the following pattern in the decay rates: 
\be
Br(\Lambda_b\to\Lambda_c V(A))
>
Br(\Lambda_b\to \Lambda_c(2765) V(A))
\non\\
\gtrsim
Br(\Lambda_b\to\Lambda_c(2595) V(A))
>
Br(\Lambda_b\to\Lambda_c(2940) V(A)).
\en
For the case of $V=\rho^-$, we have $Br(\Lambda_b\to \Lambda_c(2595) \rho)
\gtrsim
Br(\Lambda_b\to\Lambda_c(2765) \rho)$ instead. 
For the $\Xi_b\to\Xi_c M$ mode, we have
\be
Br(\Xi_b\to\Xi_c V(A))
>Br(\Xi_b\to\Xi_c(2790) P).
\en
Finally for $\Omega_b$ decays, we have
\be
Br(\Omega_b\to\Omega_c V(A))>Br(\Omega_b\to\Omega_c(3090) V(A)).
\en
These patterns are similar to those in $\B_b\to\B_c P$ decays.
The above patterns reflect the fact that 
$\Lambda_c(2595)$, $\Xi_c(2790)$ and  $\Lambda_c(2940)$ are $p$-wave states,
and $\Lambda_c(2765)$, $\Lambda_c(2940)$ and $\Omega_c(3090)$ are radial excited states.
Larger mis-match between initial and final state configurations, 
usually lead to smaller rates.
Note that in $\B_b\to\B_c V, \B_c A$ decays, 
rates in type~(ii) transition are much smaller than those in type (i) transition and are also smaller than those in type (iii) transition.

\begin{table}[t!]
\caption{\label{tab:rate M compare} Various theoretical results on the branching ratios of $\Lambda_b\to \Lambda_c M$, $\Xi_b\to\Xi_c M$ and $\Omega_b\to\Omega_c M$ decays are compared. The branching ratios are given in the unit of $10^{-3}$. These are to be compared to the experimental branching ratios for $\Lambda_b\to \Lambda_c \pi^-,  \Lambda_c K^-, \Lambda_c D^-, \Lambda_c D_s^-$ decays, which are $4.9\pm 0.4$, $0.359\pm0.030$, $0.46\pm0.06$ and $11.0\pm1.0$ in unit of $10^{-3}$, respectively. See text for the results in ref.~\cite{Cheng97a}. 
}
\begin{ruledtabular}
\begin{tabular}{lccccccccc}
Mode          
          & This work
          & \cite{Mannel:1992ti} 
          & \cite{Cheng97a}
          & \cite{Ivanov:1997ra,Ivanov:1997hi} 
          & \cite{Giri:1997te}
          & \cite{Fayyazuddin:1998ap}
          & \cite{Mohanta:1998iu}
          & \cite{Zhu:2018jet}
          & \cite{Gutsche:2018utw}
          \\
\hline  
$\Lambda_b\to \Lambda_c \pi^-$
          & $4.19^{+1.94}_{-1.44}$
          & $4.6^{+2.0}_{-3.1}$
          & $4.6$
          & $5.62$
          & 3.91 
          & $-$
          & $1.75$
          & $4.96$
          & $-$
          \\    
$\Lambda_b\to \Lambda_c K^-$      
          & $0.32^{+0.15}_{-0.11}$
          & $-$
          & $-$
          & $-$
          & $-$
          & $-$
          & $0.13$
          & $0.393$
          & $-$
          \\
$\Lambda_b\to \Lambda_c D^-$
          & $0.53^{+0.32}_{-0.22}$
          & $-$
          & $-$
          & $-$
          & $-$
          & $-$
          & $0.30$
          & $0.522$
          & $-$
          \\          
$\Lambda_b\to \Lambda_c D_s^-$
          & $13.58^{+8.15}_{-5.63}$
          & $23^{+3}_{-4}$
          & $13.7$
          & $-$
          & $12.91$
          & $22.3$
          & 7.70
          & $12.4$
          & 14.78          
          \\  
$\Lambda_b\to \Lambda_c \rho^-$
          & $12.39^{+5.79}_{-4.28}$
          & $6.6^{+2.4}_{-4.0}$
          & $12.9$
          & $-$
          & $10.82$
          & $-$
          & $4.91$
          & $8.65$
          & $-$          
          \\  
$\Lambda_b\to \Lambda_c K^{*-}$
          & $0.63^{+0.30}_{-0.22}$
          & $-$
          & $-$
          & $-$
          & $-$
          & $-$
          & $0.27$
          & $0.441$
          & $-$
          \\                      
$\Lambda_b\to \Lambda_c D^{*-}$
          & $0.79^{+0.38}_{-0.28}$
          & $-$
          & $-$
          & $-$
          & $-$
          & $-$
          & $0.49$
          & $0.520$
          & $-$
          \\ 
$\Lambda_b\to \Lambda_c D_s^{*-}$
          & $16.33^{+7.94}_{-5.81}$
          & $17.3^{+2.0}_{-3.0}$
          & $21.8$
          & $-$
          & $19.83$
          & $32.6$
          & 14.14
          & $10.5$
          & $25.16$
          \\     
$\Lambda_b\to \Lambda_c a_1^-$
          & $11.53^{+5.44}_{-4.01}$
          & $-$
          & $-$
          & $-$
          & $-$
          & $-$
          & $5.32$
          & $-$
          & $-$
          \\
$\Xi^0_b\to \Xi^+_c \pi^-$ 
          & $3.08^{+1.43}_{-1.06}$
          & $-$
          & $4.9$
          & $7.08$
          & $-$
          & $-$          
          & $-$
          & $-$
          & $-$
          \\    
$\Xi^-_b\to \Xi^0_c \pi^-$ 
          & $3.27^{+1.52}_{-1.12}$
          & $-$
          & $5.2$
          & 10.13
          & $-$
          & $-$
          & $-$
          & $-$
          & $-$
          \\  
$\Xi^0_b\to \Xi^-_c D^-$
          & $0.43^{+0.26}_{-0.18}$
          & $-$
          & $-$
          & $-$
          & $-$
          & $-$
          & $-$
          & $-$
          & $0.45$
          \\ 
$\Xi^0_b\to \Xi^-_c D_s^-$
          & $11.20^{+6.72}_{-4.65}$
          & $-$
          & $14.6$
          & $-$
          & $-$
          & $-$
          & $-$
          & $-$
          & $-$
          \\           
$\Xi^0_b\to \Xi^-_c D^{*-}$
          & $0.65^{+0.32}_{-0.23}$
          & $-$
          & $-$
          & $-$
          & $-$
          & $-$
          & $-$
          & $-$
          & $0.95$
          \\  
$\Xi^0_b\to \Xi^-_c D_s^{*-}$
          & $13.50^{+6.62}_{-4.83}$
          & $-$
          & $23.1$
          & $-$
          & $-$
          & $-$
          & $-$
          & $-$
          & $-$
          \\                                         
$\Omega_b\to \Omega_c \pi^-$ 
          & $1.33^{+0.62}_{-0.46}$
          & $-$
          & $4.92$
          & 5.81
          & $-$
          & $-$
          & $-$
          & $-$
          & $1.88$
          \\ 
$\Omega_b\to \Omega_c D_s^-$ 
          & $4.75^{+2.85}_{-1.97}$
          & $-$
          & $17.9$
          & $-$
          & $-$
          & $-$
          & $-$
          & $-$
          & $-$
          \\           
$\Omega_b\to \Omega_c \rho^-$ 
          & $1.63^{+1.17}_{-0.75}$
          & $-$
          & $12.8$
          & $-$
          & $-$
          & $-$
          & $-$
          & $-$
          & $5.43$
          \\
$\Omega_b\to \Omega_c D_s^{*-}$ 
          & $1.84^{+1.11}_{-0.71}$
          & $-$
          & $11.5$
          & $-$
          & $-$
          & $-$
          & $-$
          & $-$
          & $-$
          \\                                                          
\end{tabular}
\end{ruledtabular}
\end{table}

In Tables~\ref{tab:rate M compare}, we compare our results on the branching ratios of $\Lambda_b\to \Lambda_c M$, $\Xi_b\to\Xi_c M$ and $\Omega_b\to\Omega_c M$ decays to those obtained in other works. 
Note that in the table the results of ref.~\cite{Cheng97a} are obtained by using Table II in \cite{Cheng97a} with $a_1\simeq 1$,
while for the $\B_b\to \B_c V$ rates the numerics are corrected by a factor of two, see footnote~7 in \cite{Cheng:2018hwl}. 
Overall speaking our results agree reasonably well with most of the results obtained in other works.
Note that in $\Omega_b\to\Omega_c M^-$ decays, the predicted rates are in general smaller than those obtained in other works,
except that the predicted $Br(\Omega_b\to\Omega_c \pi^-)$ is close to the one in ref.~\cite{Gutsche:2018utw}.

\begin{table}[t!]
\caption{\label{tab:alpha P} The predicted up-down asymmetries of $\B_b\to \B_c P$ decays. The asymmetries are given in unit of $\%$. The asterisks in the first column indicate that the baryons in the final states are radial excited.}
\begin{ruledtabular}
\begin{tabular}{lccccc}
Type
          & Mode
          & $P=\pi^-$ 
          & $P=K^-$ 
          & $P=D^-$ 
          & $P=D_s^-$ 
          \\
\hline  
(i)  
          & $\alpha(\Lambda_b\to \Lambda_c P)$
          & $-99.99^{+2.24}_{-0.00}$%
          & $-99.98^{+2.41}_{-0.00}$%
          & $-98.47^{+8.91}_{-1.52}$%
          & $-98.06^{+9.41}_{-1.87}$%
          \\    
(i)
          & $\alpha(\Xi^0_b\to \Xi^+_c P)$
          & $-99.99^{+2.24}_{-0.00}$%
          & $-99.97^{+2.41}_{-0.00}$%
          & $-98.40^{+9.01}_{-1.59}$%
          & $-97.96^{+9.52}_{-1.96}$%
          \\ 
(i)
          & $\alpha(\Xi^-_b\to \Xi^0_c P)$
          & $-99.99^{+2.24}_{-0.00}$%
          & $-99.97^{+2.41}_{-0.00}$%
          & $-98.39^{+9.01}_{-1.59}$%
          & $-97.96^{+9.53}_{-1.96}$%
          \\                        
(i)$^*$
          & $\alpha[\Lambda_b\to \Lambda_c(2765) P]$
          & $-100.00^{+2.14}_{-0.00}$%
          & $-99.98^{+2.39}_{-0.00}$%
          & $-96.61^{+10.76}_{-3.32}$%
          & $-95.54^{+11.49}_{-4.46}$%
          \\
(ii)          
          & $\alpha(\Omega_b\to \Omega_c P)$
          & $59.92^{+9.88}_{-9.22}$%
          & $59.93^{+9.88}_{-9.22}$%
          & $59.95^{+14.95}_{-13.54}$%
          & $59.90^{+14.95}_{-13.53}$%
          \\   
(ii)$^*$
          & $\alpha[\Omega_b\to \Omega_c(3090) P]$
          & $60.02^{+9.88}_{-9.23}$%
          & $60.02^{+9.88}_{-9.23}$%
          & $59.49^{+14.93}_{-13.47}$%
          & $59.23^{+14.92}_{-13.43}$%
          \\ 
(iii)
          & $\alpha[\Lambda_b\to \Lambda_c(2595) P]$
          & $-98.86^{+4.77}_{-1.04}$%
          & $-98.84^{+4.79}_{-1.05}$%
          & $-97.86^{+9.63}_{-2.03}$%
          & $-97.57^{+9.93}_{-2.25}$%
          \\  
(iii)
          & $\alpha[\Xi^0_b\to \Xi^+_c(2790) P]$
          & $-99.13^{+4.44}_{-0.84}$%
          & $-99.12^{+4.44}_{-0.84}$%
          & $-98.58^{+8.77}_{-1.42}$%
          & $-98.39^{+9.02}_{-1.59}$%
          \\  
(iii)
          & $\alpha[\Xi^-_b\to \Xi^0_c(2790) P]$
          & $-99.13^{+4.44}_{-0.84}$%
          & $-99.12^{+4.44}_{-0.84}$%
          & $-98.58^{+8.76}_{-1.42}$%
          & $-98.39^{+9.02}_{-1.59}$%
          \\                                
(iii)$^*$
          & $\alpha[\Lambda_b\to \Lambda_c(2940) P]$
          & $-98.86^{+4.76}_{-1.03}$%
          & $-98.84^{+4.78}_{-1.05}$%
          & $-97.04^{+10.41}_{-2.81}$%
          & $-96.36^{+10.94}_{-3.60}$%
          \\                         
\end{tabular}
\end{ruledtabular}
\end{table}

\begin{table}[t!]
\caption{\label{tab:alpha V A} The predicted up-down asymmetries of $B_b\to \B_c V$ and $B_b\to \B_c A$ decays. The asymmetries are given in unit of $\%$. The asterisks in the first column indicate that the baryons in the final states are radial excited.}
\begin{ruledtabular}
\begin{tabular}{lcccccc}
Type
          & Mode
          & $M=\rho^-$ 
          & $M=K^{*-}$ 
          & $M=D^{*-}$ 
          & $M=D_s^{*-}$ 
          & $M=a^-_1$ 
          \\
\hline   
(i)
          & $\alpha(\Lambda_b\to \Lambda_c M)$
          & $-90.50^{+2.07}_{-0.23}$%
          & $-87.50^{+2.34}_{-0.30}$%
          & $-48.19^{+4.21}_{-2.75}$%
          & $-44.10^{+4.19}_{-2.94}$%
          & $-77.40^{+3.15}_{-0.74}$
          \\ 
(i)
          & $\alpha(\Xi^0_b\to \Xi^+_c M)$
          & $-90.86^{+2.04}_{-0.27}$%
          & $-87.97^{+2.33}_{-0.35}$%
          & $-49.52^{+4.41}_{-2.90}$%
          & $-45.45^{+4.41}_{-3.12}$%
          & $-78.18^{+3.19}_{-0.78}$%
          \\   
(i)
          & $\alpha(\Xi^-_b\to \Xi^0_c M)$
          & $-90.86^{+2.04}_{-0.27}$%
          & $-87.97^{+2.33}_{-0.35}$%
          & $-49.53^{+4.41}_{-2.90}$%
          & $-45.46^{+4.41}_{-3.12}$%
          & $-78.18^{+3.19}_{-0.78}$%
          \\                             
(i)$^*$
          & $\alpha[\Lambda_b\to \Lambda_c(2765) M]$
          & $-88.29^{+2.32}_{-0.26}$%
          & $-84.65^{+2.69}_{-0.34}$%
          & $-38.47^{+4.54}_{-3.80}$%
          & $-33.83^{+4.36}_{-3.87}$%
          & $-72.48^{+3.78}_{-1.23}$%
          \\
(ii)
          & $\alpha(\Omega_b\to \Omega_c M)$
          & $85.23^{+11.84}_{-14.93}$%
          & $85.99^{+11.33}_{-14.89}$%
          & $89.18^{+6.02}_{-8.69}$%
          & $87.80^{+5.74}_{-7.38}$%
          & $88.34^{+9.18}_{-14.43}$%
          \\   
(ii)$^*$
          & $\alpha[\Omega_b\to \Omega_c(3090) M]$
          & $84.20^{+11.75}_{-14.41}$%
          & $85.12^{+11.12}_{-14.29}$%
          & $83.23^{+5.78}_{-6.34}$%
          & $79.90^{+6.23}_{-7.25}$%
          & $87.70^{+8.44}_{-13.37}$%
          \\  
(iii)
          & $\alpha[\Lambda_b\to \Lambda_c(2595) M]$
          & $-83.26^{+7.00}_{-4.51}$%
          & $-80.37^{+6.59}_{-4.14}$%
          & $-39.26^{+4.24}_{-3.26}$%
          & $-34.49^{+4.50}_{-3.76}$%
          & $-70.45^{+5.21}_{-2.89}$%
          \\
(iii)
          & $\alpha[\Xi^0_b\to \Xi^+_c(2790) M]$
          & $-83.09^{+7.00}_{-4.52}$%
          & $-80.16^{+6.58}_{-4.13}$%
          & $-37.67^{+4.42}_{-3.63}$%
          & $-32.69^{+4.64}_{-4.11}$%
          & $-70.02^{+5.16}_{-2.86}$%
          \\     
(iii)
          & $\alpha[\Xi^-_b\to \Xi^0_c(2790) M]$
          & $-83.10^{+7.00}_{-4.52}$%
          & $-80.17^{+6.58}_{-4.14}$%
          & $-37.72^{+4.42}_{-3.63}$%
          & $-32.74^{+4.64}_{-4.11}$%
          & $-70.04^{+5.16}_{-2.86}$%
          \\                                      
(iii)$^*$
          & $\alpha[\Lambda_b\to \Lambda_c(2940) M]$
          & $-82.69^{+6.30}_{-3.64}$%
          & $-79.33^{+5.74}_{-3.10}$%
          & $-29.73^{+4.94}_{-4.67}$%
          & $-24.03^{+4.80}_{-4.86}$%
          & $-67.60^{+3.87}_{-1.88}$%
          \\ 
\end{tabular}
\end{ruledtabular}
\end{table}

\begin{table}[t!]
\caption{\label{tab:alpha M compare} Various theoretical results on the up-down asymmetries ($\alpha$) of $\Lambda_b\to \Lambda_c M$, $\Xi_b\to\Xi_c M$ and $\Omega_b\to\Omega_c M$ decays are compared. The asymmetries are given in the unit of \%. 
}
\begin{ruledtabular}
\begin{tabular}{lcccccccc}
Mode          
          & This work
          & \cite{Mannel:1992ti} 
          & \cite{Cheng97a}
          & \cite{Ivanov:1997ra,Ivanov:1997hi} 
          & \cite{Fayyazuddin:1998ap}
          & \cite{Mohanta:1998iu}
          & \cite{Zhu:2018jet}
          & \cite{Gutsche:2018utw}
          \\
\hline  
$\Lambda_b\to \Lambda_c \pi^-$
          & $-99.99^{+2.24}_{-0.00}$%
          & $-100$%
          & $-99$%
          & $-99$%
          & $-$%
          & $-99.9$%
          & $-99.8$%
          & $-$
          \\    
$\Lambda_b\to \Lambda_c K^-$      
          & $-99.98^{+2.41}_{-0.00}$%
          & $-$%
          & $-$%
          & $-$%
          & $-$%
          & $-100$%
          & $-100$%
          & $-$
          \\
$\Lambda_b\to \Lambda_c D^-$
          & $-98.47^{+8.91}_{-1.52}$%
          & $-$%
          & $-$%
          & $-$%
          & $-$%
          & $-98.7$%
          & $-99.9$%
          & $-98.9$%
          \\          
$\Lambda_b\to \Lambda_c D_s^-$
          & $-98.06^{+9.41}_{-1.87}$%
          & $-99.1$%
          & $-99$%
          & $-$%
          & $-98$%
          & $-98.4$%
          & $-100$%
          & $-98.6$%
          \\  
$\Lambda_b\to \Lambda_c \rho^-$
          & $-90.50^{+2.07}_{-0.23}$%
          & $-90.3$%
          & $-88$%
          & $-$%
          & $-$%
          & $-89.8$%
          & $-88.8$%
          & $-$%
          \\  
$\Lambda_b\to \Lambda_c K^{*-}$
          & $-87.50^{+2.34}_{-0.30}$%
          & $-$%
          & $-$%
          & $-$%
          & $-$%
          & $-86.5$%
          & $-85.9$%
          & $-$%
          \\                      
$\Lambda_b\to \Lambda_c D^{*-}$
          & $-48.19^{+4.21}_{-2.75}$%
          & $-$%
          & $-$%
          & $-$%
          & $-$%
          & $-45.9$%
          & $-47.8$%
          & $-$
          \\ 
$\Lambda_b\to \Lambda_c D_s^{*-}$
          & $-44.10^{+4.19}_{-2.94}$%
          & $-43.7$%
          & $-36$%
          & $-$%
          & $-40$%
          & $-41.9$%
          & $-43.9$%
          & $-36.4$%
          \\     
$\Lambda_b\to \Lambda_c a_1^-$
          & $-77.40^{+3.15}_{-0.74}$%
          & $-$%
          & $-$%
          & $-$%
          & $-$%
          & $-75.8$%
          & $-$%
          & $-$%
          \\
$\Xi^0_b\to \Xi^+_c \pi^-$ 
          & $-99.99^{+2.24}_{-0.00}$%
          & $-$%
          & $-100$%
          & $-100$
          & $-$
          & $-$
          & $-$
          & $-$
          \\      
$\Xi^-_b\to \Xi^0_c \pi^-$ 
          & $-99.99^{+2.24}_{-0.00}$%
          & $-$%
          & $-100$%
          & $-97$%
          & $-$%
          & $-$%
          & $-$
          & $-$
          \\             
$\Xi^0_b\to \Xi^+_c D_s^-$ 
          & $-97.96^{+9.52}_{-1.96}$%
          & $-$%
          & $-99$%
          & $-$%
          & $-$%
          & $-$%
          & $-$%
          & $-$%
          \\                      
$\Xi^0_b\to \Xi^-_c D_s^{*-}$
          & $-45.45^{+4.41}_{-3.12}$%
          & $-$%
          & $-36$%
          & $-$%
          & $-$%
          & $-$%
          & $-$
          & $-$
          \\             
$\Omega_b\to \Omega_c \pi^-$ 
          & $59.92^{+9.88}_{-9.22}$%
          & $-$%
          & $51$%
          & $60$%
          & $-$%
          & $-$%
          & $-$
          & $-$
          \\ 
$\Omega_b\to \Omega_c D_s^-$ 
          & $59.90^{+14.95}_{-13.53}$%
          & $-$%
          & $42$%
          & $-$%
          & $-$%
          & $-$%
          & $-$%
          & $-$%
          \\           
$\Omega_b\to \Omega_c \rho^-$ 
          & $85.23^{+11.84}_{-14.93}$%
          & $-$%
          & $53$%
          & $-$%
          & $-$%
          & $-$%
          & $-$%
          & $-$%
          \\
$\Omega_b\to \Omega_c D_s^{*-}$ 
          & $87.80^{+5.74}_{-7.38}$%
          & $-$%
          & $64$%
          & $-$%
          & $-$%
          & $-$%
          & $-$%
          & $-$%
          \\                                                                                
\end{tabular}
\end{ruledtabular}
\end{table}

In Tables~\ref{tab:alpha P} and \ref{tab:alpha V A}, we show the predicted up-down asymmetries. The signs are mostly negative, except for those in the type (ii) transitions [$\B_b({\bf 6_f},1/2^+)\to\B_c({\bf 6_f},1/2^+)$ transitions]. 
These signs can be easily traced to the relative signs of their form factors. 
Most of these asymmetries are large in sizes.
Note that in type (i) and (ii) cases, the asymmetries $|\alpha(\B_b\to B_c D^*_{(s)})|$ are smaller than 
$|\alpha(\B_b\to B_c \rho)|$, $|\alpha(\B_b\to B_c K^*)|$ and $|\alpha(\B_b\to B_c a^-)|$.

In Tables~\ref{tab:alpha M compare}, we compare our results on the up-down asymmetries of $\Lambda_b\to \Lambda_c M$, $\Xi_b\to\Xi_c M$ and $\Omega_b\to\Omega_c M$ decays to those obtained in other works. 
Our results agree well in signs and magnitudes of the asymmetries with those in other works, except in 
$\Omega_b\to\Omega_c D_s^-, \Omega_c \rho^-, \Omega_c D_s^{*-}$ decays the predicted asymmetries are larger than those in ref.~\cite{Cheng97a}, but nevertheless the signs agree.

The predictions on rates and asymmetries presented in Tables~\ref{tab:rate P}, \ref{tab:rate V A}, \ref{tab:alpha P} and \ref{tab:alpha V A} can be verified experimentally.
These information may shed light on the quantum numbers of $\Lambda_c(2765)$, $\Lambda_c(2940)$
and $\Omega_c(3090)$.

\section{Conclusions}

We began with a brief overview of the charmed and bottom baryon spectroscopy and discussed their possible structure and $J^P$ assignment in the quark model.
As a working assumption we follow ref.~\cite{Cheng:2017ove} to assign the quantum numbers of some singled charmed states.
It is known that among low lying singly bottom baryons, only $\Lambda_b$, $\Xi_b$ and $\Omega_b$ decay weakly.
Consequently, we study $\Lambda_b\to \Lambda^{(*,**)}_c M^-$, $\Xi_b\to\Xi_c^{(**)} M^-$ and $\Omega_b\to\Omega^{(*)}_c M^-$ decays with $M=\pi, K,\rho, K^*, D, D_s, D^*, D^*_s, a_1$,
$\Lambda^{(*,**)}_c=\Lambda_c, \Lambda_c(2595), \Lambda_c(2765), \Lambda_c(2940)$,
$\Xi_c^{(**)}=\Xi_c, \Xi_c(2790)$ and
$\Omega^{(*)}_c=\Omega_c, \Omega_c(3090)$, in this work. 
There are three types of transitions, namely
${\cal B}_b({\bf \bar 3_f},1/2^+)$ to ${\cal B}_c({\bf \bar 3_f},1/2^+)$,
${\cal B}_b({\bf 6_f},1/2^+)$ to ${\cal B}_c({\bf 6_f},1/2^+)$
and
${\cal B}_b({\bf \bar 3_f},1/2^+)$ to ${\cal B}_c({\bf \bar 3_f},1/2^-)$ transitions.
The bottom baryon to charmed baryon form factors are calculated using the light-front quark model.
The formulas for the ${\cal B}_b({\bf 6_f},1/2^+)$ to ${\cal B}_c({\bf 6_f},1/2^+)$
and
${\cal B}_b({\bf \bar 3_f},1/2^+)$ to ${\cal B}_c({\bf \bar 3_f},1/2^-)$ transition form factors are new results.
Those with an excited state in the ${\cal B}_b({\bf \bar 3_f},1/2^+)$ to ${\cal B}_c({\bf \bar 3_f},1/2^+)$ transition are also new.

Numerical results of form factors, decay rates and up-down asymmetries in these decays are shown.
We see that rates of $\B_b({\bf \bar 3_f},1/2^+)\to\B_c({\bf \bar 3_f},1/2^+)$ transitions [type (i)] are greater than those of $\B_b({\bf \bar 3_f},1/2^+)\to\B_c({\bf \bar 3_f},1/2^-)$ [type (iii)] transitions, and decay rates involving excited states are smaller within the same type of transition. 
These are reasonable as the configurations of the final states in excited $\B_c({\bf \bar 3_f},1/2^+)$ state and low lying or excited $\B_c({\bf \bar 3_f},1/2^-)$ $p$-wave states differ from those in the low lying $\B_{b,c}({\bf \bar 3_f},1/2^+)$ states.
Larger mis-match between initial and final state configurations, usually lead to smaller form factors, and, consequently, smaller rates.
Furthermore, we find that in ${\cal B}_b\to {\cal B}_cP$ decays, rates in ${\cal B}_b({\bf 6_f},1/2^+)\to {\cal B}_c({\bf 6_f},1/2^+)$ [type~(ii)] transition are smaller than those in ${\cal B}_b({\bf \bar 3_f},1/2^+)\to {\cal B}_c({\bf \bar 3_f},1/2^+)$ [type (i)] transition, but similar to those in ${\cal B}_b({\bf \bar 3_f},1/2^+)\to {\cal B}_c({\bf \bar 3_f},1/2^-)$ [type (iii)] transition, while in $\B_b\to\B_c V, \B_c A$ decays, 
rates in ${\cal B}_b({\bf 6_f},1/2^+)\to {\cal B}_c({\bf 6_f},1/2^+)$ [type~(ii)] transition are much smaller than those in ${\cal B}_b({\bf \bar 3_f},1/2^+)\to {\cal B}_c({\bf \bar 3_f},1/2^+)$ [type (i)] transition and are also smaller than those in ${\cal B}_b({\bf \bar 3_f},1/2^+)\to {\cal B}_c({\bf \bar 3_f},1/2^-)$ [type (iii)] transition. 

For the up-down asymmetries,
the signs are mostly negative, except for those in the $\B_b({\bf 6_f},1/2^+)\to\B_c({\bf 6_f},1/2^+)$ [type (ii)] transition. 
These asymmetries are large in sizes.
Note that in type (i) and (ii) cases, the asymmetries $|\alpha(\B_b\to B_c D^*_{(s)})|$ are smaller than 
$|\alpha(\B_b\to B_c \rho)|$, $|\alpha(\B_b\to B_c K^*)|$ and $|\alpha(\B_b\to B_c a^-)|$.

We compare our results of rates and asymmetries of $\Lambda_b\to \Lambda_c M$, $\Xi_b\to\Xi_c M$ and $\Omega_b\to\Omega_c M$ decays to existing results in other works. 
In most cases the agreements are reasonably well.

Predictions on rates and asymmetries can be checked experimentally.
The study on these decay modes may shed light on the quantum numbers of the charmed baryons, as the decays depend on bottom baryon to charmed baryon form factors, which are sensitive to the configurations of the final state charmed baryons. 
This work can be further extended by including transitions having spin-3/2 baryons in the final states. 
The result will be reported elsewhere.

\section{Acknowledgments}
The author likes to thank Hai-Yang Cheng for discussion. 
This research was supported in part by the Ministry of Science and Technology of R.O.C. under Grant
No. 106-2112-M-033-004-MY3.

\appendix

\section{Vertex functions}\label{appendix: vertex}

\subsection{Some useful identities}

We collect some useful identities for the derivation of vertex functions in the following parts.
Relations involving Melosh transform for spin-1/2 and spin-1 particles are given by 
\be
\la\lambda_1|{\cal R}_M^\dagger(x_1,k_{1\bot}, m_1)|s_1\rangle\bar u_D(k_1,s_1)
&=&
       \bar u(k_1,\lambda_1) \frac{u_D(k_1,s_1)\bar u_D(k_1,s_1)}{2 m_1}
      =\bar u(k_1,\lambda_1),       
\label{eq: meloshspin1/2}
\\
\la\lambda_2|{\cal R}_M^\dagger(x_2,k_{2\bot}, m_2)|s_2\rangle \varepsilon_I^*(k_2,s_2)
&=&
      - \varepsilon^*_{LF}(k_2,\lambda_2) \cdot \varepsilon_I(k_2,s_2) \varepsilon_I^*(k_2,s_2)
      = \varepsilon^*_{LF}(k_2,\lambda_2),
\label{eq: meloshspin1}      
\en
where the familiar formulas of polarization sums are used, $u_D$ and $\varepsilon_I$ are the spinor and polarization vector in the instant form, while $u$ and $\varepsilon_{LF}$ are the ones in the light-front form. 
Note that in the particle rest frame, $\varepsilon_I$ and $\varepsilon_{LF}$ are identical, and likewise $u_D$ and $u$ are identical.

The relevant Clebsch-Gordan coefficients can be expressed in compact forms:
\be
\la \frac{1}{2} 0; s_1 0|\frac{1}{2} 0;\frac{1}{2} J_z\ra
&=&
      \chi^\dagger_{s_1}\cdot\chi_{_{J_z}}
\non\\
&=&
      \frac{1}{\sqrt {(M_0+m_1)^2-m_2^2}}\,
      \bar u_D(k_1,s_1) u(k_1+k_2,J_z),
\label{eq: CG1/201/2}     
\\
\la \frac{1}{2} 1; s_1 s_2|\frac{1}{2}1 ;\frac{1}{2} J_z\ra
&=&
      \frac{1}{\sqrt3}\chi^\dagger_{s_1}\vec\sigma\cdot
      \vec\varepsilon_I^*(k_1+k_2,s_2)\chi_{_{J_z}}
\non\\
&=&
       \frac{1}{\sqrt {3 [(M_0+m_1)^2-m_2^2]}}
\non\\
&&
       \times  \bar u_D(k_1,s_1)\gamma_5\not\!\varepsilon_I^*(k_1+k_2,s_2) u(k_1+k_2,J_z).
\label{eq: CG1/211/2}
\en
The following relation of the polarization vectors will be needed,
\be
\varepsilon_I^\mu(k_2,s_2)
&=& 
       \varepsilon_I^\mu(\bar P, s_2)
       -\frac{M_0k_2^\mu+m_2 \bar P^\mu}{m_2 M_0} \frac{\varepsilon_I(\bar P, s_2)\cdot k_2}{e_2+m_2},
\label{eq: ep2}
\\
\varepsilon^*_{I\mu}(\bar P, m)\varepsilon_{I\nu}(\bar P, m)
&=&
       -g_{\mu\nu}+\frac{\bar P_\mu \bar P_\nu}{M_0}.
\label{eq: epsum}       
\en
Derivations of some of the above relations will be given briefly in the following discussion.

The relations in Eqs. (\ref{eq: CG1/201/2}) and (\ref{eq: CG1/211/2})
can be easily proved by using the explicit expression of the Dirac
spinors. 
Explicitly, we use
\be
u_D(k_1,s)
&=&\frac{k_1\cdot\gamma+m_1}{\sqrt{e_1+m_1}}
\Bigg(
\begin{array}{c}
\chi_s\\
0
\end{array}
\Bigg)
=\frac{1}{\sqrt{e_1+m_1}}
\Bigg(
\begin{array}{c}
(e_1+m_1)\chi_s\\
\vec\sigma\cdot\vec p\chi_s
\end{array}
\Bigg),
\non\\
u(k_1+k_2,\lambda)
&=&\frac{(k_1+k_2)\cdot\gamma+M_0}{\sqrt{2 M_0}}\gamma^+\gamma^0
\Bigg(
\begin{array}{c}
\chi_\lambda\\
0
\end{array}
\Bigg)
=\sqrt{2 M_0}
\Bigg(
\begin{array}{c}
\chi_\lambda\\
0\\
\end{array}
\Bigg),
\en 
the standard Dirac representation of $\gamma^\mu$, $\gamma_5$
and $\varepsilon_I(k_1+k_2,s)=(0,\vec \varepsilon(s))$ with $ \vec \varepsilon (\pm 1)=\mp(1,\pm i,0)/\sqrt{2}$, $\vec \varepsilon(0)=(0,0,1)$.

The derivation of Eq. (\ref{eq: ep2}) is a bit tricky. 
We want to express $\varepsilon_I(k_2, s_2)$ in terms of $\varepsilon_I(k_1+k_2, s_2)$, 
which are polarization vectors in the instant form and in the rest frame of $\bar P=k_1+k_2=(M_0,\vec 0)$.
It is useful to note that $\varepsilon_I(k_1+k_2, s_2)$ and the polarization vector of particle 2 in its rest fame, 
$\varepsilon_I((m_2,\vec 0),s_2)$, are indeed identical,
as both are equal to $(0,\vec \varepsilon(s_2))$ 
with $ \vec \varepsilon (\pm 1)=\mp(1,\pm i,0)/\sqrt{2}$, $\vec \varepsilon(0)=(0,0,1)$, 
i.e.
\be
\varepsilon_I\left(\bar P=(M_0,\vec 0), s_2\right)
=\varepsilon_I\left((m_2,\vec 0),s_2\right)
=(0,\vec \varepsilon(s_2)).
\label{eq: ePbar=em2}
\en
Therefore, $\varepsilon_I(k_2, s_2)$ and $\varepsilon_I(k_1+k_2, s_2)$ [or $\varepsilon_I((m_2,\vec 0),s_2)]$
can be related by a suitable Lorentz boost. 

When the Lorentz boost, which brings particle 2 with momentum from $(m_2,\vec 0)$ to $k_2=(e_2,\vec k_2)$, acts on a generic four vector $A^\mu$, we have the following transformations:
\be
A^0 \to A^0\frac{e_2}{m_2}+\vec A\cdot \frac{\vec k_2}{m_2},
\qquad
\vec A \to \vec A+\frac{\vec k_2}{m_2}\left(\frac{\vec k_2\cdot \vec A}{e_2+m_2}\right)+ A^0\frac{\vec k_2}{m_2}. 
\en
One can easily check that it indeed brings $(m_2,\vec 0)$ to $k_2$.
Now by boosting the diquark polarization vector, 
$\varepsilon_I\left((m_2,\vec 0),s_2\right)=\left(0, \vec\varepsilon(s_2)\right)$, 
from its rest frame to $\varepsilon_I(k_2,s_2)$, which is in the $k_1+k_2$ rest frame,  
we obtain
\be
\varepsilon_I^0(k_2, s_2)
&=&\frac{1}{m_2}\vec\varepsilon(s_2)\cdot\vec k_2,
\non\\
\vec\varepsilon_I(k_2, s_2)
&=&\vec\varepsilon_I(s_2)
+\frac{\vec k_2}{m_2}\left(\frac{\vec\varepsilon_I(s_2)\cdot\vec k_2}{e_2+m_2}\right).
\en
We can express the above results in a compact form:
\be
\varepsilon_I^\mu(k_2,s_2)
&=&\varepsilon_I^\mu(\bar P, s_2)
-\frac{M_0k_2^\mu+m_2 \bar P^\mu}{m_2 M_0} \frac{\varepsilon_I(\bar P, s_2)\cdot k_2}{e_2+m_2}.
\en
Note that we have made use of Eq.~(\ref{eq: ePbar=em2}) in the above equation, and, consequently, Eq.~(\ref{eq: ep2}) is obtained.
One can easily check that the above expression for $\varepsilon_I(k_2,s_2)$ satisfies the well-known relations, $k_2\cdot\varepsilon_I(k_2,s_2)=0$ and 
$\varepsilon_I^*(k_2,s_2)\cdot\varepsilon_I(k_2,s'_2)=-\delta_{s_2,s'_2}$.

\subsection{$\Gamma$ for the $(n,L_K, S_{[qq]}^P, J_l^P,J^P)=(n,0, 0^+,0^+, \frac{1}{2}^+)$ configuration}

From Eq. (\ref{eq: Psi}) the  corresponding momentum-space wave-function $\Psi^{JJ_z}_{nL_K S_{[qq]} J_l}$ is given by
\be
\Psi^{1/2 J_z}_{ns00}(\tilde p_1,\tilde p_2,\lambda_1,\lambda_2)
&=& \langle \lambda_1|{\cal R}_M^\dagger(p_1^+,\vec p_{1\bot}, m_1)|s_1\rangle
         \langle 0|{\cal R}_M^\dagger(p_2^+,\vec p_{2\bot}, m_2)|0\rangle
\non\\ 
&&
               \la \frac{1}{2} 0; s_1 0|\frac{1}{2} 0; \frac{1}{2} J_z\ra
                \la 0 0; 0 0|0 0;0 0\ra
\non\\
&&          
                  ~\phi_{n 0 0}(x_1,x_2,k_{1\bot},k_{2\bot}).
\en
It can be expressed
as
\be
\Psi^{1/2 J_z}_{ns00}(\tilde p_1,\tilde p_2,\lambda_1,\lambda_2)
&=& \langle \lambda_1|{\cal R}_M^\dagger(p_1^+,\vec p_{1\bot}, m_1)|s_1\rangle
       \la \frac{1}{2} 0; s_1 0|\frac{1}{2} 0; \frac{1}{2} J_z\ra
\non\\
&&       
        ~\phi_{n 0 0}(x_1,x_2,k_{1\bot},k_{2\bot})
\non\\    
&=&  
      \frac{1}{\sqrt{(M_0+m_1)^2-m_2^2}}
        ~\bar u(p_1,\lambda_1)\Gamma_{s00} u(\bar P,J_z)
\non\\
&&
         ~\phi_{n s}(x_1,x_2,k_{1\bot},k_{2\bot}),           
\label{eq: Psi1/2s00}
\en
with
\be
\Gamma_{s00}=1,
\label{eq: Gammas00}
\en
where we have used Eqs.~(\ref{eq: meloshspin1/2}) and (\ref{eq: CG1/201/2}).
Putting everything together and
boosting $k_i\to p_i$ in the LF boost we obtain Eqs.~(\ref{eq: Psi1/2s00}) and (\ref{eq: Gammas00}).

\subsection{$\Gamma$ for the
     $(n,L_K, S_{[qq]}^P, J_l^P,J^P)=(n,0, 1^+,1^+, \frac{1}{2}^+)$ configuration}

From Eq. (\ref{eq: Psi}) the corresponding momentum-space wave-function $\Psi^{1/2J_z}_{nL_K S_{[qq]} J_l}$ is given by
as
\be
\Psi^{1/2J_z}_{ns11}(\tilde p_1,\tilde p_2,\lambda_1,\lambda_2)
&=& \langle \lambda_1|{\cal R}_M^\dagger(p_1^+,\vec p_{1\bot}, m_1)|s_1\rangle
         \langle \lambda_2|{\cal R}_M^\dagger(p_2^+,\vec p_{2\bot}, m_2)|s_2\rangle
\non\\ 
          && \la \frac{1}{2} 1; s_1 J_{lz}|\frac{1}{2} 1; \frac{1}{2} J_z\ra
                \la 0 1; 0 s_2|0 1;1 J_{l z}\ra
\non\\
&&          
                  ~\phi_{n 0 0}(x_1,x_2,k_{1\bot},k_{2\bot}).
\en
It can be expressed as
\be
\Psi^{1/2J_z}_{ns11}(\tilde p_1,\tilde p_2,\lambda_1,\lambda_2)
&=&\frac{1}{\sqrt{(M_0+m_1)^2-m_2^2}}
        ~\bar u(p_1,\lambda_1)\Gamma_{s11} u(\bar P,J_z)
\non\\
&&         
        \phi_{n s}(x_1,x_2,k_{1\bot},k_{2\bot}),
\label{eq: Psi1/2s11}
\en
with
\begin{eqnarray}
\Gamma_{s11}=\frac{\gamma_5}{\sqrt3}
\bigg(\not\!\varepsilon_{LF}^*(p_2,\lambda_2)
        -\frac{M_0 \not\! p_2+m_2 \not\!\bar P}{\bar P\cdot p_2+m_2 M_0}
        \frac{\varepsilon_{LF}^*(p_2,\lambda_2)\cdot \bar P}{M_0}\bigg),
\label{eq: Gammas11}        
\end{eqnarray}
where we have made use of 
Eqs.~(\ref{eq: meloshspin1/2}), (\ref{eq: meloshspin1}), (\ref{eq: CG1/211/2}), (\ref{eq: ep2}) and (\ref{eq: epsum}).
In particular using Eqs. (\ref{eq: meloshspin1}), (\ref{eq: ep2}) and (\ref{eq: epsum}), we have
\be
\la \lambda_2|{\cal R}^\dagger_M (x_2,k_{2\bot},m_2)|s_2\ra \varepsilon_I^{*\nu}(\bar P,s_2)
&=&\varepsilon_{LF}^{*\nu}(k_2,\lambda_2)
-\frac{M_0 k^\nu_2+m_2 \bar P^\nu}{(\bar P\cdot k_2+m_2 M_0)}\frac{\varepsilon_{LF}^*(k_2,\lambda_2)\cdot \bar P}{M_0}.
\label{eq: meloshspin1ep2}
\en
Putting everything together and
boosting $k_i\to p_i$ in the LF boost we obtain Eqs.~(\ref{eq: Psi1/2s11}) and (\ref{eq: Gammas11}).
Using equation of motion, we finally obtain $\Gamma_{s11}$ as shown in Eq. (\ref{eq: Gamma}).

\subsection{$\Gamma$ for the 
     $(n,L_K, S_{[qq]}^P, J_l^P,J^P)= (n,1, 0^+,1^-, \frac{1}{2}^-)$ configuration}

From Eq. (\ref{eq: Psi}) the corresponding momentum-space wave-function $\Psi^{JJ_z}_{n L_K S_{[qq]} J_l}$ 
is given by
\be
\Psi^{\frac{1}{2}J_z}_{np01}(\tilde p_1,\tilde p_2,\lambda_1,\lambda_2)
&=& \langle \lambda_1|{\cal R}_M^\dagger(p_1^+,\vec p_{1\bot}, m_1)|s_1\rangle
\non\\ 
          && \la \frac{1}{2} 1; s_1 J_{lz}|\frac{1}{2} 1; \frac{1}{2} J_z\ra
                \la 1 0; L_z 0|1 0;1 J_{l z}\ra
\non\\
&&          
                  ~\phi_{n 1 L_z}(x_1,x_2,k_{1\bot},k_{2\bot}).
\en
It can be expressed
as
\be
\Psi^{\frac{1}{2}J_z}_{np01}(\tilde p_1,\tilde p_2,\lambda_1,\lambda_2)
&=&\frac{1}{\sqrt{(M_0+m_1)^2-m_2^2}}
        ~\bar u(p_1,\lambda_1)\Gamma_{p01} u(\bar P,J_z)
\non\\
&&         
        \phi_{np}(x_1,x_2,k_{1\bot},k_{2\bot}),
\en
with
\begin{eqnarray}
\Gamma_{p01}=\frac{\gamma_5}{2\sqrt3}
\bigg(\not\! p_1-\not \! p_2-\frac{m_1^2-m_2^2}{M_0^2} \not\! \bar P\bigg),
\end{eqnarray}
where Eqs. (\ref{eq: meloshspin1/2}), (\ref{eq: CG1/211/2}), (\ref{eq: varphi}) and (\ref{eq: epsum}) have been used.
Using equation of motion, we finally obtain $\Gamma_{p01}$ as shown in Eq. (\ref{eq: Gamma}).

\section{Obtaining Transition Form Factors}

We shall follow~\cite{Schlumpf, Cheng:2004cc} to project out various form
factors from the transition matrix elements. As in
\cite{CCH,Schlumpf, Cheng:2004cc}, we consider the $q^+=0$, $\vec q_\bot\not=\vec 0$ case.
With the help of the following identities,
 \be
 \frac{\bar u(P',J'_z)\gamma^+ u(P, J_z)}{2\sqrt{P^+ P^{\prime
 +}}}&=&\delta_{J'_z J_z},
 \qquad
 i\frac{\bar u(P',J'_z)\sigma^{+\nu} q_\nu u(P, J_z)}{2\sqrt{P^+
P^{\prime
 +}}}=(\vec\sigma\cdot \vec q_\bot\sigma^3)_{J'_z J_z},
  \non\\
  \frac{\bar u(P',J'_z)\gamma^+\gamma_5 u(P, J_z)}{2\sqrt{P^+ P^{\prime
 +}}}&=&(\sigma^3)_{J'_z J_z},
 \quad
 i\frac{\bar u(P',J'_z)\sigma^{+\nu} q_\nu\gamma_5 u(P, J_z)}{2\sqrt{P^+
P^{\prime
 +}}}=(\vec\sigma\cdot \vec q_\bot)_{J'_z J_z},
 \label{eq:spinorprojection}
  \en
the matrix elements of $\B_b(1/2^+)\to\B_c(1/2^+)$ transition can be expressed as
 \be
 \la \B_{c}(P',J'_z)|V^+|\B_b(P,J_z)\ra
 &=&2\sqrt{P^+ P^{\prime +}}
 \left[f^V_1(q^2)~\delta_{J'_z J_z}+\frac{f^V_2(q^2)}{M+M'}(\vec\sigma\cdot \vec q_\bot\sigma^3)_{J'_z
 J_z}\right],
 \non\\
 \la \B_{c}(P',J'_z)|A^+|\B_b(P,J_z)\ra
 &=&2\sqrt{P^+P^{\prime +}}
 \left[ g^A_1(q^2)~(\sigma^3)_{J'_z J_z}+\frac{g^A_2(q^2)}{M+M'}(\vec\sigma\cdot \vec q_\bot)_{J'_z
 J_z}\right],
 \label{eq:projection}
  \en
and similar expressions for the $\B_b(1/2^+)\to\B_c(1/2^-)$ case with suitable replacement of $V$ and $A$.  
Various form factors can be projected out by applying the
orthogonality of the corresponding matrices, $\delta_{J_z J'_z}$,
$(\sigma^3\sigma_{\bot}^i)_{J_z J'_z}$, $(\sigma^3)_{J_z J'_z}$
and $(\sigma_\bot^i)_{J_z J'_z}$, under the trace operation:
\be
f^V_1(q^2)
&=&\frac{1}{2}\sum_{J_z, J'_z} \delta_{J_z J'_z}
                                   \frac{\la \B_{c}(P',J'_z)|V^+|\B_b(P,J_z)\ra}{2\sqrt{P^+ P^{\prime +}}},
\non\\
\frac{f^V_2(q^2)q^i_\bot}{M+M'} 
&=&\frac{1}{2}\sum_{J_z, J'_z} (\sigma^3\sigma_{\bot}^i)_{J_z J'_z}
                                   \frac{\la \B_{c}(P',J'_z)|V^+|\B_b(P,J_z)\ra}{2\sqrt{P^+ P^{\prime +}}},
\non\\
g^A_1(q^2)
&=&\frac{1}{2}\sum_{J_z, J'_z} (\sigma^3)_{J_z J'_z}
                                   \frac{\la \B_{c}(P',J'_z)|A^+|\B_b(P,J_z)\ra}{2\sqrt{P^+ P^{\prime +}}},
\non\\
\frac{g^A_2(q^2)q^i_\bot}{M+M'} 
&=&\frac{1}{2}\sum_{J_z, J'_z} (\sigma_{\bot}^i)_{J_z J'_z}
                                   \frac{\la \B_{c}(P',J'_z)|A^+|\B_b(P,J_z)\ra}{2\sqrt{P^+ P^{\prime +}}},
\en 
and similar equations for $f^{A}_{1,2}$ and $g^V_{1,2}$ in the $\B_b(1/2^+)\to\B_c(1/2^-)$ case 
by suitably replacing $V$ and $A$.
Note
that due to the condition $q^+=0$ we have imposed, the
form factors $f^{V,A}_3(q^3)$ and $g^{A,V}_3(q^3)$ cannot be extracted in this
manner. 
Substituting Eq. (\ref{eq: B->B'}) to the right-hand-side of the above equations, expressions of  
$\sum_{J_z,J'_z} \delta_{J_z J'_z}\bar u(\bar P',J'_z)(\dots) u(\bar P, J_z)$ and so on occur.
They can be further simplified by using the following identities:~\footnote{These identities can be easily proved by using Eq.~(\ref {eq:spinorprojection}), but with $P$ and $P'$ replaced by
$\bar P$ and $\bar P'$ and with suitable replacements of $J_z$ and $J'_z$.}
  \be
 \frac{1}{2}\sum_{J_z,J'_z} u(\bar P, J_z)\delta_{J_z J'_z}\bar u(\bar P',J'_z)
  &=&\frac{1}{4\sqrt{P^+P^{\prime+}}}(\not\!\bar P+M_0)\gamma^+(\not\!\bar P'+M'_0),
 \non\\
 \frac{1}{2}\sum_{J_z,J'_z} u(\bar P, J_z)(\sigma^3\sigma_{\bot}^i )_{J_z J'_z}\bar u(\bar P',J'_z)
  &=&-\frac{i}{4\sqrt{P^+P^{\prime+}}}(\not\!\bar P+M_0)\sigma^{i+}(\not\!\bar P'+M'_0),
 \non\\
 \frac{1}{2}\sum_{J_z,J'_z} u(\bar P, J_z)(\sigma^3)_{J_z J'_z}\bar u(\bar P',J'_z)
  &=&\frac{1}{4\sqrt{P^+P^{\prime+}}}(\not\!\bar P+M_0)\gamma^+\gamma_5(\not\!\bar P'+M'_0),
 \non\\
 \frac{1}{2}\sum_{J_z,J'_z} u(\bar P, J_z)(\sigma_\bot^i)_{J_z J'_z}\bar u(\bar P',J'_z)
  &=&\frac{i}{4\sqrt{P^+P^{\prime+}}}(\not\!\bar P+M_0)\sigma^{i+}\gamma_5(\not\!\bar P'+M'_0).
  \label{eq:spinorprojectionbar}
 \en
With the above generic discussions on $\B_b\to \B_{c}$
transition, we are ready to extract the transition form factors:
for $\B_b(1/2^+)\to\B_c(1/2^+)$ transition, we have
\be
f^V_1(q^2) & =&\int \{d\tilde p_2\}~
      \frac{\phi^{\prime*}_{nL'_k}(\{x'\},\{k'_{\bot}\})\phi_{1 L_K}(\{x\},\{k_{\bot}\})}
             {16 P^+P^{\prime+}\sqrt{p_1^{\prime +} p_1^+ (p'_1\cdot\bar P'+m'_1 M'_0)(p_1\cdot \bar P+m_1 M_0)}}
\non\\
&&
     \qquad\times~Tr[(\not\!\bar P+M_0)\gamma^+(\not\!\bar P'+M'_0)\bar \Gamma_{L'_K S_{[qq]} J'_l} 
     (\not\! p'_1+m'_1)\gamma^+(\not\! p_1+m_1)\Gamma_{L_K S_{[qq]} J_l}],
\non\\
\frac{f^V_2(q^2) q^i_\bot}{M+M'} 
&=&-i\int \{d\tilde p_2\}~
      \frac{\phi^{\prime*}_{nL'_k}(\{x'\},\{k'_{\bot}\})\phi_{1 L_K}(\{x\},\{k_{\bot}\})}
             {16 P^+P^{\prime+}\sqrt{p_1^{\prime +} p_1^+ (p'_1\cdot\bar P'+m'_1 M'_0)(p_1\cdot \bar P+m_1 M_0)}}
\non\\
&&
     \qquad\times~Tr[(\not\!\bar P+M_0)\sigma^{i+}(\not\!\bar P'+M'_0)\bar \Gamma_{L'_K S_{[qq]} J'_l} 
     (\not\! p'_1+m'_1)\gamma^+(\not\! p_1+m_1)\Gamma_{L_K S_{[qq]} J_l}],
\non\\
g^A_1(q^2) & =&\int \{d\tilde p_2\}~
      \frac{\phi^{\prime*}_{nL'_k}(\{x'\},\{k'_{\bot}\})\phi_{1 L_K}(\{x\},\{k_{\bot}\})}
             {16 P^+P^{\prime+}\sqrt{p_1^{\prime +} p_1^+ (p'_1\cdot\bar P'+m'_1 M'_0)(p_1\cdot \bar P+m_1 M_0)}}
\non\\
&&
     \qquad\times~Tr[(\not\!\bar P+M_0)\gamma^+\gamma_5(\not\!\bar P'+M'_0)\bar \Gamma_{L'_K S_{[qq]} J'_l} 
     (\not\! p'_1+m'_1)\gamma^+\gamma_5(\not\! p_1+m_1)\Gamma_{L_K S_{[qq]} J_l}],
\non\\
\frac{g^A_2(q^2) q^i_\bot}{M+M'} 
&=&i\int \{d\tilde p_2\}~
      \frac{\phi^{\prime*}_{nL'_k}(\{x'\},\{k'_{\bot}\})\phi_{1 L_K}(\{x\},\{k_{\bot}\})}
             {16 P^+P^{\prime+}\sqrt{p_1^{\prime +} p_1^+ (p'_1\cdot\bar P'+m'_1 M'_0)(p_1\cdot \bar P+m_1 M_0)}}
\non\\
&&
     \qquad\times~Tr[(\not\!\bar P+M_0)\sigma^{i+}\gamma_5(\not\!\bar P'+M'_0)\bar \Gamma_{L'_K S_{[qq]} J'_l} 
     (\not\! p'_1+m'_1)\gamma^+\gamma_5(\not\! p_1+m_1)\Gamma_{L_K S_{[qq]} J_l}],
\non\\
\label{eq: figi (i) and (ii)}
\en
with $q_\bot^ i=q_\bot^1$ or $q_\bot^2$ (no sum over $i$),
and similarly, for $\B_b(1/2^+)\to\B_c(1/2^-)$ transition, we have
\be
f^A_1(q^2) & =&\int \{d\tilde p_2\}~
      \frac{\phi^{\prime*}_{nL'_k}(\{x'\},\{k'_{\bot}\})\phi_{1 L_K}(\{x\},\{k_{\bot}\})}
             {16 P^+P^{\prime+}\sqrt{p_1^{\prime +} p_1^+ (p'_1\cdot\bar P'+m'_1 M'_0)(p_1\cdot \bar P+m_1 M_0)}}
\non\\
&&
     \qquad\times~Tr[(\not\!\bar P+M_0)\gamma^+(\not\!\bar P'+M'_0)\bar \Gamma_{L'_K S_{[qq]} J'_l} 
     (\not\! p'_1+m'_1)\gamma^+\gamma_5(\not\! p_1+m_1)\Gamma_{L_K S_{[qq]} J_l}],
\non\\
\frac{f^A_2(q^2) q^i_\bot}{M+M'} 
&=&-i\int \{d\tilde p_2\}~
      \frac{\phi^{\prime*}_{nL'_k}(\{x'\},\{k'_{\bot}\})\phi_{1 L_K}(\{x\},\{k_{\bot}\})}
             {16 P^+P^{\prime+}\sqrt{p_1^{\prime +} p_1^+ (p'_1\cdot\bar P'+m'_1 M'_0)(p_1\cdot \bar P+m_1 M_0)}}
\non\\
&&
     \qquad\times~Tr[(\not\!\bar P+M_0)\sigma^{i+}(\not\!\bar P'+M'_0)\bar \Gamma_{L'_K S_{[qq]} J'_l} 
     (\not\! p'_1+m'_1)\gamma^+\gamma_5(\not\! p_1+m_1)\Gamma_{L_K S_{[qq]} J_l}],
\non\\
g^V_1(q^2) & =&\int \{d\tilde p_2\}~
      \frac{\phi^{\prime*}_{nL'_k}(\{x'\},\{k'_{\bot}\})\phi_{1 L_K}(\{x\},\{k_{\bot}\})}
             {16 P^+P^{\prime+}\sqrt{p_1^{\prime +} p_1^+ (p'_1\cdot\bar P'+m'_1 M'_0)(p_1\cdot \bar P+m_1 M_0)}}
\non\\
&&
     \qquad\times~Tr[(\not\!\bar P+M_0)\gamma^+(\not\!\bar P'+M'_0)\bar \Gamma_{L'_K S_{[qq]} J'_l} 
     (\not\! p'_1+m'_1)\gamma^+\gamma_5(\not\! p_1+m_1)\Gamma_{L_K S_{[qq]} J_l}],
\non\\
\frac{g^A_2(q^2) q^i_\bot}{M+M'} 
&=&i\int \{d\tilde p_2\}~
      \frac{\phi^{\prime*}_{nL'_k}(\{x'\},\{k'_{\bot}\})\phi_{1 L_K}(\{x\},\{k_{\bot}\})}
             {16 P^+P^{\prime+}\sqrt{p_1^{\prime +} p_1^+ (p'_1\cdot\bar P'+m'_1 M'_0)(p_1\cdot \bar P+m_1 M_0)}}
\non\\
&&
     \qquad\times~Tr[(\not\!\bar P+M_0)\sigma^{i+}\gamma_5(\not\!\bar P'+M'_0)\bar \Gamma_{L'_K S_{[qq]} J'_l} 
     (\not\! p'_1+m'_1)\gamma^+(\not\! p_1+m_1)\Gamma_{L_K S_{[qq]} J_l}],
\non\\     
\label{eq: figi (iii)}
\en
with $q_\bot^ i=q_\bot^1$ or $q_\bot^2$ (no sum over $i$).
We are now ready to discuss various transitions in more detail.

\subsection{Form factors for $\B_b({\bf \bar 3_f}, 1/2^+)\to \B_c({\bf \bar 3_f}, 1/2^+)$ transition [type (i)]}

We will discuss how to obtain the formulas of form factors of the type (i) transition in this subsection. 
The $\B_b({\bf \bar 3_f},{1}/{2}^+)\to\B_c({\bf \bar 3_f},{1}/{2}^+)$ transitions involve initial states in
$(n, L_K, S_{[qq]}^P, J_l^P,J^P)_b
 =(1, 0, 0^+,0^+, \frac{1}{2}^+)$ configuration and final states in
$(n, L_K, S_{[qq]}^P, J_l^P,J^P)_c=(n, 0, 0^+,0^+, \frac{1}{2}^+)$ configurations (with $n$=1,2). 
In these transitions the scalar diquarks are spectators.
 
Following Eq.~(\ref{eq: Gamma}),
\be
\Gamma_{L'_K S'_{[qq]} J'_l}=\Gamma_{L_K S_{[qq]} J_l}=\Gamma_{s00}=1,
\en 
and (\ref{eq: B->B'}),
we have
\be
&&
      \la \B_{c}(P',J'_z)|\bar c\gamma^+ (1-\gamma_5) b|\B_{b}(P,J_z)\ra
\non\\
&&\qquad
       =\int \{d\tilde p_2\}~
      \frac{\phi^{\prime*}_{nL'_k}(\{x'\},\{k'_{\bot}\})\phi_{1 L_K}(\{x\},\{k_{\bot}\})}
             {2\sqrt{p_1^{\prime +} p_1^+ (p'_1\cdot\bar P'+m'_1 M'_0)(p_1\cdot \bar P+m_1 M_0)}}
\non\\
&&
    \qquad \qquad\times~\bar u(\bar P',J'_z)
     (\not\! p'_1+m'_1)\gamma^\mu(1-\gamma_5)(\not\! p_1+m_1) u(\bar P,J_z),
\label{eq: (i) B->B'}
\en
for the type (i) transition. By using
Eq. (\ref{eq: figi (i) and (ii)}) the transition form
factors are given by
\be
f^V_1(q^2) & =&\int \frac{dx_2 d^2 k_{2\bot}}{2 (2\pi)^3}~
      \frac{\phi^{\prime*}_{ns}(\{x'\},\{k'_{\bot}\})\phi_{1s}(\{x\},\{k_{\bot}\})}
             {16 P^+P^{\prime+}\sqrt{x_1^{\prime +} x_1^+ (p'_1\cdot\bar P'+m'_1 M'_0)(p_1\cdot \bar P+m_1 M_0)}}
\non\\
&&
     \qquad\times~Tr[
      (\not\!\bar P+M_0)\gamma^+(\not\!\bar P'+M'_0)(\not\! p'_1+m'_1)\gamma^+(\not\! p_1+m_1)],
\non\\
\frac{f^V_2(q^2) q^i_\bot}{M+M'} 
&=&-i\int \frac{dx_2 d^2 k_{2\bot}}{2 (2\pi)^3}~
      \frac{\phi^{\prime*}_{ns}(\{x'\},\{k'_{\bot}\})\phi_{1s}(\{x\},\{k_{\bot}\})}
             {16 P^+P^{\prime+}\sqrt{x_1^{\prime +} x_1^+ (p'_1\cdot\bar P'+m'_1 M'_0)(p_1\cdot \bar P+m_1 M_0)}}
\non\\
&&
     \qquad\times~Tr[
     (\not\!\bar P+M_0)\sigma^{i+}(\not\!\bar P'+M'_0)(\not\! p'_1+m'_1)\gamma^+(\not\! p_1+m_1)],
\non\\
g^A_1(q^2) & =&\int \frac{dx_2 d^2 k_{2\bot}}{2 (2\pi)^3}~
      \frac{\phi^{\prime*}_{ns}(\{x'\},\{k'_{\bot}\})\phi_{1s}(\{x\},\{k_{\bot}\})}
             {16 P^+P^{\prime+}\sqrt{x_1^{\prime +} x_1^+ (p'_1\cdot\bar P'+m'_1 M'_0)(p_1\cdot \bar P+m_1 M_0)}}
\non\\
&&
     \qquad\times~Tr[
      (\not\!\bar P+M_0)\gamma^+\gamma_5(\not\!\bar P'+M'_0)(\not\! p'_1+m'_1)\gamma^+\gamma_5(\not\! p_1+m_1)],
\non\\
\frac{g^A_2(q^2) q^i_\bot}{M+M'} 
&=&i\int \frac{dx_2 d^2 k_{2\bot}}{2 (2\pi)^3}~
      \frac{\phi^{\prime*}_{ns}(\{x'\},\{k'_{\bot}\})\phi_{1s}(\{x\},\{k_{\bot}\})}
             {16 P^+P^{\prime+}\sqrt{x_1^{\prime +} x_1^+ (p'_1\cdot\bar P'+m'_1 M'_0)(p_1\cdot \bar P+m_1 M_0)}}
\non\\
&&
     \qquad\times~Tr[
     (\not\!\bar P+M_0)\sigma^{i+}\gamma_5(\not\!\bar P'+M'_0) (\not\! p'_1+m'_1)\gamma^+\gamma_5(\not\! p_1+m_1)].
\en
with $q_\bot^ i=q_\bot^1$ or $q_\bot^2$ (no sum over $i$).

It is straightforward to work out the traces in $f^V_{1,2}(q^2)$ as
shown in the above equation and obtain~\cite{Cheng:2004cc}
 \be
 &&\frac{1}{8 P^+ P^{\prime+}}{\rm Tr}[(\not\!\bar P+M_0)\gamma^+(\not\!\bar P'+M'_0) (\not\!
 p'_1+m'_1)\gamma^+(\not\! p_1+m_1)]
 \non\\
 &&\qquad\qquad=-(p_1-x_1\bar P)\cdot (p'_1-x'_1 \bar P')+(x_1 M_0+m_1)(x'_1
 M'_0+m'_1),
 \non\\
&&\frac{i}{8 P^+ P^{\prime+}}{\rm Tr}[(\not\!\bar
P+M_0)\sigma^{i+}(\not\!\bar P'+M'_0) (\not\!
 p'_1+m'_1)\gamma^+(\not\!
 p_1+m_1)]
 \non\\
 &&\qquad\qquad=(m'_1+x'_1 M'_0) (p^i_\bot-x_1 \bar P^i_\bot)
 -(m_1+x_1 M_0) (p^{\prime i}_\bot-x'_1 \bar P^{\prime i}_\bot),
 \label{eq:trace1}
 \en
for $i=1,2$, where uses of  $\bar P^{(\prime)+}=P^{(\prime)+}$,
$\bar P_\bot^{(\prime)i}=P^{(\prime)i}_\bot$, $p^{(\prime)
+}_1=x^{(\prime)} P^{(\prime)+}$, $p^{(\prime)
i}_{1\bot}=x^{(\prime)} P^{(\prime)i}_\bot+k^{(\prime)i}_{1\bot}$
have been made. 
Similarly the traces in $g^A_1(q^2)$ and $g^A_2(q^2)$ can be obtained
by replacing $m'_1\to-m'_1$, $M'_0\to -M'_0$ in the above traces and with an
additional overall minus sign.
With the help of Eq.~(\ref{eq: kinematics}) the above form factors can be expressed in
terms of the internal variables via~\cite{Cheng:2004cc}
 \be
 &&p_1\cdot \bar P=e_1 M_0=\frac{m_1^2+x^2_1 M_0^2+k^2_{1\bot}}{2 x_1},\quad
 p'_1\cdot \bar P'=e'_1 M'_0=\frac{m_1^{\prime 2}+x^2_1 M_0^{\prime 2}+k^{\prime 2}_{1\bot}}{2 x_1},
 \non\\
 &&(p_1-x_1\bar P)\cdot (p'_1-x'_1 \bar P')=-k_{1\bot}\cdot
 k'_{1\bot},\quad
 p^{(\prime)i}_\bot-x_1 \bar
 P^{(\prime)i}_\bot=k_{1\bot}^{(\prime)i},
 \label{eq:internal1}
 \en
where $k_{1\bot}\cdot
 k'_{1\bot}$ is a scalar product in two-dimensional space.
The obtained form factors are shown in Eq.~(\ref{eq:ff type i}).

\subsection{Form factors for $\B_b({\bf 6_f}, 1/2^+)\to \B_c({\bf 6_f}, 1/2^+)$ transition [type (ii)]}

We will discuss how to obtain the formulas of form factors of the type (ii) transition in this subsection. 
The $\B_b({\bf 6_f}, 1/2^+)\to \B_c({\bf 6_f}, 1/2^+)$ transitions involve initial states in
$(n, L_K, S_{[qq]}^P, J_l^P,J^P)_b
 =(1, 0, 1^+,1^+, \frac{1}{2}^+)$ configuration and final states in
$(n, L_K, S_{[qq]}^P, J_l^P,J^P)_c=(n, 0, 1^+,1^+, \frac{1}{2}^+)$ configurations (with $n$=1,2). 
In these transitions the axial-vector diquarks are spectators.

Following Eqs.~~(\ref{eq: Gamma}) and (\ref{eq: B->B'})
we have
\be
&&
      \la \B_{c}(P',J'_z)|\bar c\gamma^\mu (1-\gamma_5) b|\B_{b}(P,J_z)\ra
\non\\
&&\qquad
       =\int \{d\tilde p_2\}~
      \frac{\phi^{\prime*}_{nL'_k}(\{x'\},\{k'_{\bot}\})\phi_{1 L_K}(\{x\},\{k_{\bot}\})}
             {2\sqrt{p_1^{\prime +} p_1^+ (p'_1\cdot\bar P'+m'_1 M'_0)(p_1\cdot \bar P+m_1 M_0)}}
\non\\
&&
     \qquad\qquad\times~\bar u(\bar P',J'_z)\bar \Gamma'_{s 1 1} 
     (\not\! p'_1+m'_1)\gamma^\mu(1-\gamma_5)(\not\! p_1+m_1)\Gamma_{s 1 1} u(\bar P,J_z),
\label{eq: (ii)B->B'}
\en
with
\be
\Gamma_{s11}
       =\frac{\gamma_5}{\sqrt3}
          \bigg(\not\!\varepsilon_{LF}^*(p_2,\lambda_2)
        -
        \frac{M_0+m_1+m_2}{\bar P\cdot p_2+m_2 M_0}\varepsilon_{LF}^*(p_2,\lambda_2)\cdot \bar P\bigg)
        \equiv\varepsilon_{LF\mu}^*(p_2,\lambda_2) \Gamma^\mu_{s11},
        \non\\
\bar\Gamma'_{s11}
       =\frac{\gamma_5}{\sqrt3}
          \bigg(\not\!\varepsilon_{LF}(p_2,\lambda_2)
        +
        \frac{M'_0+m'_1+m_2}{\bar P'\cdot p_2+m_2 M'_0}\varepsilon_{LF}(p_2,\lambda_2)\cdot \bar P'\bigg)
        \equiv\varepsilon_{LF\mu}(p_2,\lambda_2)\bar\Gamma^{\prime\mu}_{s11},      
\en
where we have made use of the fact that the diquark is a spectator of the transition.
By using
Eq.~(\ref{eq: figi (i) and (ii)})
the transition form
factors for the $\B_b({\bf 6_f}, 1/2^+)\to \B_c({\bf 6_f}, 1/2^+)$ case are given by:
\be
f^V_1(q^2) & =&\int \{d\tilde p_2\}~
      \frac{\phi^{\prime*}_{nL'_k}(\{x'\},\{k'_{\bot}\})\phi_{1 L_K}(\{x\},\{k_{\bot}\})}
             {16 P^+P^{\prime+}\sqrt{p_1^{\prime +} p_1^+ (p'_1\cdot\bar P'+m'_1 M'_0)(p_1\cdot \bar P+m_1 M_0)}}
             ]\left(-g_{\mu\nu}+\frac{p_{2\mu}p_{2\nu}}{m_2^2}\right)
\non\\
&&
     \qquad\times~Tr[(\not\!\bar P+M_0)\gamma^+(\not\!\bar P'+M'_0)\bar\Gamma^{\prime\mu}_{s11}
     (\not\! p'_1+m'_1)\gamma^+(\not\! p_1+m_1)\Gamma^\nu_{s11},
\non\\
\frac{f^V_2(q^2) q^i_\bot}{M+M'} 
&=&-i\int \{d\tilde p_2\}~
      \frac{\phi^{\prime*}_{nL'_k}(\{x'\},\{k'_{\bot}\})\phi_{1 L_K}(\{x\},\{k_{\bot}\})}
             {16 P^+P^{\prime+}\sqrt{p_1^{\prime +} p_1^+ (p'_1\cdot\bar P'+m'_1 M'_0)(p_1\cdot \bar P+m_1 M_0)}}
             ]\left(-g_{\mu\nu}+\frac{p_{2\mu}p_{2\nu}}{m_2^2}\right)
\non\\
&&
     \qquad\times~Tr[(\not\!\bar P+M_0)\sigma^{i+}(\not\!\bar P'+M'_0)\bar\Gamma^{\prime\mu}_{s11}
     (\not\! p'_1+m'_1)\gamma^+(\not\! p_1+m_1)\Gamma^\nu_{s11},
\non\\
g^A_1(q^2) & =&\int \{d\tilde p_2\}~
      \frac{\phi^{\prime*}_{nL'_k}(\{x'\},\{k'_{\bot}\})\phi_{1 L_K}(\{x\},\{k_{\bot}\})}
             {16 P^+P^{\prime+}\sqrt{p_1^{\prime +} p_1^+ (p'_1\cdot\bar P'+m'_1 M'_0)(p_1\cdot \bar P+m_1 M_0)}}
             ]\left(-g_{\mu\nu}+\frac{p_{2\mu}p_{2\nu}}{m_2^2}\right)
\non\\
&&
     \qquad\times~Tr[(\not\!\bar P+M_0)\gamma^+\gamma_5(\not\!\bar P'+M'_0)\bar\Gamma^{\prime\mu}_{s11}
     (\not\! p'_1+m'_1)\gamma^+\gamma_5(\not\! p_1+m_1)\Gamma^\nu_{s11}],
\non\\
\frac{g^A_2(q^2) q^i_\bot}{M+M'} 
&=&i\int \{d\tilde p_2\}~
      \frac{\phi^{\prime*}_{nL'_k}(\{x'\},\{k'_{\bot}\})\phi_{1 L_K}(\{x\},\{k_{\bot}\})}
             {16 P^+P^{\prime+}\sqrt{p_1^{\prime +} p_1^+ (p'_1\cdot\bar P'+m'_1 M'_0)(p_1\cdot \bar P+m_1 M_0)}}
             ]\left(-g_{\mu\nu}+\frac{p_{2\mu}p_{2\nu}}{m_2^2}\right)
\non\\
&&
     \qquad\times~Tr[(\not\!\bar P+M_0)\sigma^{i+}\gamma_5(\not\!\bar P'+M'_0)\bar\Gamma^{\prime\mu}_{s11}
     (\not\! p'_1+m'_1)\gamma^+\gamma_5(\not\! p_1+m_1)\Gamma^\nu_{s11}],
\en
with $q_\bot^ i=q_\bot^1$ or $q_\bot^2$ (no sum over $i$).
As one can see the traces are rather complicate. To simplify the derivations, we choose to work in the $\vec P_\bot=0$ frame.
After working out the traces and making use of Eq. (\ref{eq: kinematics}), we obtain the form factors as shown in Eq.~(\ref{eq:ff type ii}).

\subsection{Form factors for $\B_b({\bf \bar 3_f}, 1/2^+)\to \B_c({\bf \bar 3_f}, 1/2^-)$ transition [type (iii)]}

We will discuss how to obtain the formulas of form factors of the type (iii) transition in this subsection. 
The $\B_b({\bf \bar 3_f},{1}/{2}^+)\to\B_c({\bf \bar 3_f},{1}/{2}^-)$ transitions
involve initial states in
$(n, L_K, S_{[qq]}^P, J_l^P,J^P)_b
 =(1, 0, 0^+,0^+, \frac{1}{2}^+)$ configuration and final states in
$(n, L_K, S_{[qq]}^P, J_l^P,J^P)_c=(n, 1, 0^+,1^-, \frac{1}{2}^-)$ configurations (with $n$=1,2). 
In these transitions, the scalar diquarks are spectators.

Following Eqs.~~(\ref{eq: Gamma}) and (\ref{eq: B->B'}),
we have, 
\be
&&
      \la \B_{c}(P',J'_z)|\bar c\gamma^+ (1-\gamma_5) b|\B_{b}(P,J_z)\ra
\non\\
&&\qquad
       =\int \{d\tilde p_2\}~
      \frac{\phi^{\prime*}_{nL'_k}(\{x'\},\{k'_{\bot}\})\phi_{1 L_K}(\{x\},\{k_{\bot}\})}
             {2\sqrt{p_1^{\prime +} p_1^+ (p'_1\cdot\bar P'+m'_1 M'_0)(p_1\cdot \bar P+m_1 M_0)}}
\non\\
&&
     \qquad\qquad\times~\bar u(\bar P',J'_z)
     \bar\Gamma'_{p01}
     (\not\! p'_1+m'_1)\gamma^\mu(1-\gamma_5)(\not\! p_1+m_1) u(\bar P,J_z),
\label{eq: (i) B->B'}
\en
with
\be
\Gamma_{L_K S_{[qq]} J_l}=\Gamma_{s00}=1,
\quad
\bar\Gamma_{L'_K S'_{[qq]} J'_l}
=\bar\Gamma'_{p01}=\frac{\gamma_5}{2\sqrt3}
     \bigg(\not\! p'_1-\not \! p_2+\frac{m_1^{\prime 2}-m_2^2}{M'_0} \bigg),
\en 
for the $\B_b({\bf \bar 3_f},{1}/{2}^+)\to\B_c({\bf \bar 3_f},{1}/{2}^-)$ transition [type (iii)]. 
By using Eq. (\ref{eq: figi (iii)}) we obtain the transition form factors:
\be
f^A_1(q^2) & =&\int \frac{dx_2 d^2 k_{2\bot}}{2 (2\pi)^3}~
      \frac{\phi^{\prime*}_{np}(\{x'\},\{k'_{\bot}\})\phi_{1s}(\{x\},\{k_{\bot}\})}
             {16 P^+P^{\prime+}\sqrt{x_1^{\prime +} x_1^+ (p'_1\cdot\bar P'+m'_1 M'_0)(p_1\cdot \bar P+m_1 M_0)}}
\non\\
&&
     \qquad\times~Tr[
      (\not\!\bar P+M_0)\gamma^+(\not\!\bar P'+M'_0)\bar\Gamma'_{p01}(\not\! p'_1+m'_1)\gamma^+\gamma_5(\not\! p_1+m_1)],
\non\\
\frac{f^A_2(q^2) q^i_\bot}{M+M'} 
&=&-i\int \frac{dx_2 d^2 k_{2\bot}}{2 (2\pi)^3}~
      \frac{\phi^{\prime*}_{np}(\{x'\},\{k'_{\bot}\})\phi_{1s}(\{x\},\{k_{\bot}\})}
             {16 P^+P^{\prime+}\sqrt{x_1^{\prime +} x_1^+ (p'_1\cdot\bar P'+m'_1 M'_0)(p_1\cdot \bar P+m_1 M_0)}}
\non\\
&&
     \qquad\times~Tr[
     (\not\!\bar P+M_0)\sigma^{i+}(\not\!\bar P'+M'_0)\bar\Gamma'_{p01}(\not\! p'_1+m'_1)\gamma^+\gamma_5(\not\! p_1+m_1)],
\non\\
g^V_1(q^2) & =&\int \frac{dx_2 d^2 k_{2\bot}}{2 (2\pi)^3}~
      \frac{\phi^{\prime*}_{np}(\{x'\},\{k'_{\bot}\})\phi_{1s}(\{x\},\{k_{\bot}\})}
             {16 P^+P^{\prime+}\sqrt{x_1^{\prime +} x_1^+ (p'_1\cdot\bar P'+m'_1 M'_0)(p_1\cdot \bar P+m_1 M_0)}}
\non\\
&&
     \qquad\times~Tr[
      (\not\!\bar P+M_0)\gamma^+\gamma_5(\not\!\bar P'+M'_0)\bar\Gamma'_{p01}(\not\! p'_1+m'_1)\gamma^+(\not\! p_1+m_1)],
\non\\
\frac{g^V_2(q^2) q^i_\bot}{M+M'} 
&=&i\int \frac{dx_2 d^2 k_{2\bot}}{2 (2\pi)^3}~
      \frac{\phi^{\prime*}_{np}(\{x'\},\{k'_{\bot}\})\phi_{1s}(\{x\},\{k_{\bot}\})}
             {16 P^+P^{\prime+}\sqrt{x_1^{\prime +} x_1^+ (p'_1\cdot\bar P'+m'_1 M'_0)(p_1\cdot \bar P+m_1 M_0)}}
\non\\
&&
     \qquad\times~Tr[
     (\not\!\bar P+M_0)\sigma^{i+}\gamma_5(\not\!\bar P'+M'_0) \bar\Gamma'_{p01}(\not\! p'_1+m'_1)\gamma^+(\not\! p_1+m_1)].
\en
with $q_\bot^ i=q_\bot^1$ or $q_\bot^2$ (no sum over $i$).
To simplify the derivations, we choose to work in the $\vec P_\bot=0$ frame.
After working out traces and making use of Eq. (\ref{eq: kinematics}), we obtain the form factors as shown in Eq.~(\ref{eq:ff type iii}).

\newcommand{\bi}{\bibitem}


\begin{thebibliography}{199}


\bibitem{PDG}
M. Tanabashi et al. (Particle Data Group), 
``Review of Particle Physics,"
Phys. Rev. D 98, 030001 (2018).

\bibitem{Aaij:2017vbw} 
  R.~Aaij {\it et al.} [LHCb Collaboration],
  ``Study of the $D^0 p$ amplitude in $\Lambda_b^0\to D^0 p \pi^-$ decays,''
  JHEP {\bf 1705}, 030 (2017)
  doi:10.1007/JHEP05(2017)030
  [arXiv:1701.07873 [hep-ex]].

\bibitem{Aaij:2017nav} 
  R.~Aaij {\it et al.} [LHCb Collaboration],
  ``Observation of five new narrow $\Omega_c^0$ states decaying to $\Xi_c^+ K^-$,''
  Phys.\ Rev.\ Lett.\  {\bf 118}, no. 18, 182001 (2017)
  doi:10.1103/PhysRevLett.118.182001
  [arXiv:1703.04639 [hep-ex]].

\bibitem{Cheng:2017ove} 
  H.~Y.~Cheng and C.~W.~Chiang,
  ``Quantum numbers of $\Omega_c$ states and other charmed baryons,''
  Phys.\ Rev.\ D {\bf 95}, no. 9, 094018 (2017)
  doi:10.1103/PhysRevD.95.094018
  [arXiv:1704.00396 [hep-ph]].

\bibitem{Chen:2017sci} 
  H.~X.~Chen, Q.~Mao, W.~Chen, A.~Hosaka, X.~Liu and S.~L.~Zhu,
  ``Decay properties of $P$-wave charmed baryons from light-cone QCD sum rules,''
  Phys.\ Rev.\ D {\bf 95}, no. 9, 094008 (2017)
  doi:10.1103/PhysRevD.95.094008
  [arXiv:1703.07703 [hep-ph]].
  
\bibitem{Karliner:2017kfm} 
  M.~Karliner and J.~L.~Rosner,
  ``Very narrow excited $\Omega_c$ baryons,''
  Phys.\ Rev.\ D {\bf 95}, no. 11, 114012 (2017)
  doi:10.1103/PhysRevD.95.114012
  [arXiv:1703.07774 [hep-ph]].
 
\bibitem{Wang:2017hej} 
  K.~L.~Wang, L.~Y.~Xiao, X.~H.~Zhong and Q.~Zhao,
  ``Understanding the newly observed $\Omega_c$ states through their decays,''
  Phys.\ Rev.\ D {\bf 95}, no. 11, 116010 (2017)
  doi:10.1103/PhysRevD.95.116010
  [arXiv:1703.09130 [hep-ph]].
    
\bibitem{Padmanath:2017lng} 
  M.~Padmanath and N.~Mathur,
  ``Quantum Numbers of Recently Discovered $\Omega^{0}_{c}$ Baryons from Lattice QCD,''
  Phys.\ Rev.\ Lett.\  {\bf 119}, no. 4, 042001 (2017)
  doi:10.1103/PhysRevLett.119.042001
  [arXiv:1704.00259 [hep-ph]].

\bibitem{Wang:2017zjw} 
  Z.~G.~Wang,
  ``Analysis of $\Omega _c(3000)$ , $\Omega _c(3050)$ , $\Omega _c(3066)$ , $\Omega _c(3090)$ and $\Omega _c(3119)$ with QCD sum rules,''
  Eur.\ Phys.\ J.\ C {\bf 77}, no. 5, 325 (2017)
  doi:10.1140/epjc/s10052-017-4895-5
  [arXiv:1704.01854 [hep-ph]].

\bibitem{Chen:2017gnu} 
  B.~Chen and X.~Liu,
  ``New $\Omega_c^0$ baryons discovered by LHCb as the members of $1P$ and $2S$ states,''
  Phys.\ Rev.\ D {\bf 96}, no. 9, 094015 (2017)
  doi:10.1103/PhysRevD.96.094015
  [arXiv:1704.02583 [hep-ph]].

\bibitem{Agaev:2017jyt} 
  S.~S.~Agaev, K.~Azizi and H.~Sundu,
  EPL {\bf 118}, no. 6, 61001 (2017)
  doi:10.1209/0295-5075/118/61001
  [arXiv:1703.07091 [hep-ph]].

\bibitem{Aaij:2014jyk} 
  R.~Aaij {\it et al.} [LHCb Collaboration],
  ``Study of the kinematic dependences of $\Lambda_{b}^{0}$ production in pp collisions and a measurement of the $\Lambda_{b}^{0} \to \Lambda_{c}^{+}$ $\pi^{-}$ branching fraction,''
  JHEP {\bf 1408}, 143 (2014)
  doi:10.1007/JHEP08(2014)143
  [arXiv:1405.6842 [hep-ex]].

\bibitem{Aaij:2014lpa} 
  R.~Aaij {\it et al.} [LHCb Collaboration],
  ``Searches for $\Lambda^0_{b}$ and $\Xi^{0}_{b}$ decays to $K^0_{\rm S} p \pi^{-}$ and $K^0_{\rm S}p K^{-}$ final states with first observation of the $\Lambda^0_{b} \rightarrow K^0_{\rm S}p \pi^{-}$ decay,''
  JHEP {\bf 1404}, 087 (2014)
  doi:10.1007/JHEP04(2014)087
  [arXiv:1402.0770 [hep-ex]].

\bibitem{Aaij:2014pha} 
  R.~Aaij {\it et al.} [LHCb Collaboration],
  ``Study of beauty hadron decays into pairs of charm hadrons,''
  Phys.\ Rev.\ Lett.\  {\bf 112}, 202001 (2014)
  doi:10.1103/PhysRevLett.112.202001
  [arXiv:1403.3606 [hep-ex]].
  
      
\bibitem{Cheng:2004cc} 
  H.~Y.~Cheng, C.~K.~Chua and C.~W.~Hwang,
  ``Light front approach for heavy pentaquark transitions,''
  Phys.\ Rev.\ D {\bf 70}, 034007 (2004)
  doi:10.1103/PhysRevD.70.034007
  [hep-ph/0403232].


\bibitem{Mannel:1992ti} 
  T.~Mannel and W.~Roberts,
  ``Nonleptonic Lambda(b) decays at colliders,''
  Z.\ Phys.\ C {\bf 59}, 179 (1993).
  doi:10.1007/BF01555853

\bibitem{Cheng97a} 
H.~Y.~Cheng,
  ``Nonleptonic weak decays of bottom baryons,''
  Phys.\ Rev.\ D {\bf 56}, 2799 (1997)
  doi:10.1103/PhysRevD.56.2799
  [hep-ph/9612223].

\bibitem{Ivanov:1997hi} 
  M.~A.~Ivanov, J.~G.~Korner, V.~E.~Lyubovitskij and A.~G.~Rusetsky,
  ``Exclusive nonleptonic bottom to charm baryon decays including nonfactorizable contributions,''
  Mod.\ Phys.\ Lett.\ A {\bf 13}, 181 (1998)
  doi:10.1142/S0217732398000231
  [hep-ph/9709325].

\bibitem{Ivanov:1997ra} 
  M.~A.~Ivanov, J.~G.~Korner, V.~E.~Lyubovitskij and A.~G.~Rusetsky,
  ``Exclusive nonleptonic decays of bottom and charm baryons in a relativistic three quark model: Evaluation of nonfactorizing diagrams,''
  Phys.\ Rev.\ D {\bf 57}, 5632 (1998)
  doi:10.1103/PhysRevD.57.5632
  [hep-ph/9709372].
 
\bibitem{Giri:1997te} 
  A.~K.~Giri, L.~Maharana and R.~Mohanta,
  ``Two-body nonleptonic $\Lambda_b$ decays with $1/M_Q$ corrections,''
  Mod.\ Phys.\ Lett.\ A {\bf 13}, 23 (1998).
  doi:10.1142/S021773239800005X
 
  
\bibitem{Fayyazuddin:1998ap} 
  Fayyazuddin and Riazuddin,
  ``Two-body nonleptonic $\Lambda_b$ decays in quark model with factorization ansatz,''
  Phys.\ Rev.\ D {\bf 58}, 014016 (1998)
  doi:10.1103/PhysRevD.58.014016
  [hep-ph/9802326].
  
\bibitem{Mohanta:1998iu} 
  R.~Mohanta, A.~K.~Giri, M.~P.~Khanna, M.~Ishida, S.~Ishida and M.~Oda,
  ``Hadronic weak decays of $\Lambda_b$ baryon in the covariant oscillator quark model,''
  Prog.\ Theor.\ Phys.\  {\bf 101}, 959 (1999)
  doi:10.1143/PTP.101.959
  [hep-ph/9904324].
  
\bibitem{Shih:1999yh} 
  H.~H.~Shih, S.~C.~Lee and H.~n.~Li,
  ``Applicability of perturbative QCD to $\Lambda_b\to\Lambda_c$ decays,''
  Phys.\ Rev.\ D {\bf 61}, 114002 (2000)
  doi:10.1103/PhysRevD.61.114002
  [hep-ph/9906370].

\bibitem{Albertus:2004wj} 
  C.~Albertus, E.~Hernandez and J.~Nieves,
  ``Nonrelativistic constituent quark model and HQET combined study of semileptonic decays of Lambda(b) and Xi(b) baryons,''
  Phys.\ Rev.\ D {\bf 71}, 014012 (2005)
  doi:10.1103/PhysRevD.71.014012
  [nucl-th/0412006].
  
\bibitem{Ke:2007tg} 
  H.~W.~Ke, X.~Q.~Li and Z.~T.~Wei,
  ``Diquarks and Lambda(b) $\to$ Lambda(c) weak decays,''
  Phys.\ Rev.\ D {\bf 77}, 014020 (2008)
  doi:10.1103/PhysRevD.77.014020
  [arXiv:0710.1927 [hep-ph]].

\bibitem{Ke:2012wa} 
  H.~W.~Ke, X.~H.~Yuan, X.~Q.~Li, Z.~T.~Wei and Y.~X.~Zhang,
  ``$\Sigma_{b}\to\Sigma_c$ and $\Omega_b\to\Omega_c$ weak decays in the light-front quark model,''
  Phys.\ Rev.\ D {\bf 86}, 114005 (2012)
  doi:10.1103/PhysRevD.86.114005
  [arXiv:1207.3477 [hep-ph]].

\bibitem{Detmold:2015aaa} 
  W.~Detmold, C.~Lehner and S.~Meinel,
  ``$\Lambda_b \to p \ell^- \bar{\nu}_\ell$ and $\Lambda_b \to \Lambda_c \ell^- \bar{\nu}_\ell$ form factors from lattice QCD with relativistic heavy quarks,''
  Phys.\ Rev.\ D {\bf 92}, no. 3, 034503 (2015)
  doi:10.1103/PhysRevD.92.034503
  [arXiv:1503.01421 [hep-lat]].

\bibitem{Zhu:2018jet} 
  J.~Zhu, Z.~T.~Wei and H.~W.~Ke,
  ``The semi-leptonic and non-leptonic weak decays of $\Lambda_b^0$,''
  arXiv:1803.01297 [hep-ph].

\bibitem{Zhao:2018zcb} 
  Z.~X.~Zhao,
  ``Weak decays of heavy baryons in the light-front approach,''
  Chin.\ Phys.\ C {\bf 42}, no. 9, 093101 (2018)
  doi:10.1088/1674-1137/42/9/093101
  [arXiv:1803.02292 [hep-ph]].

  
\bibitem{Gutsche:2018utw} 
  T.~Gutsche, M.~A.~Ivanov, J.~G.~Körner and V.~E.~Lyubovitskij,
  ``Nonleptonic two-body decays of single heavy baryons  $\Lambda_Q$, $\Xi_Q$, and $\Omega_Q$ $(Q=b,c)$ induced by $W$ emission in the covariant confined quark model,''
  Phys.\ Rev.\ D {\bf 98}, no. 7, 074011 (2018)
  doi:10.1103/PhysRevD.98.074011
  [arXiv:1806.11549 [hep-ph]].
  
  
\bibitem{Cheng:2006dk} 
  H.~Y.~Cheng and C.~K.~Chua,
  ``Strong Decays of Charmed Baryons in Heavy Hadron Chiral Perturbation Theory,''
  Phys.\ Rev.\ D {\bf 75}, 014006 (2007)
  doi:10.1103/PhysRevD.75.014006
  [hep-ph/0610283].

\bibitem{Cheng:2015naa} 
  H.~Y.~Cheng and C.~K.~Chua,
  ``Strong Decays of Charmed Baryons in Heavy Hadron Chiral Perturbation Theory: An Update,''
  Phys.\ Rev.\ D {\bf 92}, no. 7, 074014 (2015)
  doi:10.1103/PhysRevD.92.074014
  [arXiv:1508.05653 [hep-ph]].


\bibitem{Aaij:2018yqz} 
  R.~Aaij {\it et al.} [LHCb Collaboration],
  ``Observation of a new $\Xi_b^-$ resonance,''
  Phys.\ Rev.\ Lett.\  {\bf 121}, 072002 (2018)
  doi:10.1103/PhysRevLett.121.072002
  [arXiv:1805.09418 [hep-ex]].



 \bibitem{CCH}
 H.~Y.~Cheng, C.~K.~Chua and C.~W.~Hwang,
  ``Covariant light front approach for s wave and p wave mesons: Its application to decay constants and form-factors,''
  Phys.\ Rev.\ D {\bf 69}, 074025 (2004)
  doi:10.1103/PhysRevD.69.074025
  [hep-ph/0310359].


 \bibitem{Cheng97} 
        H.~Y.~Cheng, C.~Y.~Cheung and C.~W.~Hwang,
  ``Mesonic form-factors and the Isgur-Wise function on the light front,''
  Phys.\ Rev.\ D {\bf 55}, 1559 (1997)
  doi:10.1103/PhysRevD.55.1559
  [hep-ph/9607332].

 \bibitem{Jaus90} 
 W.~Jaus,
  ``Semileptonic Decays of B and D Mesons in the Light Front Formalism,''
  Phys.\ Rev.\ D {\bf 41}, 3394 (1990).
  doi:10.1103/PhysRevD.41.3394

\bibitem{deAraujo:1999ugw} 
  W.~R.~B.~de Araujo, M.~Beyer, T.~Frederico and H.~J.~Weber,
  ``Feynman versus Bakamjian-Thomas in light front dynamics,''
  J.\ Phys.\ G {\bf 25}, 1589 (1999)
  doi:10.1088/0954-3899/25/8/303
  [hep-ph/9904307].

  \bibitem{Brodsky:1997de}
  S.~J.~Brodsky, H.~C.~Pauli, and S.~S.~Pinsky,
  ``Quantum chromodynamics and other field theories on the light cone,''
  Phys.\ Rept.\  {\bf 301}, 299 (1998).

\bibitem{Wang:2017mqp} 
  W.~Wang, F.~S.~Yu and Z.~X.~Zhao,
  ``Weak decays of doubly heavy baryons: the $1/2\rightarrow 1/2$ case,''
  Eur.\ Phys.\ J.\ C {\bf 77}, no. 11, 781 (2017)
  doi:10.1140/epjc/s10052-017-5360-1
  [arXiv:1707.02834 [hep-ph]].


 \bibitem{Jaus91}
     W.~Jaus,
     ``Relativistic Constituent Quark Model Of Electroweak Properties Of Light Mesons,''
     Phys.\ Rev.\ D {\bf 44}, 2851 (1991).




  \bibitem{Schlumpf}
 F.~Schlumpf,
 ``Relativistic constituent quark model of electroweak properties of baryons,''
 Phys.\ Rev.\ D {\bf 47}, 4114 (1993); {\bf 49},
 6246(E) (1994)].


\bibitem{Ebert:2010af} 
  D.~Ebert, R.~N.~Faustov and V.~O.~Galkin,
  ``Masses of tetraquarks with open charm and bottom,''
  Phys.\ Lett.\ B {\bf 696}, 241 (2011)
  doi:10.1016/j.physletb.2010.12.033
  [arXiv:1011.2677 [hep-ph]].

\bibitem{Jaus96}
       W.~Jaus,
       ``Semileptonic, Radiative, And Pionic Decays of $B$, $B^*$ and $D$, $D^*$ Mesons,''
       Phys.\ Rev.\ D {\bf 53}, 1349 (1996); {\bf 54},
       5904(E) (1996).
         
\bibitem{CC2004}
H.~Y.~Cheng and C.~K.~Chua,
  ``Covariant light front approach for $B \to K^* \gamma, K_1 \gamma, K^*_2 \gamma$ decays,''
  Phys.\ Rev.\ D {\bf 69}, 094007 (2004)
  Erratum: [Phys.\ Rev.\ D {\bf 81}, 059901 (2010)]
  doi:10.1103/PhysRevD.69.094007, 10.1103/PhysRevD.81.059901
  [hep-ph/0401141].




\bibitem{BBNS} 
  M.~Beneke, G.~Buchalla, M.~Neubert and C.~T.~Sachrajda,
  ``QCD factorization in $B \to \pi K, \pi \pi$ decays and extraction of Wolfenstein parameters,''
  Nucl.\ Phys.\ B {\bf 606}, 245 (2001)
  doi:10.1016/S0550-3213(01)00251-6
  [hep-ph/0104110].

\bibitem{Beneke:2000ry} 
  M.~Beneke, G.~Buchalla, M.~Neubert and C.~T.~Sachrajda,
  ``QCD factorization for exclusive, nonleptonic B meson decays: General arguments and the case of heavy light final states,''
  Nucl.\ Phys.\ B {\bf 591}, 313 (2000)
  doi:10.1016/S0550-3213(00)00559-9
  [hep-ph/0006124].


\bibitem{Cheng:2018hwl} 
  H.~Y.~Cheng, X.~W.~Kang and F.~Xu,
  ``Singly Cabibbo-suppressed hadronic decays of $\Lambda_c^+$,''
  Phys.\ Rev.\ D {\bf 97}, no. 7, 074028 (2018)
  doi:10.1103/PhysRevD.97.074028
  [arXiv:1801.08625 [hep-ph]].

\bibitem{ckmfitter}
J.~Charles {\it et al.} [CKMfitter Group],
  ``CP violation and the CKM matrix: Assessing the impact of the asymmetric $B$ factories,''
  Eur.\ Phys.\ J.\ C {\bf 41}, no. 1, 1 (2005)
  doi:10.1140/epjc/s2005-02169-1
  [hep-ph/0406184];
  updated results available at: http://ckmfitter.in2p3.fr.
  

  
\end{thebibliography}
\end{document}